\documentclass[11pt,a4paper]{article}
\pdfoutput=1
\usepackage{jheppub}
\usepackage{latexsym,amsfonts,amsmath,amssymb}
\usepackage{bbm}
\usepackage{empheq}
\usepackage{graphicx}
\usepackage{color}
\usepackage[normalem]{ulem}
\usepackage{comment}
\usepackage{url}
\usepackage{slashed}
\usepackage{tabu}
\usepackage{multirow}
\usepackage{extarrows}
\usepackage{amsmath}
\usepackage{mathrsfs}
\usepackage{bm}
\usepackage{stmaryrd}
\usepackage[dvipsnames]{xcolor}
\usepackage[utf8]{inputenc}
\usepackage[T1]{fontenc}
\usepackage{slashed}
\usepackage{upgreek}
\usepackage{orcidlink}

\usepackage[super]{nth}
\usepackage{mathbbol}

\usepackage{mathtools}
\DeclarePairedDelimiter{\ceil}{\lceil}{\rceil}
\DeclarePairedDelimiter\floor{\lfloor}{\rfloor}

\usepackage{amsxtra,graphics,epsfig,bm,tikz,xfrac,lscape}
\usetikzlibrary{decorations.pathmorphing}
\usetikzlibrary{decorations.markings}
\usetikzlibrary{arrows, decorations.markings, calc, fadings, decorations.pathreplacing, patterns, decorations.pathmorphing, positioning}

\usetikzlibrary{positioning,shapes}
\usetikzlibrary{chains}
\usetikzlibrary{arrows,fit,decorations.pathreplacing}
\tikzstyle{every picture}+=[remember picture]
\tikzstyle{na} = [baseline=-.5ex]

\definecolor{islamicgreen}{rgb}{0.0, 0.56, 0.0}

\newcommand\cC{{\cal C}}

\newcommand\cN{{\cal N}}

\newcommand\cV{{\cal V}}

\newcommand{\Tr}{\mbox{Tr}}

\newcommand{\half}{\frac{1}{2}}

\def\i{\mathrm{i}}

\def\beqa{\begin{eqnarray}}
\def\eeqa{\end{eqnarray}}
\def\be{\begin{equation}}
\def\ee{\end{equation}}
\def\bse{\begin{subequations}}
\def\ese{\end{subequations}}

\newcommand{\bem}{\begin{pmatrix}}
\newcommand{\eem}{\end{pmatrix}}

\renewcommand{\=}{\;  = \;}
\def\+{\, + \,}

\def\wt{\widetilde}

\def\bar{\overline}
\def\ads2{$AdS_2$}
\def\ss2{S$^2$}

\def\rt2{\sqrt{2}}

\newcommand\qeq{Q_{eq}}

\def\eps{\epsilon}

\def\r{ r}

\def\S{{\sigma}}
\def\P{{\rho}}

\newcommand{\ba}{\begin{array}}
\newcommand{\ea}{\end{array}}

\newcommand\redsout{\bgroup\markoverwith{\textcolor{red}{\rule[0.5ex]{2pt}{0.4pt}}}\ULon}

\title{Localisation of $\mathcal{N} = (2,2)$ theories on spindles of both twists}

\author{Imtak Jeon\orcidlink{0000-0001-5249-3538}$^{a}$,} 
\author{Hyojoong Kim\orcidlink{0000-0002-5838-6853}$^b$,}
\author{Nakwoo Kim\orcidlink{0000-0002-4182-8459}$^c$\,,}
\author{Aaron Poole\orcidlink{0000-0002-4966-4874}$^c$}
\author{and Augniva Ray\orcidlink{0000-0003-1880-9193}$^b$}
\affiliation{$^a$ School of Science, Huzhou Normal University, Huzhou 313000, Zhejiang, China }
\affiliation{$^b$ Department of Physics and Astronomy \& Center for Theoretical Physics, Seoul National University, Seoul 08826, Korea }
\affiliation{$^c$ Department of Physics and Research Institute of Basic Science, Kyung Hee University, 26 Kyungheedae-ro, Seoul 02447, Korea}
\emailAdd{imtakjeon@gmail.com}
\emailAdd{hjkim1996@gmail.com}
\emailAdd{nkim@khu.ac.kr}
\emailAdd{apoole@khu.ac.kr}
\emailAdd{augniva@gmail.com}

\abstract{We consider two-dimensional $\mathcal{N}=(2,2)$ supersymmetric field theories living on a spindle $\mathbb{WCP}_{[n_1,n_2]}^1$. Starting from the spindle solutions of five-dimensional STU gauged supergravity, we construct theories on a spindle which preserve supersymmetry via either the twist or anti-twist mechanism and admit two Killing spinors of opposite R-charge. While the study of field theories on anti-twisted spindles has already been undertaken in some detail, the advantage of our approach allows for the derivation of analogous results in the twist case. We apply the technique of supersymmetric localisation to compute the exact partition function for a theory consisting of an abelian vector multiplet and a charged chiral multiplet in the presence of a Fayet-Iliopoulos term. We compare and contrast the results for the twisted and anti-twisted spindle and find a general formula which encompasses the partition function for both cases simultaneously.}

\begin{document}

\maketitle

\section{Introduction and summary of results}

Supersymmetric localisation \cite{Duistermaat:1982vw, Atiyah:1984px, Witten:1988ze, Witten:1988xj, Nekrasov:2002qd, Pestun:2007rz} is a powerful tool to compute observables for quantum field theories, including those defined on curved manifolds \cite{Festuccia:2011ws} of various dimensions. Such techniques have also been applied extensively to study theories in two dimensions \cite{Benini:2012ui, Doroud:2012xw, Benini:2013nda, Honda:2013uca, Hori:2013ika, Benini:2013xpa, Closset:2015rna, Benini:2016qnm, Ohta:2019odi, GonzalezLezcano:2023cuh, Leeb-Lundberg:2023jsj}, with deep insights being made into their connection with string theory \cite{Witten:1993yc} and the understanding of supersymmetric dualities \cite{Hori:2000kt, Hori:2002fa, Ueda:2016wfa, Kim:2016jye, Closset:2017vvl}. From the mathematical perspective, localisation provides an efficient method to compute invariants of interest \cite{Gromov, Witten:1988xj, Jockers:2012dk, Gomis:2012wy, Bonelli:2013mma}. 

Inspired by the discovery of several new supergravity solutions corresponding to wrapping branes on \textit{orbifolds} \cite{Ferrero:2020laf, Ferrero:2020twa, Hosseini:2021fge, Boido:2021szx, Bah:2021mzw, Ferrero:2021wvk, Bah:2021hei, Ferrero:2021ovq, Couzens:2021rlk, Faedo:2021nub, Ferrero:2021etw, Suh:2022olh, Arav:2022lzo, Couzens:2022yiv, Couzens:2022lvg, Faedo:2022rqx, Suh:2022pkg, Suh:2023xse, Amariti:2023mpg, Hristov:2023rel, Amariti:2023gcx, Faedo:2024upq, Boisvert:2024jrl, Ferrero:2024vmz, Hristov:2024qiy, Suh:2024fru, Crisafio:2024fyc, Suh:2025cii, Conti:2025rfd}, there has been recent interest in applying localisation techniques to field theories defined on orbifold backgrounds \cite{Inglese:2023wky, Inglese:2023tyc, Pittelli:2024ugf, Mauch:2024uyt, Ruggeri:2025kmk, Mauch:2025irx, Jeon:2025rfc}. Such work has allowed for precision checks of the AdS/CFT correspondence and provided a microscopic derivation of four-dimensional accelerating black hole entropy via the dual field theory \cite{Colombo:2024mts}. 

The most frequently occuring orbifold in the aforementioned solutions is the so called \textit{spindle}, $\mathbb{\Sigma} \cong \mathbb{WCP}^1_{[n_1,n_2]}$, a space parameterised by coprime integers $n_{1,2}$\footnote{See however \cite{Arav:2025jee, Arav:2026unc} for recent constructions of spindle solutions with $\text{gcd}(n_1,n_2) \neq 1$.}. This background shares many features in common with the sphere $S^2$, although now the smooth structure at the poles is replaced by conical singularities with deficit angles $2\pi(1-1/n_{1,2})$, making the local structure at the poles homeomorphic to $\mathbb{C}/\mathbb{Z}_{n_{1,2}}$ respectively. Despite the appearance of such singularities, spindles may possess well-behaved Killing spinors \cite{Ferrero:2021etw} and thus serve as suitable backgrounds upon which to define SQFTs. 

A fascinating result found in the early literature concerning spindle solutions is that they preserve supersymmetry while admitting Killing spinors which are neither constant or chiral on $\mathbb{\Sigma}$ \cite{Ferrero:2020laf}. In fact, as the general analysis of \cite{Ferrero:2021etw} shows, a necessary condition for the realisation of supersymmetry on spindles is that the R-symmetry flux must satisfy one of the following quantisation conditions 
\begin{equation}
        \frac{1}{2\pi} \int_{\mathbb{\Sigma}} \textrm{d} A^{\text{R}} = \frac{1}{2} \cdot \frac{\eta n_2 - n_1}{n_1 n_2} = \eta \frac{\chi_{-\eta}}{2}\,,
\end{equation}
where we introduced the sign parameter $\eta$
\begin{equation} \label{eq: sign_parameter}
    \eta = \begin{cases} -1\,, \qquad \text{twist}\,, \\
    1 \,, \qquad \text{anti-twist}\,,
    \end{cases}
\end{equation}
and we note that the twist case has an R-symmetry flux proportional to the (orbifold) Euler characteristic $\chi = \chi_{+1} = \chi_+$ of the spindle. This feature relates this case to the well-known ``topological twist'' \cite{Maldacena:2000mw}, although we note that in the twist case considered above, the R-symmetry gauge field does not necessarily pointwise cancel the spin connection in the Killing spinor equation, and thus the Killing spinors will not be constant.

In this work we will consider two-dimensional $\mathcal{N} = (2,2)$ theories on spindles of both twist and anti-twist, consisting of a single vector multiplet and a single charged chiral multiplet. This setup extends the scope of our previous work  \cite{Jeon:2025rfc} which focused on anti-twist only. We will apply supersymmetric localisation to compute observables of these theories, with our main result being a unified form of the partition function for the two cases \eqref{eq: sign_parameter}, which for unit gauge charge reads
\begin{align}
\begin{split} \label{eq: full_partition_both_twists}
    Z^{\eta}_{\mathbb{\Sigma}}  & = \frac{L}{L_0} \textrm{e}^{-\frac{r}{2} \left( \frac{2\pi \xi - \i \eta \theta}{n_1} +  \frac{2\pi \xi + \i  \theta}{n_2} \right)} \textrm{e}^{-\i (1+\eta) \frac{L}{L_0} \textrm{e}^{-2\pi \xi} \sin \theta} \\
  &  \qquad \times  \left( \frac{1-\textrm{e}^{-(2\pi \xi - \i \eta \theta)}}{1-\textrm{e}^{-(2\pi \xi - \i \eta \theta)/n_1}} \right)\left( \frac{1-\textrm{e}^{-(2\pi \xi + \i \theta)}}{1-\textrm{e}^{-(2\pi \xi + \i \theta)/n_2}} \right)\,,
  \end{split}
\end{align}
where $\xi, \theta$ are respectively the Fayet-Iliopoulos and topological parameters of the theory and $L/L_0$ is the ratio of the spindle size relative to a reference length scale.
For the anti-twist case of $\eta = 1$, the formula above gives the result already derived in \cite{Jeon:2025rfc}. The twist case of $\eta =-1$ is a new result, which we will derive in explicit detail in this work. 

In order to derive this result we will make use of the spindle solutions of $D=5$ STU gauged supergravity \cite{Ferrero:2020laf, Hosseini:2021fge, Boido:2021szx, Ferrero:2021etw}. As was first shown in \cite{Ferrero:2021etw}, this theory admits spindle solutions of both types of twist, and thus we can use these solutions as a tool to construct two-dimensional $\mathcal{N} = (2,2)$ theories defined on spindles of either twist. We note that \eqref{eq: full_partition_both_twists} does not depend on the explicit background derived from the STU model \cite{Festuccia:2011ws, Closset:2014pda}, but rather we employ this as an explicit choice of metric and R-symmetry gauge field on the spindle, making the localisation calculations more tractable in the process.

The structure of the paper is as follows: in section \ref{sec: STU_solutions} we review the spindle solutions of STU gauged supergravity, following closely the style of \cite{Ferrero:2021etw}. In section \ref{sec: 2d_2,2} we use the aforementioned solutions to construct supersymmetric theories on spindles of both twists by solving the two-dimensional Killing spinor equations, as well as giving explicit details of the field content and classical action for the theories of interest. In section \ref{sec: BPS_locus} we compute the localisation locus for the theory by solving the vector and chiral multiplet BPS equations, as well as evaluating the classical contributions to the partition function on the locus. In section \ref{sec: one_loop_determinants} we compute the one-loop determinant contributions to the partition functions using both the method of unpaired eigenvalues and the (orbifold) fixed point theorem, finding exact agreement between the two approaches. In section \ref{sec: partition_function} we put all of the previous pieces together and perform the computation of the full partition function via integration over the surviving vector multiplet moduli on the BPS locus, before comparing the results for the twist and anti-twist cases. We conclude and discuss further directions in section \ref{sec: Conclusions}. We also present several appendices where we present our conventions, provide comparison to related works in the present literature, and give several technical details relating to the calculations in the main text.
 
\section{Spindles from \texorpdfstring{$D=5$}{D=5} STU gauged supergravity} \label{sec: STU_solutions}

\subsection{\texorpdfstring{$D=5$}{D=5} gauged supergravity model}

We consider the $D=5$ STU, $U(1)^3$ gauged supergravity model as considered in \cite{Ferrero:2021etw} although we alter a few aspects of the presentation, mainly via the inclusion of the cosmological constant $\Lambda = -6/L^2$, where $L$ is the radius of the five-dimensional anti-de Sitter spacetime (\cite{Ferrero:2021etw} uses $L=1$). Working with a general $L$ will be useful when we dimensionally reduce down to the two-dimensional theory, as we will reinterpret this length scale as an overall scaling for the spindle. 

The Lagrangian of the $D=5$ STU, $U(1)^3$ theory is (see \textit{e.g.} \cite{Behrndt:1998ns})
\begin{align}
\begin{split} \label{eq: STU_Lagrangian}
    \mathcal{L}^{\text{STU}} = \sqrt{-g} &\left[  R - \frac{\mathcal{V}}{L^2} - \frac{1}{2}\sum_{I =1}^{3} \big(X^{(I)}\big)^{-2} \big( \partial X^{(I)} \big)^2 - \sum_{I =1}^3 \big( X^{(I)} \big)^{-2} \big( F^{(I)} \big)^2   \right] \\
    & \quad  - 8 F^{(1)} \wedge F^{(2)} \wedge A^{(3)} \,,
    \end{split}
\end{align}
where $A^{(I)}$ are $U(1)$ gauge fields, $I = 1,2,3,$ with field strengths $F^{(I)} = \textrm{d} A^{(I)}$. $X^{(I)} >0$ are scalar fields satisfying $X^{(1)} X^{(2)} X^{(3)} = 1$ and can be written in terms of canonically normalised scalars $\varphi_{1,2}$ as 
\begin{equation}
    X^{(1)} =  \textrm{e}^{-\frac{\varphi_1}{\sqrt{6}}-\frac{\varphi_2}{\sqrt{2}}}\,, \quad X^{(2)} = \textrm{e}^{-\frac{\varphi_1}{\sqrt{6}}+\frac{\varphi_2}{\sqrt{2}}}\,,\quad X^{(3)} = \textrm{e}^{\frac{2 \varphi_1}{\sqrt{6}}}\,,
\end{equation}
 and the potential is 
\begin{equation}
    \mathcal{V} = - 4 \sum_{I =1}^3 \big( X^{(I)} \big)^{-1}.
\end{equation}

In order for a bosonic solution of the theory to preserve supersymmetry we require a solution to the Killing spinor equations 
\begin{align} 
&\left[\nabla_M-\dfrac{\i}{L}  \sum_{I=1}^{3}A_{M}^{(I)}-\dfrac{1}{6L}\sum_{I=1}^3 X^{(I)}\Gamma_M
-\dfrac{\i}{12 }\sum_{I=1}^3\left(X^{(I)}\right)^{-1}\left(\Gamma_{M}^{\phantom{M}N P}-4\delta_{M}^{N}\Gamma^{P}\right)F_{NP}^{(I)} \right] \varepsilon=0\,,\\
&   \left[  \slashed{\partial} \varphi_i + \frac{2}{L} \sum_{I=1}^{3} \partial_{\varphi_i} X^{(I)} - \i \sum_{I=1}^3 \partial_{\varphi_i} (X^{(I)})^{-1} F_{M N}^{(I)} \Gamma^{M N} \right] \varepsilon = 0\,,
\end{align}
 which arise from demanding the vanishing of the gravitino and gaugino variations of $D=5$, $\mathcal{N} =2$ gauged supergravity coupled to two vector multiplets. $\Gamma^{\mu}$ are $D=5$ gamma matrices satisfying $\{ \Gamma_{\mu}, \Gamma_{\nu}\} = 2 g_{\mu \nu}$, the Levi-Civita connection is $\nabla_{\mu} = \partial_{\mu} + \frac{1}{4} \omega_{\mu}^{ab} \Gamma_{ab}$, and $\varepsilon$ is a Dirac spinor. The R-symmetry gauge field is then defined as 
\begin{equation} \label{eq: A^R}
    A^{\text{R}} \equiv \sum_{I=1}^3 A^{(I)}\,,
\end{equation}
which will be a crucial ingredient in constructing an $\mathcal{N} = (2,2)$ theory on a spindle.

Finally, we note that upon setting 
\begin{equation} \label{eq: stu_to_minimal}
    A^{(1)} = A^{(2)} = A^{(3)} = \frac{1}{3} A^{\text{R}}\,, \qquad X^{(1)} = X^{(2)} =X^{(3)} =1\,,
\end{equation}
we recover the theory of $D=5$, $U(1)$ minimal gauged supergravity \cite{Gunaydin:1983bi}
\begin{align} \label{eq: minimal_gauged_lagrangian}
    \begin{split}
    \mathcal{L}^{\text{min}} = \sqrt{-g} &\left[  R - \frac{12}{L^2} - \frac{1}{3} \big( F^{\text{R}} \big)^2   \right]  - \frac{8}{9} F^{\text{R}} \wedge F^{\text{R}} \wedge A^{\text{R}} \,,
    \end{split}
\end{align}
with gravitino Killing spinor equation
\begin{align} 
\left[\nabla_M-\dfrac{\i}{L} A^{\text{R}} -\dfrac{1}{2L} \Gamma_M
-\dfrac{\i}{12 }\left(\Gamma_{M}^{\phantom{M}N P}-4\delta_{M}^{N}\Gamma^{P}\right)F_{NP}^{\text{R}} \right] \varepsilon=0\,.
\end{align}
Spindle solutions of this theory were first constructed in \cite{Ferrero:2020laf} and were used to provide the metric and $U(1)_{\text{R}}$ symmetry gauge field for two-dimensional $\mathcal{N}=(2,2)$ theory on $\mathbb{\Sigma}$ in \cite{Jeon:2025rfc}. We will now discuss the analogous spindle solutions to the full STU theory \eqref{eq: STU_Lagrangian}, including an analysis of the advantages that these have when used to construct field theories on $\mathbb{\Sigma}$.

\subsection{The spindle solutions}

 We consider the spindle solutions of $D=5$ STU gauged supergravity, following closely the presentation in \cite{Ferrero:2021etw} (see also \cite{Boido:2021szx}). The local form of the solutions is 
\begin{align} 
    \textrm{d}s_5^2 & = L^2 H^{1/3} \left[ \textrm{d}s_{\textrm{AdS}_3}^2 + \frac{1}{4P}  \textrm{d}y^2 + \frac{P}{H} \textrm{d}z^2 \right] = L^2 H^{1/3} \textrm{d}s_{\textrm{AdS}_3}^2 + \textrm{d}s_{\mathbb{\Sigma}}^2 \label{eq: 5d_spindle_metric} \,, \\ \label{eq: 5d_gauge_field}
    A^{(I)} & = \frac{L}{2} \left(\frac{y}{h_I} - \frac{2}{3}\right) \textrm{d} z\,, \qquad X^{(I)} = \frac{H^{1/3}}{h_I}\,,
\end{align}
where $\textrm{d}s^2_{\textrm{AdS}_3}$ is a unit radius $\textrm{AdS}_3$ metric and $h_I$, $H$ and $P$ are functions of $y$ given by 
\begin{equation} \label{eq: P-H_defs}
    h_I = y+ q_I\,, \qquad H = \prod_{I=1}^3 h_I = h_1 h_2 h_3\,, \qquad P = H -y^2\,,
\end{equation}
with $q_I$ being constants. We note the field strength can be written as 
\begin{equation} \label{eq: field_strength_volume_relation}
    \big( X^{(I)} \big)^{-2}  F^{(I)} = \frac{1}{L} \frac{q_I}{H^{1/2}} \textrm{vol}_{\mathbb{\Sigma}} \quad \textrm{for}~ I=1,2,3 \,,
\end{equation}
and note from \eqref{eq: A^R} and \eqref{eq: 5d_gauge_field} that the R-symmetry gauge field is 
\begin{equation} \label{eq: 5d_R_gauge_field}
    A^{\text{R}} = \sum_{I=1}^3 A^{(I)} = \frac{L}{2} \left[ \sum_{I=1}^3  \left(\frac{y}{h_I}\right) - 2 \right] \textrm{d} z\,.
\end{equation}   

Now we recall the results from \cite{Ferrero:2021etw} which allow the metric $\textrm{d}s^2_{\mathbb{\Sigma}}$ appearing in \eqref{eq: 5d_spindle_metric} to describe a compact spindle. For a positive-definite metric, we want both $P,H>0$  and note that $P>0$ automatically gives $H>0$ by \eqref{eq: P-H_defs}. Since $X^{(I)}>0$ we must have $h_I>0$ by \eqref{eq: 5d_gauge_field}. We assume $P$ has three real roots, $y_1<y_2<y_3$ and focus on the interval $y \in [y_1, y_2]$, for which $P \geq 0$. Expanding near the boundaries of this interval and setting $\varrho_i = 2|y-y_i|^{1/2}$ for $i=1,2$ we have 
\begin{equation} \label{eq: metric_near_poles}
    \textrm{d}s^2_{\mathbb{\Sigma}} \simeq L^2 \frac{|y_i|^{1/3}}{4|P'(y_i)|} \big( \textrm{d}\varrho_i^2 + \kappa_i^2 \varrho_i^2 \textrm{d}z^2 \big) \,, \quad \textrm{where} \quad \kappa_i = \left| \frac{P'(y_i)}{y_i} \right|\,, \quad i = 1,2\,.
\end{equation}
Now following \cite{Ferrero:2021etw}, one can remove the absolute value in $\kappa_i$ via use of the signs 
\begin{equation}
    \eta_1 y_1 < 0 \, , \quad \eta_2y_2 > 0\,,
\end{equation}
where as shown in \cite{Ferrero:2021etw}, only two of the three possible cases are realised
\begin{align}
    \text{Case A}: \quad (\eta_1,\eta_2) = (-1, +1)\,, \label{eq: case_A} \\
     \text{Case B}: \quad (\eta_1,\eta_2) = (+1, +1)\,, \label{eq: case_B}
\end{align}
which correspond to both roots being positive ($0<y_1<y_2$) and one positive and one negative ($y_1<0<y_2$) respectively. We can thus write 
\begin{equation} \label{eq: kappa_no_mod}
    \kappa_i = - \frac{\eta_i P'(y_i)}{y_i}\,,
\end{equation}
and note that the metric $\textrm{d}s^2_{\mathbb{\Sigma}}$ is that of a spindle when one imposes the periodicity $z \sim z + \Delta z$, where 
\begin{equation}
    \Delta z = \frac{2\pi}{\kappa_1 n_1} = \frac{2\pi}{\kappa_2 n_2}\,,
\end{equation}
with $n_{1,2} \in \mathbb{N}$, $\textrm{gcd}(n_1,n_2) =1$. We note that in both cases A \& B we have $\eta_2 = +1$, so to avoid clutter, we can just keep $\eta_1 =\eta$ as the parameter which determines which case we are in.

Following \cite{Ferrero:2021etw, Boido:2021szx}, we demand that the gauge field fluxes on the spindle are quantised. Our conventions for the gauge fields are such that the $D=5$ spinors carry charge $1$  with respect to each of the gauge fields and thus we demand
\begin{align}\label{eq:defp}
\begin{split}
    Q_I & \equiv \frac{1}{2\pi L} \int_{\mathbb{\Sigma}} F^{(I)} = \frac{1}{2} \left( \frac{y_2}{h_I(y_2)} - \frac{y_1}{h_I(y_1)} \right) \frac{\Delta z}{2\pi} \,, \\
    & \equiv \frac{1}{2} \frac{p_I}{n_1 n_2} \,, \qquad p_I \in \mathbb{Z}\,,
    \end{split}
\end{align} where the second line imposes the charge quantisation condition via the integer-valued $p_I$.
One can now solve for all quantities $y_{i}$, $q_{I}$ and $\Delta z$ in terms of the ``spindle data'' $n_{i}, \eta$ and the quantised fluxes $p_{I}$. Before doing this, we will first derive some simple topological results for the spindle. 

We start with the Euler number $\chi(\mathbb{\Sigma})$. Using the metric
\begin{equation} \label{eq: metric_spindle}
    \textrm{d}s^2_{\mathbb{\Sigma}} = L^2 H^{1/3} \left(\frac{1}{4P} \textrm{d}y^2 + \frac{P}{H} \textrm{d}z^2\right)\,,
\end{equation}
the Ricci scalar can conveniently be expressed as a total derivative
\begin{equation}
\sqrt{g}_{\mathbb{\Sigma}} R_{\mathbb{\Sigma}} = \left( \frac{4 P H' -6HP'}{3H^{3/2}} \right)'\,,
\end{equation} with primes denoting derivative with respect to $y$.
Thus the Euler number is given by 
\begin{equation}
    \chi(\mathbb{\Sigma}) = \frac{1}{4\pi} \int_{\mathbb{\Sigma}} \sqrt{g}_{\mathbb{\Sigma}} R_{\mathbb{\Sigma}} \, \textrm{d}y \textrm{d}z = \frac{\Delta z}{2\pi} \left( \frac{P'(y_1)}{|y_1|} - \frac{P'(y_2)}{|y_2|} \right) = \frac{n_1+n_2}{n_1 n_2}\,,
\end{equation}
where we note that despite our metric on $\mathbb{\Sigma}$ being a conformal rescaling of the one considered in \cite{Ferrero:2021etw}, we still have the same Euler number $\chi(\mathbb{\Sigma})$. This should of course be the case as a Weyl transformation does not alter the underlying topology. As in \cite{Ferrero:2021etw}, we can equivalently compute the Euler number using the spin connection via the following choice of orthonormal frame on the spindle 
\begin{equation}
    \textrm{e}^1 = L \frac{ H^{1/6}}{2 P^{1/2}} \textrm{d}y\,, \qquad \textrm{e}^2 = L \frac{P^{1/2}}{H^{1/3}} \textrm{d}z\,, 
\end{equation}
from which one can compute the spin connection 
\begin{equation} \label{eq: spindle_spin_connection}
    \omega^{12}_{\mathbb{\Sigma}} = \frac{2 P H' - 3H P'}{3H^{3/2}}  \textrm{d}z\,,
\end{equation}
which evaluates at the roots to 
\begin{equation}
    \omega_{\mathbb{\Sigma}}^{12}(y_1) = - \frac{2\pi}{\Delta z} \frac{\textrm{d}z}{n_1}\,, \qquad \omega_{\mathbb{\Sigma}}^{12}(y_2) = \frac{2\pi}{\Delta z} \frac{\textrm{d}z}{n_2}\,,
\end{equation}
and from which one can compute
\begin{equation} \label{eq: Euler_spin_connection}
    \chi(\mathbb{\Sigma}) = \frac{1}{2\pi} \int_{\mathbb{\Sigma}} \textrm{d} \omega^{12}_{\mathbb{\Sigma}} = \frac{n_1 + n_2}{n_1 n_2}\,,
\end{equation}
in precise agreement with the Ricci scalar integral. Notice again that the spin connection \eqref{eq: spindle_spin_connection} differs from that of \cite{Ferrero:2021etw}, but the topological number \eqref{eq: Euler_spin_connection} is unchanged. 

Now we consider the total R-symmetry flux $Q^{\text{R}} \equiv Q_1 + Q_2 + Q_3$. One can follow the arguments given in \cite{Ferrero:2021etw} in order to derive a simple expression for this quantity. First one may use the definitions \eqref{eq: P-H_defs} and the fact that $P(y_i)=0$ in order to show   
\begin{align}\label{eq:identities}
\begin{split}
    P^{'} & = h_1h_2 + h_2h_3+ h_3h_1 - 2y\,, \\
 y^2_i & = h_1(y_i)h_2(y_i)h_{3}(y_i)\,, \qquad  i=1,2\,, 
 \end{split}
\end{align}
and then applying these results to \eqref{eq: kappa_no_mod} we obtain 
\begin{equation}
    \frac{\eta_i}{n_i} = \frac{\Delta z}{2\pi} \left(2- y_i \sum_I \frac{1}{h_I (y_i)}\right) = -\left. \frac{\Delta z}{2\pi} \frac{2}{L} A^{\text{R}} \right|_{y_i} \,, \qquad i=1,2\,,
\end{equation}
and thus one can immediately compute
\begin{equation} \label{eq: Q^R}
    Q^{\text{R}} \equiv \frac{1}{2\pi L}  \int_{\mathbb{\Sigma}} \textrm{d} A^{\text{R}} = \frac{1}{2}\cdot \frac{\eta n_2 - n_1}{n_1 n_2} \quad \Leftrightarrow \quad p_1 + p_2 + p_3 = \eta n_2 - n_1\,,
\end{equation}
with $p_I$ defined in \eqref{eq:defp}. We see that case A \eqref{eq: case_A} is the twist case, and case B \eqref{eq: case_B} is the anti-twist.\footnote{Our sign convention for the total flux is chosen to match that of section 3.2 of \cite{Ferrero:2021etw} (see equation (3.21) there). We could equally have chosen the opposite overall sign, which at the two-dimensional level is equivalent to exchanging $(\epsilon, A, \mathcal{H},\mathcal{G}) \leftrightarrow (\widetilde{\epsilon}, -A, \mathcal{H}, -\mathcal{G})$ in \eqref{kse}, leaving the system of equations unchanged. For more discussion on this point, see p17 of \cite{Ferrero:2021etw}.} 

One may express the values of the parameters $(q_I, y_i, \Delta z)$ entirely in terms of the flux and spindle data $(p_I, n_i, \eta)$. Adapting the results of \cite{Ferrero:2021etw} to our notation, we first have the constants $q_I$ given by
\begin{align}
\begin{split}
    q_1 & = \frac{8}{s^3} p_2 p_3 (n_1 + p_2)(n_1 + p_3)(n_2 - \eta p_2)(n_2 -\eta p_3) \,, \\
    q_2 & = \left. q_1 \right|_{p_1 \leftrightarrow p_2}\,,\\
    q_3 & = \left. q_1 \right|_{p_1 \leftrightarrow p_3} \,, 
    \end{split}
\end{align}
where 
\begin{equation}
    s \equiv n_1^2 + n_2^2 - (p_1^2 + p_2^2 +p_3^2)\,.  
\end{equation}
Next, the roots $y_{i}$ are given by 
\begin{align}
\begin{split} \label{eq: y_spindle_data}
    y_1 & = - \frac{8}{s^3} p_1 p_2 p_3 (n_1 + p_1)(n_1 + p_2)(n_1+ p_3)\,, \\
    y_2 &  =  \frac{8\eta}{s^3} p_1 p_2 p_3 (n_2 -\eta p_1)(n_2 -\eta p_2)(n_2-\eta p_3) \,,
    \end{split}
\end{align}
and finally 
\begin{equation} \label{eq: Delta_z_spindle_data}
    \frac{\Delta z}{2\pi} = \frac{s}{2n_1 n_2 (n_1 + \eta n_2)}\,.
\end{equation} The upshot of the above expressions is the fact that the geometry is completely fixed by the set of flux and spindle data $(p_I, n_{i}, \eta)$. 

We conclude this section by noting that one of the main advantages of working with the STU model \eqref{eq: STU_Lagrangian} is that it admits solutions of both twist and anti-twist type, whereas minimal gauged supergravity \eqref{eq: minimal_gauged_lagrangian} only admits the anti-twist. In order to see this, one may first follow the argument given in \cite{Ferrero:2021etw} that the conditions of 
\begin{equation}
    h_I > 0\,, \quad y_1 < y_2 < y_3\,, \quad \eta_1 y_1 < 0\,,\quad \eta_2 y_2 > 0\,,
\end{equation}
forces $n_2>n_1$ for both cases, with all $p_I > 0$ in the anti-twist case, whereas for the twist case two of the $p_I$ are positive and the third negative. In order to study minimal gauged supergravity, we note that the minimal limit of the STU model \eqref{eq: stu_to_minimal} simply corresponds to 
\begin{equation} \label{eq: stu_min_par_limt}
    q_I = q  \quad \Leftrightarrow \quad h_I = h \quad \Leftrightarrow \quad p_I = p  \,, \qquad \quad I = 1,2,3\,,
\end{equation}
at the level of the local solutions \eqref{eq: 5d_gauge_field}.\footnote{We show in detail that one recovers the solutions of \cite{Ferrero:2020laf} in appendix \ref{app: STU_to_minimal}.} Importantly, the minimal limit forces all $p_I$ to be equivalent (and thus one cannot have two positive and one negative) so the twist case is not allowed. 

\subsection{Killing spinor equations} 

We start with the presentation of the Killing spinor equations for the STU model in our conventions:
\begin{equation} \label{eq: Gravitino_KSE}
\left[\nabla_M-\dfrac{\i}{L}  \sum_{I=1}^{3}A_{M}^{(I)}-\dfrac{1}{6L}\sum_{I=1}^3 X^{(I)}\Gamma_M
-\dfrac{\i}{12}\sum_{I=1}^3\left(X^{(I)}\right)^{-1}\left(\Gamma_{M}^{\phantom{M}N P}-4\delta_{M}^{N}\Gamma^{P}\right)F_{NP}^{(I)} \right] \varepsilon=0\,,
\end{equation}
and 
\begin{equation} \label{eq: Gaugino_KSE}
   \left[  \slashed{\partial} \varphi_i + \frac{2}{L}\sum_{I=1}^{3} \partial_{\varphi_i} X^{(I)} - \i  \sum_{I=1}^3 \partial_{\varphi_i} (X^{(I)})^{-1} F_{M N}^{(I)} \Gamma^{M N} \right] \varepsilon = 0\,.
\end{equation}
We note that there are several subtle differences between the presentation of the Killing spinor equations as written above and those in \cite{Ferrero:2021etw}, more detail of which is given in appendix \ref{App: Gamma-matrix_comparison}. 

In order to solve the equations \eqref{eq: Gravitino_KSE}, \eqref{eq: Gaugino_KSE}, we first make an explicit choice of frame and five-dimensional gamma matrices. For the frame we choose
\begin{equation}
    \textrm{e}^a = H^{1/6} \bar{\textrm{e}}^a\,, \quad a =0,1,2\,, \qquad \textrm{e}^3 = L \frac{ H^{1/6}}{2 P^{1/2}} \textrm{d}y\,, \qquad \textrm{e}^4 = L \frac{P^{1/2}}{H^{1/3}} \textrm{d}z\,,
\end{equation}
where $\bar{\textrm{e}}^a$ is an orthonormal frame for the metric on the radius $L$ AdS$_3$.
For the gamma matrices, we follow our conventions in \cite{Jeon:2025rfc} and set
\begin{equation} \label{eq: 5d_Gamma}
    \Gamma^a = \rho^a \otimes \gamma_3\,, \quad a = \{ 0,1,2\} \,, \qquad \Gamma^3 = - 1 \otimes \gamma_2\,, \qquad \Gamma^4 = 1 \otimes \gamma_1\,.
\end{equation}
We denote the usual Pauli matrices as $\sigma_{1,2,3}$ and then make the choice of $\rho^a = (\i \sigma_{2}, -\sigma_1, -\sigma_3)$, for the three-dimensional gamma matrices along the AdS$_3$ directions and $\gamma_{m} = (\sigma_1, \sigma_2)$ for the two-dimensional gamma matrices along the spindle directions. The chirality matrix along the spindle directions is defined via 
\begin{equation}
\gamma_3 \equiv -\i \gamma_{12} = - \i \sigma_1 \sigma_2 = \sigma_3\,, 
\end{equation}
and we note that the choice \eqref{eq: 5d_Gamma} ensures that the following Clifford algebra 
\begin{eqnarray}
        \left \lbrace\Gamma^A,\Gamma^B\right \rbrace  = 2 \eta^{AB} \textrm{I}_4\,, \quad A,B = \{0,1,2,3,4\}\,,
\end{eqnarray} is satisfied. 
    
In order to solve the Killing spinor equations \eqref{eq: Gravitino_KSE}, \eqref{eq: Gaugino_KSE}, we make the following ansatz for the $D=5$ Killing spinors
\begin{equation} \label{eq: spinor_ansatz}
    \varepsilon = \vartheta \otimes \chi\,,
\end{equation}
where $\vartheta$ is a Killing spinor for AdS$_3$ satisfying 
\begin{equation}
    \bar{\nabla}_{\alpha} \vartheta = -\frac{1}{2L} \bar{\rho}_{\alpha} \vartheta\,,
\end{equation}
where as in \cite{Jeon:2025rfc} the barred quantities denote radius $L$ AdS$_3$ quantities, with $\bar{\rho}_{\alpha} = \bar{\textrm{e}}^a_{\alpha} \rho_a$. Even with the simplifying ansatz \eqref{eq: spinor_ansatz}, it is still a non-trivial process to solve the Killing spinor equations, but fortunately the calculation has already been performed in \cite{Ferrero:2021etw}. Utilising the mapping between our conventions and those of that work as derived in appendix \ref{App: Gamma-matrix_comparison}, we obtain the following form of the Killing spinor along the spindle directions   
\begin{equation} \label{eq: chi_kse}
\chi= \frac{1}{\sqrt{2}H^{1/6}} \begin{pmatrix}
[ \sqrt{H} - y]^{1/2} \\
-\i [\sqrt{H} + y]^{1/2}
\end{pmatrix}  = H^{1/12} \begin{pmatrix}
\sin \frac{\alpha}{2} \\
-\i \cos \frac{\alpha}{2}
\end{pmatrix}\,,
\end{equation}
where we note that the intermediate expression makes it manifest that the for the twist case $\eta = -1$, the first component vanishes at both $y=y_i$ and thus $\chi$ is anti-chiral at both poles. For the anti-twist case $\eta = 1$, the bottom component vanishes at $y=y_1$ and the top component at $y=y_2$, demonstrating that $\chi$ is chiral at the North Pole and anti-chiral at the South Pole. In the second equality we have borrowed the notation of $\alpha$ from \cite{Ferrero:2021etw}, given by 
\begin{equation}
    \sin \alpha = \frac{P^{1/2}}{H^{1/2}} \,, \qquad \cos \alpha = \frac{y}{H^{1/2}}\,.
\end{equation}

We can also construct a spinor of opposite R-charge which is related to \eqref{eq: spinor_ansatz} via charge conjugation. Defining the charge conjugate as $\widetilde{\varepsilon} = B^{-1} \varepsilon^*$ where $B$ is defined in \eqref{5dgammaLor}, we have
\begin{equation} \label{eq: Gravitino_KSE_conj}
\left[\nabla_M+\dfrac{\i}{L}  \sum_{I=1}^{3}A_{M}^{(I)}+\dfrac{1}{6L}\sum_{I=1}^3 X^{(I)}\Gamma_M
-\dfrac{\i}{12}\sum_{I=1}^3\left(X^{(I)}\right)^{-1}\left(\Gamma_{M}^{\phantom{M}N P}-4\delta_{M}^{N}\Gamma^{P}\right)F_{NP}^{(I)} \right] \widetilde{\varepsilon}=0\,,
\end{equation}
and 
\begin{equation} \label{eq: Gaugino_KSE_conj}
   \left[  \slashed{\partial} \varphi_i - \frac{2}{L} \sum_{I=1}^{3} \partial_{\varphi_i} X^{(I)} + \i  \sum_{I=1}^3 \partial_{\varphi_i} (X^{(I)})^{-1} F_{M N}^{(I)} \Gamma^{M N} \right] \widetilde{\varepsilon} = 0\,.
\end{equation}

In order to compute the solution to this equation, we need to compute the charge conjugation matrix $B$ corresponding to our choice of gamma matrices. Solving equation \eqref{5dgammaLor} allows us to identify the charge conjugation matrix $\mathcal{C}$ as
\begin{equation}
    \mathcal{C} = \textrm{e}^{\i \theta} (\sigma_2 \otimes \sigma_1)\,, \quad \theta \in [0,2\pi)\,,
\end{equation}
and thus using \eqref{5dgammaLor} with $A = \Gamma_0 = -\i \sigma_2 \otimes \sigma_3$ we find $B$ to be 
\begin{equation}
    B = - \textrm{e}^{\i \theta} ( 1 \otimes \sigma_2)\,,
\end{equation}
which allows us to immediately construct solutions to the conjugate equations \eqref{eq: Gravitino_KSE_conj} and \eqref{eq: Gaugino_KSE_conj}. In particular, we find the solutions 
\begin{equation}
    \widetilde{\varepsilon} = \widetilde{\vartheta} \otimes \widetilde{\chi} = B^{-1} \varepsilon^* = - \textrm{e}^{-\i \theta} (\vartheta^* \otimes \sigma_2 \chi^*)\,,
\end{equation}
and thus focusing on the spindle part we have
\begin{equation} \label{eq: chi_symplectic_majorana_relations}
    \widetilde{\chi} = - \textrm{e}^{-\i \theta} \sigma_2 \chi^* \qquad \Leftrightarrow \qquad \chi^{\dagger} = \textrm{e}^{\i \theta} \widetilde{\chi}^{T} \sigma_2\,, \quad \widetilde{\chi}^{\dagger} = -\textrm{e}^{\i \theta} \chi^T \sigma_2\,,
\end{equation}
and thus a choice of $\theta = - \pi/2$ allows us to recover the same condition as in equation (2.19) of \cite{Jeon:2025rfc}. From now on, we will fix this value of the phase angle $\theta$ and obtain the conjugate Killing spinor on the spindle as 
\begin{equation}
    \widetilde{\chi} = - \i \sigma_2 \chi^*  = H^{1/12} \begin{pmatrix}
-\i \cos \frac{\alpha}{2}  \\
\sin \frac{\alpha}{2}
\end{pmatrix}\,,
\end{equation}
which we note is chiral at both poles for the twist case, and anti-chiral at the North Pole and chiral at the South Pole for the anti-twist case.

\section{Two-dimensional \texorpdfstring{$\cN =(2,2)$}{cn=(2,2)} theory on spindles of both twists} \label{sec: 2d_2,2}

\subsection{Killing spinors}

As in \cite{Jeon:2025rfc}, we now reinterpret the spindle components from the five-dimensional Killing spinor equation \eqref{eq: Gravitino_KSE} as the Killing spinor equation for a two-dimensional $\mathcal{N} = (2,2)$ theory living on the spindle. We recall that the Killing spinor equations for such a theory take the generic form \cite{Closset:2014pda}
\begin{subequations} \label{kse}
\begin{align} 
    D_\mu\epsilon &= - \frac{1}{2} \mathcal{H} \gamma_\mu \epsilon - \frac{1}{2} \mathcal{G} \gamma_\mu \gamma_{12} \epsilon\,, \label{kse1}\\
    D_\mu\widetilde{\epsilon} &= - \frac{1}{2} \mathcal{H} \gamma_\mu \widetilde{\epsilon} + \frac{1}{2} \mathcal{G} \gamma_\mu \gamma_{12} \widetilde{\epsilon}\,,
    \label{kse2}
\end{align}
\end{subequations}
where $D_\mu=\nabla_\mu + \i A_\mu$ for $\epsilon$ and $D_\mu=\nabla_\mu-\i A_\mu$ for $\widetilde{\epsilon}$ as the R-charge of $\epsilon$ and $\widetilde{\epsilon}$ are $-1$ and $+1$ respectively. Therefore, by comparison with equations \eqref{eq: Gravitino_KSE} and \eqref{eq: Gravitino_KSE_conj} we identify 
\begin{equation} \label{eq: A_2d}
    A = \frac{1}{L} A^{\text{R}} =  \frac{1}{2}\left[ \sum_{I=1}^3  \left(\frac{y}{h_I}\right) - 2 \right] \textrm{d} z 
\, ,
\end{equation}
and 
\begin{equation}
\label{eq:curlyGval}
    \mathcal{H} =  \frac{2\i}{3} \sum_{I =1}^{3} \big( X^{(I)} \big)^{-1} \mathscr{F}^{(I)} \, , \qquad \mathcal{G} = - \frac{1}{3 L} \sum_{I =1}^{3} X^{(I)} = - \frac{1}{3L} \frac{H'}{H^{2/3}}\,,
\end{equation}
where we introduced 
\begin{equation}\label{eq:Rflux}
    \mathscr{F}^{(I)} = \frac{1}{2} \epsilon^{\mu \nu} F^{(I)}_{\mu \nu} = \frac{1}{\sqrt{g_{\mathbb{\Sigma}}}} F^{(I)}_{yz} = \big( X^{(I)} \big)^2 \frac{1}{L} \frac{q_I}{H^{1/2}}\,,
\end{equation}
using \eqref{eq: field_strength_volume_relation} in the final equality. Recalling \eqref{eq: P-H_defs}, this allows us to write the function $\mathcal{H}$ as 
\begin{equation}
\label{eq:curlyHval}
    \mathcal{H} =  \frac{2\i}{3L}\frac{1}{H^{1/2}} \sum_{I =1}^3 X^{(I)} q_I = \frac{2\i}{3L} \frac{1}{H^{1/6}} \left( 3 - \frac{y H'}{H} \right) \,,
\end{equation}
which will be a useful way to write the function. We note that the functions  $\mathcal{H}$, $\mathcal{G}$ are all written entirely in terms of the spindle coordinate $y$ and the warp factor $H$ (and its derivatives).

Based on the comparison of equations \eqref{kse} with the 5d Killing spinor equations \eqref{eq: Gravitino_KSE} and \eqref{eq: Gravitino_KSE_conj}, we identify 
\begin{equation}
\label{eq:killingspinors}
    \epsilon = \sqrt{k_0} \widetilde{\chi} = \sqrt{k_0} H^{1/12} \begin{pmatrix}
-\i \cos \frac{\alpha}{2}  \\
\sin \frac{\alpha}{2}
\end{pmatrix}\,, \qquad \widetilde{\epsilon} = \sqrt{k_0} \chi = \sqrt{k_0} H^{1/12} \begin{pmatrix}
\sin \frac{\alpha}{2} \\
-\i \cos \frac{\alpha}{2}
\end{pmatrix}\,,
\end{equation}
and thus these spinors satisfy the symplectic Majorana relation 
\begin{equation}
    \widetilde{\epsilon}^\dagger = - \i \epsilon^T \sigma_2 \,, \qquad \epsilon^{\dagger} = \i \widetilde{\epsilon}^T \sigma_2\,,
\end{equation}
which follows directly from \eqref{eq: chi_symplectic_majorana_relations}.

We note that the spinors $\epsilon$, $\widetilde{\epsilon}$ are charged under $U(1)_{\text{R}}$, and the gauge field \eqref{eq: A_2d} is currently singular at the poles $y_{1,2}$ of the spindle. As in \cite{Jeon:2025rfc} we can move into regular gauges at the poles via constant gauge transformations of the form 
\begin{equation} \label{eq: A_gauge_transformation}
    A \rightarrow A' = A + \alpha \textrm{d} z\,,
\end{equation}
where different choices of $\alpha$ make the gauge field $A$ regular in the North ($y_1$) and South ($y_2$) poles respectively: 
\begin{equation}
    A \rightarrow A + \left[1 - \frac{1}{2}\sum_{I =1}^3 \left( \frac{y_i}{h_I(y_i)} \right) \right]\textrm{d}z = \frac{1}{2} \left[ \sum_{I =1}^3 \left( \frac{y}{h_I(y)}  - \frac{y_i}{h_I(y_i)}\right) \right] \textrm{d}z\,,
\end{equation}
which is regular via the vanishing of the gauge field at both poles $y=y_i$, $i=1,2$. Since the spinors $\epsilon$, $\widetilde{\epsilon}$ are charged under the $U(1)_{\text{R}}$ gauge field, under transformations of the form \eqref{eq: A_gauge_transformation} the spinors transform as
\begin{equation}
    \epsilon \rightarrow \textrm{e}^{-\i \alpha z} \epsilon\,, \qquad \widetilde{\epsilon} \rightarrow \textrm{e}^{\i \alpha z} \widetilde{\epsilon}\,,
\end{equation}
and thus we have the following expressions for the spinors in regular gauges at the poles of the spindle
\begin{align}
    \left. \epsilon \right|_{\mathcal{U}_{i}} & = \textrm{e}^{-\i \left[1 - \frac{1}{2}\sum_{I =1}^3 \left( \frac{y_i}{h_I(y_i)} \right) \right] z} \sqrt{k_0} H^{1/12} \begin{pmatrix}
-\i \cos \frac{\alpha}{2}  \\
\sin \frac{\alpha}{2} 
\end{pmatrix} = \textrm{e}^{-\i \frac{\eta_i}{n_i}\frac{\pi}{\Delta z} z} \sqrt{k_0} H^{1/12} \begin{pmatrix}
-\i \cos \frac{\alpha}{2}  \\
\sin \frac{\alpha}{2} 
\end{pmatrix}\,, \\
\left. \widetilde{\epsilon} \right|_{\mathcal{U}_{i}} & = \textrm{e}^{\i \left[1 - \frac{1}{2}\sum_{I =1}^3 \left( \frac{y_i}{h_I(y_i)} \right) \right]z} \sqrt{k_0} H^{1/12} \begin{pmatrix}
\sin \frac{\alpha}{2} \\
-\i \cos \frac{\alpha}{2} 
\end{pmatrix} = \textrm{e}^{\i \frac{\eta_i}{n_i}\frac{\pi}{\Delta z} z} \sqrt{k_0} H^{1/12} \begin{pmatrix}
\sin \frac{\alpha}{2} \\
-\i \cos \frac{\alpha}{2} 
\end{pmatrix}\,,
\end{align}
recalling that $\eta_i = (\eta, +1)$, we can write the spinors in regular gauge at the South Pole as
\begin{align}
    \left. \epsilon \right|_{\mathcal{U}_{2}} & =  \textrm{e}^{-\i \frac{1}{n_2}\frac{\pi}{\Delta z} z} \sqrt{k_0} H^{1/12} \begin{pmatrix}
-\i \cos \frac{\alpha}{2}  \\
\sin \frac{\alpha}{2} 
\end{pmatrix}\,, \\
\left. \widetilde{\epsilon} \right|_{\mathcal{U}_{2}} & = \textrm{e}^{\i \frac{1}{n_2}\frac{\pi}{\Delta z} z} \sqrt{k_0} H^{1/12} \begin{pmatrix}
\sin \frac{\alpha}{2} \\
-\i \cos \frac{\alpha}{2} 
\end{pmatrix}\,,
\end{align}
and at the North Pole as 
\begin{align}
    \left. \epsilon \right|_{\mathcal{U}_{1}} = \textrm{e}^{-\i \frac{\eta}{n_1}\frac{\pi}{\Delta z} z} \sqrt{k_0} H^{1/12} \begin{pmatrix}
-\i \cos \frac{\alpha}{2}  \\
\sin \frac{\alpha}{2} 
\end{pmatrix}\,, \\
\left. \widetilde{\epsilon} \right|_{\mathcal{U}_{1}} = \textrm{e}^{\i \frac{\eta}{n_1}\frac{\pi}{\Delta z} z}\sqrt{k_0} H^{1/12} \begin{pmatrix}
\sin \frac{\alpha}{2} \\
-\i \cos \frac{\alpha}{2} 
\end{pmatrix}\,. 
\end{align}
As in \cite{Jeon:2025rfc}, the spinors become periodic up to $\mathbb{Z}_{n_i}$ orbifolding around each pole. The monodromy of the spinors around $y_i$ is 
\begin{equation}
    \widetilde{\epsilon}(z+\Delta z) = \textrm{e}^{\i \frac{\eta_i}{n_i} \pi} \widetilde{\epsilon}(z)\,, \qquad \epsilon(z+ \Delta z) = \textrm{e}^{-\i \frac{\eta_i}{n_i} \pi} \epsilon(z)\,.
\end{equation}

We are ready to compute the bilinears which will be useful objects in the localisation procedure. We define multiplication of two spinors $\psi$ and $\varphi$ as
\begin{equation}
    \psi \varphi = \psi^T \sigma_2 \varphi\,,
\end{equation}
and thus $\psi \gamma_A \varphi = \psi^T \sigma_2 \gamma_A \varphi$ etc. For $\gamma_A = (\sigma_1, \sigma_2,  \sigma_3, 1)$ we have
\begin{align}
    \begin{split} \label{eq: spinor_bilinears}
        \left.  \i \widetilde{\epsilon} \gamma_A \epsilon \right|_{\mathcal{U}_i} & = k_0 H^{1/6} \left(0 , \sin \alpha, \cos \alpha,1 \right)\,,\\
        \left. \i \epsilon \gamma_A \epsilon \right|_{\mathcal{U}_i} & = k_0 H^{1/6} \textrm{e}^{-\i \frac{2\pi}{\Delta z} \frac{\eta_i}{n_i} z } \left( -1, -\i \cos \alpha, \i \sin \alpha, 0 \right)\,, \\
        \left. \i \widetilde{\epsilon} \gamma_A \widetilde{\epsilon} \right|_{\mathcal{U}_i} & = k_0 H^{1/6} \textrm{e}^{\i \frac{2\pi}{\Delta z}\frac{\eta_i}{n_i}z} (1, - \i \cos \alpha, \i \sin \alpha, 0)\,,
    \end{split}
\end{align}
and we note that the bispinors formed from the Killing spinors \eqref{eq:killingspinors} in the singular gauge \eqref{eq: A_2d} are
\begin{align}
    \begin{split} \label{eq: spinor_bilinears_singular}
        \i \widetilde{\epsilon} \gamma_A \epsilon & = k_0 H^{1/6} \left(0 , \sin \alpha, \cos \alpha,1 \right)\,,\\
        \i \epsilon \gamma_A \epsilon & = k_0 H^{1/6}  \left( -1, -\i \cos \alpha, \i \sin \alpha, 0 \right)\,, \\
      \i \widetilde{\epsilon} \gamma_A \widetilde{\epsilon} & = k_0 H^{1/6}  (1, - \i \cos \alpha, \i \sin \alpha, 0)\,,
    \end{split}
\end{align}
the same as \eqref{eq: spinor_bilinears} with the exponential factors removed.
As in \cite{Jeon:2025rfc}, we have
\begin{equation}
    \left(  \left. \i \epsilon \gamma_A \epsilon \right|_{\mathcal{U}_i} \right)^* = - \left(  \left. \i \widetilde{\epsilon} \gamma_A \widetilde{\epsilon} \right|_{\mathcal{U}_i} \right)\,,
\end{equation}
and we can construct the $U(1)$ Killing vector on the spindle as 
\begin{equation} \label{eq: Killing_z}
    v^{\mu} \partial_{\mu} \equiv \i \widetilde{\epsilon} \gamma^{\mu} \epsilon \partial_{\mu} = \frac{k_0}{L} \partial_z \,,
\end{equation}
the normalisation of which suggests that we should make the choice of 
\begin{equation}
    k_0 = L_0 \frac{\Delta z}{2\pi}\,, 
\end{equation}
in order to normalise the Killing vector as 
\begin{equation} \label{eq: killing_phi}
    v^{\mu} \partial_{\mu} = \frac{L_0}{L} \frac{\Delta z}{2\pi} \partial_z = \frac{L_0}{L} \partial_{\phi}\,, \qquad \phi = \frac{2\pi}{\Delta z} z\,, \quad \phi \sim \phi + 2\pi\,,
\end{equation}
using the coordinate $\phi$ with period $2\pi$. This will allow us to keep the reference ratio $L_0/L$ in the calculation and see how this affects the localisation steps. Note that unlike \cite{Jeon:2025rfc} our Killing spinor solutions presented here preserve supersymmetry via both the twist and anti-twist mechanisms, with the particular case depending on the choice of the parameter $\eta = \mp 1$.       

\subsection{Equivariant supercharge and cohomological variables} 

As in \cite{Jeon:2025rfc}, we will consider $\mathcal{N} = (2,2)$ supersymmetric theories defined on $\mathbb{\Sigma}$ consisting of a single vector and a single chiral multiplet. For the generic details concerning the fields in these multiplets and their supersymmetry transformations, see appendix \ref{app: multiplets}. See also \cite{Doroud:2012xw, Benini:2012ui, Closset:2014pda, Closset:2015rna, GonzalezLezcano:2023cuh} for related discussion. 

Here we will be interested in analysing the square of the equivariant supercharge operator, whose action on a field of $U(1)_{\text{R}}$-charge $\widehat{q}_{\text{R}}$ and $U(1)_{\text{G}}$-charge $\widehat{q}_{\text{G}}$ can be written as 
\begin{equation} \label{eq: Q^2}
    Q_{eq}^2 = \mathcal{L}_{v} + \i \widehat{q}_{\text{R}} \Lambda^{\text{R}} + \i \widehat{q}_{\text{G}} \Lambda^\text{G}_0\,,
\end{equation} 
where
\begin{equation} \label{eq: Lambda_R}
    \Lambda^{\text{R}} = - v^{\mu} A_{\mu} - \frac{1}{2} \left( \mathcal{H} \widetilde{\epsilon} \epsilon + \i \mathcal{G} \widetilde{\epsilon} \gamma_3 \epsilon \right)\,,
\end{equation}
and $\Lambda^{\text{G}}_0$ is the constant part of 
\begin{equation} \label{eq: Lambda_G}
\Lambda^{\text{G}} \equiv - v^{\mu} \mathcal{A}_{\mu} - \widetilde{\epsilon} \epsilon \sigma - \i \widetilde{\epsilon}\gamma_3 \epsilon \rho\,.
\end{equation} 
If one uses \eqref{eq: A_2d} then one immediately finds 
\begin{equation} \label{eq: Lambda_R_singular}
    \Lambda^{\text{R}} = 0\,,
\end{equation}
which is using the gauge field which is singular at the poles of the spindle. If we perform the gauge transformations \eqref{eq: A_gauge_transformation} in order to make the gauge field regular at either pole we instead find 
\begin{equation} \label{eq: Lambda_R_regular}
    \left. \Lambda^{\text{R}} \right|_{\mathcal{U}_i} = - \frac{k_0}{L} \left. \alpha \right|_{\mathcal{U}_i} = - \frac{k_0}{L} \frac{\eta_i}{n_i} \frac{\pi}{\Delta z} = - \frac{L_0}{L} \frac{\eta_i}{2n_i} \,,
\end{equation}
so again we nicely recover the anti-twist result \cite{Jeon:2025rfc} as well as producing a new expression for the twist in the regular gauges. We will return to evaluate $\Lambda^{\text{G}}_0$ after computing the BPS locus in section \ref{sec: BPS_locus}. 

As in \cite{GonzalezLezcano:2023cuh, Jeon:2025rfc}, it is often advantageous to repackage the fields in the multiplet into so-called ``cohomological variables''.  We can take the same choice of cohomological variables as in \cite{Jeon:2025rfc}, first by defining the twisted variables for the spinor fields 
\begin{equation}
    \epsilon \psi\,, \quad \widetilde{\epsilon} \widetilde{\psi}\,, \quad \widetilde{\epsilon} \psi\,, \quad \epsilon \widetilde{\psi}\,,  
\end{equation}
then the inverse relation is given by 
\begin{equation}
    \psi = \frac{1}{\widetilde{\epsilon} \epsilon} \left( \epsilon \left(\widetilde{\epsilon} \psi \right) - \widetilde{\epsilon} \left(\epsilon \psi \right) \right) \,, \quad  \widetilde{\psi} = \frac{1}{\widetilde{\epsilon} \epsilon} \left( \epsilon (\widetilde{\epsilon} \widetilde{\psi} ) - \widetilde{\epsilon} (\epsilon \widetilde{\psi}) \right) \,. 
\end{equation}
This redefinition is well-defined since the Jacobian is 
\begin{equation}
    |J| = \left| \text{det} \frac{\partial(\psi, \widetilde{\psi})}{\partial(\widetilde{\epsilon} \psi, \epsilon \psi, \widetilde{\epsilon} \widetilde{\psi}, \epsilon \widetilde{\psi})} \right| = \frac{1}{\left| (\widetilde{\epsilon} \epsilon)^2 \right|} = \frac{1}{k_0^2 H^{1/3}}\,,
\end{equation}
which is non-zero since $H > 0$ for $y \in [y_1, y_2]$.\footnote{Note that there is a typo in equation (3.35) of \cite{Jeon:2025rfc}, where the factor of $k_0$ in the denominator should be $k_0^{2}$.} The fact that this map is invertible means that one can define the $Q_{eq}$-cohomology complex $(\Phi, Q_{eq} \Phi, \Psi, Q_{eq} \Psi)$, with cohomological variables defined as in \cite{Jeon:2025rfc} as
    \begin{equation}
\begin{alignedat}{2}
        \Phi & =  \{\phi\, ,\widetilde\phi\}\, ,\quad \qquad \qquad && \qeq \Phi =\{ \qeq \phi \, ,\qeq \widetilde{\phi}\}\,  \label{eq:CQarray} ,\\ 
\Psi & =  \{\epsilon \psi 
\, , \widetilde{\epsilon} \widetilde{\psi}
\}\,, \qquad\qquad\quad && \qeq \Psi = \{\qeq \left(\eps \psi\right)\, , \qeq \bigl( \widetilde{\epsilon} \widetilde{\psi}\bigr)\}\, , 
\end{alignedat}
\end{equation}
where the explicit expressions for the elements of $\qeq \Phi$ and $\qeq \Psi$ can be read off from the supersymmetry transformations of the physical chiral multiplet variables in \eqref{deltachiral}.

\subsection{Supersymmetric action}

In order to demonstrate the proof of concept of localisation on the spindle, we will consider a simple case of a supersymmetric action on $\mathbb{\Sigma}$. Following \cite{Jeon:2025rfc}, we will consider an action of the form
\begin{equation} \label{eq: classical_action}
    S = \int_{\mathbb{\Sigma}} d^2 x\, \sqrt{g}_{\mathbb{\Sigma}} \mathcal{L}\,,
\end{equation}
with Lagrangian
\begin{equation} 
    \mathcal{L} = \mathcal{L}_{\text{v.m.}} + \mathcal{L}_{\text{c.m.}} + \mathcal{L}_{\text{FI}} + \mathcal{L}_{\text{top.}}\,.
\end{equation}
The first term above is the vector multiplet Lagrangian
\begin{align}
\begin{split} \label{eq: vm_lagrangian}
    \mathcal{L}_{\text{v.m.}} = \frac{1}{2} \left( \mathcal{F} + \i(\mathcal{G} \sigma - \mathcal{H} \rho) \right)^2 + \frac{1}{2} \partial_{\mu} \sigma \partial^{\mu} \sigma + \frac{1}{2} \partial_{\mu} \rho \partial^{\mu} \rho + \frac{1}{2} D^2 + \frac{\i}{2} \widetilde{\lambda} \gamma^{\mu} D_{\mu} \lambda\,,
\end{split}
\end{align}
where as in \cite{GonzalezLezcano:2023cuh, Jeon:2025rfc} we define the field $D$ as a redefinition of the auxilliary scalar via 
\begin{equation}
    D \equiv \widehat{D} - \i (\mathcal{H} \sigma + \mathcal{G} \rho)\,,
\end{equation}
where the auxiliary scalar $\widehat{D}$ is the one that appears in the supersymmetry transformation in~\eqref{eq:deltaA}.

The chiral multiplet Lagrangian is 
\begin{equation} \label{eq: L_cm}
     \mathcal{L}_{\text{c.m.}} = D_{\mu} \widetilde{\phi} D^{\mu} \phi + M_{\phi}^2 \widetilde{\phi} \phi + \widetilde{\mathfrak{F}} \mathfrak{F} - \i \widetilde{\psi} \gamma^{\mu} D_{\mu} \psi + \widetilde{\psi} M_{\psi} \psi - \i \widetilde{\psi} \lambda \phi - \i \widetilde{\phi} \widetilde{\lambda} \psi\,,  
\end{equation}
where the mass squared of the scalar field $\phi$ and mass of the fermion $\psi$ are 
\begin{align}
\begin{split}
    M_{\phi}^2 & = \left( \sigma + \frac{r}{2} \mathcal{H}\right)^2  + \left( \rho + \frac{r}{2} \mathcal{G} \right)^2 + \frac{r}{4} R_{\mathbb{\Sigma}} + \i \widehat{D}\,, \\
     M_{\psi} & = -\i \left( \sigma + \frac{r}{2} \mathcal{H} \right) - \left(\rho + \frac{r}{2} \mathcal{G} \right) \gamma_3\,.
     \end{split}
\end{align}

The Fayet-Iliopoulos (FI) Lagrangian is 
\begin{equation} \label{eq: L_FI}
    \mathcal{L}_{\text{FI}} = - \i \xi \widehat{D}\,,
\end{equation}
where $\xi \in \mathbb{R}$ is the FI parameter. Finally, the topological term Lagrangian is 
\begin{equation} \label{eq: L_top}
    \mathcal{L}_{\text{top.}} = \i \frac{\theta}{2\pi} \mathcal{F} \,,
\end{equation}
where $\theta \in (0, 2\pi n_1n_2)$ is the topological parameter. We note that this range differs from that of the topological parameter on $S^2$ \cite{Benini:2012ui, Doroud:2012xw, Closset:2015rna, GonzalezLezcano:2023cuh} which is due to the orbifold nature of the spindle. The reason for this range will become clear in the next section when we impose the quantisation condition \eqref{eq: gauge_flux_quantisation}. 

We end this section by noting that the combination $ \mathcal{L}_{\text{FI}} +  \mathcal{L}_{\text{top.}}$ is a simple example of what is known as a \textit{twisted superpotential} term \cite{Benini:2012ui, Closset:2014pda, Closset:2015rna} (see also the discussion in appendix D of \cite{Jeon:2025rfc}.) It would be interesting to examine the topic of supersymmetric localisation on the spindle in the presence of a more general twisted superpotential, which we leave as a future direction to be pursued.

\section{BPS locus} \label{sec: BPS_locus}

\subsection{BPS equations} 

In this section we consider the process of supersymmetric localisation \cite{Duistermaat:1982vw, Atiyah:1984px, Witten:1988ze, Witten:1988xj, Nekrasov:2002qd, Pestun:2007rz, Cremonesi:2013twh} on $\mathbb{\Sigma}$ in earnest. We follow the general argument as outlined in section 4.1 of \cite{Jeon:2025rfc} and the problem of determining the localisation locus of the theory amounts to solving the BPS equations 
\begin{equation}
    \lambda = \widetilde{\lambda} = \psi = \widetilde{\psi} = 0 \,, \quad Q_{eq} \lambda = Q_{eq} \widetilde{\lambda} = Q_{eq} \psi = Q_{eq} \widetilde{\psi} = 0\,.
\end{equation}
These equations can be formulated as the equations of motion arising from an action which is a $Q_{eq}$-exact deformation of the classical action \eqref{eq: classical_action}. As in \cite{Jeon:2025rfc}, we take the vector multiplet deformation to be 
\begin{align}
\begin{split} \label{eq: QeqVvm}
  Q_{eq} \mathcal{V}^{\text{v.m.}} \bigg|_{\text{bos.}} = 2 \int _{\mathbb{\Sigma}}\textrm{d}^2 x \, \sqrt{g}_{\mathbb{\Sigma}} \,
  \epsilon^{\dagger} \epsilon \bigg[ & (D_{\mu} \rho)^2 +\left( \mathcal{F} - \i \mathcal{H} \rho \right)^2  \\
    & + \left( D_{\mu} \sigma + \i \mathcal{G} \sigma \frac{\epsilon^{\dagger} \gamma_3 \gamma_{\mu} \epsilon}{\epsilon^{\dagger} \epsilon} \right)^2 + \left( D + \mathcal{G} \sigma \frac{\epsilon^{\dagger}\gamma_3 \epsilon}{\epsilon^{\dagger} \epsilon} \right)^2\bigg]\,, 
     \end{split}
\end{align}
which, as in the purely anti-twisted case considered in \cite{Jeon:2025rfc}, is positive-definite. One can verify this via \eqref{eq:curlyGval}, where $\mathcal{H}$ is imaginary and $\mathcal{G}$ is real, making the sum-of-squares above non-negative assuming the reality condition \eqref{eq: reality_conditions_vector}. The equations of motion arising from such an action are thus simply the vanishing of all squared terms above and correspond to the BPS equations $Q_{eq} \lambda = Q_{eq} \widetilde{\lambda} = 0$. Using \eqref{eq:curlyGval} and \eqref{eq:curlyHval}, we solve the vanishing of the first two squares in \eqref{eq: QeqVvm} to be
\begin{equation} \label{eq:Rflux2}
 \rho = \rho_0 = \textrm{const.}\,, \quad \mathcal{F} = \i \mathcal{H}\rho_0 = -  \frac{2 \rho_0}{3L H^{1/2}} \sum_{I =1}^3 X^{(I)} q_I  \,,
\end{equation} while
solving the third square for $\sigma$ we obtain the equation
\begin{equation}
\partial_{\mu} \sigma - \frac{\i}{3L}  \sigma \frac{\epsilon^{\dagger} \gamma_3 \gamma_\mu \epsilon}{\epsilon^\dagger \epsilon} \sum_{I =1}^3 X^{(I)} = 0\,.
\end{equation}
Assuming $\sigma = \sigma(y)$, the $z$-component of this equation is solved automatically and using \eqref{eq:curlyGval} and \eqref{eq:curlyHval}, the $y$-component becomes  
\begin{equation}
    \sigma'(y) +\frac{H'}{6H} \sigma(y) = 0 \quad \Rightarrow \quad \sigma(y)= \left(\frac{2\pi}{\Delta z}\sigma_0 \right) H^{-1/6}\,,  
    \end{equation} where $\sigma_0$ is a (real) constant and the $2\pi/\Delta z$ scaling is chosen in order to avoid clutter in later equations. The vanishing of the final square in \eqref{eq: QeqVvm} gives 
\begin{equation}
    D = - \mathcal{G}  \frac{\epsilon^{\dagger} \gamma_3 \epsilon}{\epsilon^{\dagger} \epsilon} \sigma(y)=  \frac{2\pi}{\Delta z} \frac{\sigma_0}{3L} \frac{y H'}{H^{4/3}}\,,
\end{equation}
which completes the form of the general solution to the equations of motion from \eqref{eq: QeqVvm}. We note that the solution space is parameterised by two real constants $\rho_0, \sigma_0$. 

In order to determine the full BPS locus, we also need to apply the gauge flux quantisation condition\footnote{Our notation is $\mathcal{F}_{\mu \nu} = 2 \partial_{[\mu} A_{\nu]}$, $\mathcal{F} = \frac{1}{2} \epsilon^{\mu \nu} \mathcal{F}_{\mu \nu}= (\sqrt{g}_{\mathbb{\Sigma}})^{-1} \mathcal{F}_{yz}$.} 
\begin{equation} \label{eq: gauge_flux_quantisation}
    \mathfrak{f}_G = \frac{1}{2\pi} \int_{\mathbb{\Sigma}} \textrm{d} \mathcal{A} = \frac{1}{2\pi} \int_{\mathbb{\Sigma}} d^2x\,\sqrt{g}_{\mathbb{\Sigma}} \mathcal{F}=\frac{\mathfrak{m}}{n_1 n_2}\,, \qquad \mathfrak{m} \in \mathbb{Z}\,,
\end{equation}
which together with \eqref{eq:Rflux2} allows us to fix the value of $\rho_0$ as 
\begin{equation}
    \rho_0 L =  \frac{2\pi}{\Delta z} \frac{1}{\big(y_1^{1/3} -y_2^{1/3}\big)} \frac{\mathfrak{m}}{n_1 n_2}\,, 
\end{equation}
which can be written explicitly in terms of the flux and spindle data $(p_I, n_i, \eta)$ using the expressions \eqref{eq: y_spindle_data} and \eqref{eq: Delta_z_spindle_data}. We note that the BPS locus consists of a real number $\sigma_0$ as well as the integer $\mathfrak{m}$. The path integral will thus localise to an integral over $\sigma_0 \in \mathbb{R}$ and a sum over $\mathfrak{m} \in \mathbb{Z}$.

We can also use the BPS locus value for $\mathcal{F}$ to compute the expression for the $U(1)_{\text{G}}$ gauge field on the BPS locus. Explicitly we have 
\begin{equation}
    \mathcal{A} = \mathcal{A}_z \textrm{d}z \,,
\end{equation}
and thus using \eqref{eq:Rflux} and \eqref{eq:Rflux2} we obtain
\begin{equation}
    \partial_y \mathcal{A}_z = \mathcal{F}_{yz} =  \i \mathcal{H} \rho_0 \sqrt{g_{\mathbb{\Sigma}}}\,,
\end{equation}
which can be directly integrated to find
\begin{equation}
    \mathcal{A}_z = - L \rho_0 \frac{y}{H^{1/3}} + \beta \,,
\end{equation}
where $\beta$ is a constant of integration which can be utilised as a gauge transformation allowing one to move between gauge choices of interest. In particular, choosing 
\begin{equation} \label{eq: beta_regular}
    \left. \beta \right|_{\mathcal{U}_i} = L \rho_0 y_i^{1/3}\,,
\end{equation}
moves into gauges which are regular at the poles, and choosing 
\begin{equation} \label{eq: beta_singular}
    \beta  = L \rho_0 y_i^{1/3} + \frac{2\pi}{\Delta z} \frac{\mathfrak{m}_i}{n_i}\,,
\end{equation}
selects a gauge with the boundary conditions 
\begin{equation}
    \mathcal{A}(y_i) = \frac{2\pi}{\Delta z}\frac{\mathfrak{m}_i}{n_i}\,, \qquad \mathfrak{m}_i \in \mathbb{Z}\,,
\end{equation}
where, as in \cite{Inglese:2023tyc, Jeon:2025rfc} we have 
\begin{equation}
    \mathfrak{m} = \mathfrak{m}_2 n_1 - \mathfrak{m}_1 n_2\,,
\end{equation}
where $\mathfrak{m}_i$ can be expressed in terms of $\mathfrak{m}$ and additional $\mathfrak{a}_i \in \mathbb{Z}$ via
\begin{equation}
\mathfrak{m}_i = \mathfrak{m} \mathfrak{a}_i \,, \qquad 1 = \mathfrak{a}_2 n_1 - \mathfrak{a}_1 n_2\,,
\end{equation}
where the pair $(\mathfrak{a}_1, \mathfrak{a}_2)$ is only defined up to transformations of the form $(\mathfrak{a}_1 + n_1 \delta \mathfrak{a}, \mathfrak{a}_2 + n_2 \delta \mathfrak{a})$ for $\delta \mathfrak{a} \in \mathbb{Z}$. As in \cite{Inglese:2023wky,Inglese:2023tyc,Jeon:2025rfc}, we will see that physical observables will be independent of $\delta \mathfrak{a}$.

Using the BPS values of the vector multiplet fields, we can now return to \eqref{eq: Lambda_G} and compute
\begin{equation} \label{eq: Lambda^G}
    \Lambda^{\text{G}} = k_0 \left( \i \frac{2\pi}{\Delta z} \sigma_0 - \frac{\beta}{L} \right) := \Lambda^{\text{G}}_0\,.
\end{equation}
Thus in the singular gauge \eqref{eq: beta_singular} we have 
\begin{align}
\begin{split} \label{eq: Lambda_G_singular}
    \Lambda^{\text{G}} & = \frac{L_0}{L} \left( \i \sigma_0 L - \frac{\mathfrak{m}}{n_1 n_2} \frac{1}{1-\left( \frac{y_2}{y_1} \right)^{1/3} } - \frac{\mathfrak{m}_1}{n_1} \right) \\
    &= \frac{L_0}{L} \left( \i \sigma_0 L + \frac{\mathfrak{m}}{n_1 n_2} \frac{1}{1-\left( \frac{y_1}{y_2} \right)^{1/3} } - \frac{\mathfrak{m}_2}{n_2} \right) \equiv \Lambda^{\text{G}}_0\,,
\end{split}
\end{align}
and in regular gauges \eqref{eq: beta_regular} we have  
\begin{equation} \label{eq: Lambda^G_regular}
\left. \Lambda^{\text{G}} \right|_{\mathcal{U}_i} = \Lambda^{\text{G}} + \frac{L_0}{L} \frac{\mathfrak{m}_i}{n_i}\,
\end{equation}
both of which will be of use in the localisation calculations.

In a similar vein, we construct the chiral multiplet deformation action as in \cite{Jeon:2025rfc}, finding a bosonic part of the form 
\begin{align}
   \begin{split} \label{eq: CM_deformation_action}
  Q_{eq} \mathcal{V}^{\text{c.m.}} \bigg|_{\text{bos.}} = \int_{\mathbb{\Sigma}} & \textrm{d}^2 x \, \sqrt{g}_{\mathbb{\Sigma}} \, \bigg\{ \widetilde{\mathfrak{F}} \mathfrak{F} + \bigg| \left(\i \gamma^{\mu} D_{\mu} \phi - \i \left( \sigma + \frac{r}{2} \mathcal{H} \right) \phi + \gamma_3 \left( \rho + \frac{r}{2} \mathcal{G} \right) \phi\right) \epsilon \bigg|^2 \\
    & \qquad \qquad \: + \bigg| \left( \i \gamma^{\mu} D_{\mu} \widetilde{\phi} - \i \left( \sigma + \frac{r}{2} \mathcal{H} \right) \widetilde{\phi} - \gamma_3 \left( \rho + \frac{r}{2} \mathcal{G} \right) \widetilde{\phi} \right) \widetilde{\epsilon}  \bigg|^2   \bigg\}\,, 
    \end{split}
\end{align}
which is again positive-definite provided one uses the reality condition \eqref{eq: chiral_bosonic_reality}. Yet again, the equations of motion arising from this action are simply the vanishing of each term due to the ``sum of squares'' structure and correspond to the BPS equations $Q_{eq} \psi = Q_{eq} \widetilde{\psi} = 0$. We immediately see that the vanishing of the first term gives
\begin{equation}
 \mathfrak{F} = \widetilde{\mathfrak{F}} = 0\,,
\end{equation}
and the remaining equations can be massaged into the forms 
\begin{subequations} \label{eq: cm_BPS}
\begin{align} \label{eq: cm_BPS_1}
   ( \mathcal{L}_v  + \i r \Lambda^{\text{R}}  + \i \Lambda^{\text{G}} ) \phi & = 0\, \quad \Leftrightarrow \quad Q^2_{eq} \phi = 0\,, \\
   (\mathcal{L}_v  - \i r \Lambda^{\text{R}} - \i \Lambda^{\text{G}} )\widetilde{\phi} & = 0 \quad \Leftrightarrow \quad Q^2_{eq} \widetilde{\phi} =0 \,,
\end{align}
   \end{subequations}
which can be reached by multiplying the second term in \eqref{eq: CM_deformation_action} by $\epsilon$ and the third by $\widetilde{\epsilon}$ and using the definitions \eqref{eq: Killing_z}, \eqref{eq: Lambda_R} and \eqref{eq: Lambda_G}. We note that $\Lambda^{\text{R}}, \Lambda^{\text{G}}$ are $U(1)_{\text{R}}, U(1)_{\text{G}}$ dependent quantities respectively, though the equations \eqref{eq: cm_BPS} are gauge invariant due to the charges of the scalars $\phi, \widetilde{\phi}$. Due to the reality condition \eqref{eq: chiral_bosonic_reality}, we can focus the analysis purely on  \eqref{eq: cm_BPS_1} and for simplicity we choose the gauges \eqref{eq: Lambda_R_singular} and \eqref{eq: Lambda_G_singular}. We find the $z$-dependence of $\phi$ to be 
\begin{equation}
    \phi(y,z) = \Phi(y) \exp \left( - \i \frac{L}{k_0} \Lambda^{\text{G}}_0 z \right)\,.
\end{equation}
Now contracting the second term in \eqref{eq: CM_deformation_action} with $\epsilon$ one finds the equation
\begin{equation}
    \i \epsilon \gamma^{\mu} \epsilon D_{\mu} \phi + \epsilon \gamma_3 \epsilon \left( \rho + \frac{r}{2} \mathcal{G} \right) \phi = 0\,,
\end{equation}
which we can also write as
\begin{equation}
   L_{P_-} \phi + \epsilon \gamma_3 \epsilon \left( \rho + \frac{r}{2} \mathcal{G} \right) \phi =0 \quad \Leftrightarrow \quad  D_{10} \phi = 0\,,
\end{equation}
where we introduced $D_{10}$ as in (5.13a) of \cite{Jeon:2025rfc}. Non-trivial chiral multiplet BPS configurations will be \textit{everywhere regular} solutions to the equations above, for a generic vector multiplet configuration. In order to examine this regularity, we use the fact that the regular solutions of $D_{10} \phi = 0$ have the same regularity properties at the poles of the spindle as the solutions of $L_{P_-} \phi = 0$, an equation we examine in detail in section \ref{sec: unpaired_eigenvalues}. Jumping ahead slightly in using the results of that section, we note real solutions are only regular at the South Pole when $\sigma_0 =0$ and
\begin{equation}
    \frac{\mathfrak{m}}{n_1 n_2} \frac{1}{1-\left(\frac{y_1}{y_2}\right)^{1/3}} \geq \frac{r}{2n_2}\,,
\end{equation}
which will clearly not be valid for all $\mathfrak{m} \in \mathbb{Z}$. We can thus immediately conclude that for a generic configuration on the vector multiplet BPS locus, the only globally regular chiral multiplet solution is
\begin{equation}
    \phi = \widetilde{\phi} = 0\,,
\end{equation}
indicating that matter fields will not enter any classical contributions to the partition function and will enter at the level of the one-loop determinants only. We also note as in \cite{GonzalezLezcano:2023cuh, Jeon:2025rfc} that the vanishing of the chiral multiplet fields on the BPS locus confirms that we are using the Coulomb branch localisation scheme.

\subsection{Action on the BPS locus}

In order to compute the partition function, we need to evaluate the action \eqref{eq: classical_action} on the BPS locus that we have computed in the previous sections. Due to the $Q_{eq}$-exactness (up to boundary terms) of $\mathcal{L}_{\text{v.m.}}$ \eqref{eq: vm_lagrangian} and $\mathcal{L}_{\text{c.m.}}$ \eqref{eq: L_cm} \cite{GonzalezLezcano:2023cuh}, we immediately note that these terms give no contribution. The only non-trivial contribution to the classical part of the path integral thus comes from the FI \eqref{eq: L_FI} and topological \eqref{eq: L_top} terms, which we now compute in detail.

We begin with 
\begin{equation}
   \left. S_{\text{FI}} \right|_{\text{BPS}} = -\i \xi \int_{\mathbb{\Sigma}} \textrm{d}^2 x \, \sqrt{g}_{\mathbb{\Sigma}} \widehat{D} \,,
\end{equation}
where 
\begin{equation}
\widehat{D} = D + \i (\mathcal{H} \sigma + \mathcal{G} \rho)\,.
\end{equation}
After a bit of computation we find
\begin{equation}
     \sqrt{g}_{\mathbb{\Sigma}} \widehat{D} = - \frac{2\pi}{\Delta z} L \sigma_0 \partial_y \left( y H^{-1/2} \right) - \i L \rho_0  \partial_y \left( H^{1/6} \right) \,,  
\end{equation}
which leads to the anti-twist case result of 
\begin{equation}  \label{eq: S_anti_twist}
     \left. S^{\eta = 1}_{\text{FI}} \right|_{\text{BPS}} = 4 \pi \i \xi L \sigma_0 - 2\pi \xi \frac{\mathfrak{m}}{n_1 n_2} \frac{(y_2^{1/3}+y_1^{1/3})}{(y_1^{1/3}-y_2^{1/3})} = 4\pi \xi \gamma_{\text{G}}\,,
\end{equation}
where, as in \cite{Jeon:2025rfc}, we define $\gamma_{\text{G}}$ as 
\begin{equation} \label{eq: gamma_G}
    \gamma_{\text{G}} \equiv \frac{L}{L_0} \Lambda^{\text{G}} + \frac{1}{2} \left( \frac{\mathfrak{m}_1}{n_1} + \frac{\mathfrak{m}_2}{n_2} \right) \,.
\end{equation}
We observe that the structural form of the FI term on the BPS locus for the anti-twist case is identical to that of \cite{Jeon:2025rfc}, even though we started with the STU solution as opposed to minimal gauged supergravity. This is to be expected since the two-dimensional field theory calculation should not depend on the local form of the metric that one chooses, but just the topology of the background. We will comment further on this in section \ref{sec: partition_function}.

More interestingly, the twist case gives 
\begin{equation}
    \left. S^{\eta = -1}_{\text{FI}} \right|_{\text{BPS}} = 2\pi \xi \frac{\mathfrak{m}}{n_1 n_2}\,,
\end{equation}
this looks quite odd as the $\sigma_0$ dependence has dropped out completely! This will have a significant effect on the final partition function, giving a difference between the twist and anti-twist cases.

We also need to compute the topological action 
\begin{equation}
    \left. S_{\text{top}} \right|_{\text{BPS}} = \i \frac{\theta}{2\pi} \int_{\mathbb{\Sigma}} d^2 x \, \sqrt{g}_{\mathbb{\Sigma}} \mathcal{F} = \i \theta \frac{\mathfrak{m}}{n_1 n_2}\,, 
\end{equation}
which for the twist case shows that the parameter $\theta$ enters in exactly the same way as the parameter $\xi$, explicitly 
\begin{equation}
    \left. S^{\eta = -1}_{\text{FI}} \right|_{\text{BPS}} +   \left. S_{\text{top}} \right|_{\text{BPS}} = \frac{\mathfrak{m}}{n_1 n_2} (2\pi \xi + \i \theta) \equiv \frac{\mathfrak{m}}{n_1 n_2} \tilde{\xi}\,. 
\end{equation}
\section{One-loop determinants} \label{sec: one_loop_determinants}

\subsection{Unpaired eigenvalues} \label{sec: unpaired_eigenvalues}

Following the approach as discussed in section 5.1 of and 5.2 of \cite{Jeon:2025rfc}, see also \cite{GonzalezLezcano:2023cuh, Inglese:2023tyc, Pittelli:2024ugf} we can compute the chiral multiplet one-loop determinants entering the partition function via 
\begin{equation} \label{eq: 1-loop-det}
    Z^{ \qeq \cV}_{1\text{-loop}} =     \sqrt{\frac{ \textrm{det} ~ (\qeq^2)_{_{\Psi}} }{\textrm{det} ~ (\qeq^2)_{_{\Phi}}}} = \sqrt{\frac{\textrm{det}_{\text{Ker}(P^{(\epsilon \psi)}_+)} ~ \qeq^2 \cdot \textrm{det}_{\text{Ker}(P^{(\widetilde{\epsilon} \widetilde{\psi})}_-)} ~ \qeq^2}{\textrm{det}_{\text{Ker}(P^{(\widetilde{\phi})}_+)} ~ \qeq^2 \cdot \textrm{det}_{\text{Ker}(P^{(\phi)}_-)} ~ \qeq^2}} \,,
\end{equation}
and we thus want to examine the kernels of the operators $P_{\pm}$ acting on the cohomological variables $\Phi, \Psi$ respectively. These kernel elements will be the only contributions to the one-loop determinant as they are the ``unpaired eigenmodes'' of the $Q^2_{eq}$ operator, with all other modes cancelling due to supersymmetry. 

We first want to look for solutions to the equation
\begin{equation}
    P_-^{(\phi)} \phi = 0\,,
\end{equation}
where 
\begin{equation}
    P_- = \i \epsilon \gamma^{\mu} \epsilon D_{\mu} =  \i \epsilon \gamma^{\mu} \epsilon \left( \nabla_{\mu} - \i \widehat{q}_{\text{R}} A_{\mu} - \i \widehat{q}_{\text{G}} \mathcal{A}_{\mu} \right)\,,
\end{equation}
and thus we look to solve 
\begin{equation} \label{eq: kernel_equation}
\partial_y \phi = -\frac{\i y}{2 P} (\partial_z - \i r A_z - \i q_{\text{G}} \mathcal{A}_z ) \phi\,,
\end{equation}
we want $\phi$ to be an eigenfunction of $Q^2_{eq}$ \eqref{eq: Q^2} (equivalently an eigenfunction of $\partial_z$) and thus we demand
\begin{equation}
    \phi = \exp \left( - 2\pi \i n \frac{z}{\Delta z} \right) \phi_n (y) \, , \qquad n \in \mathbb{Z}\,,
\end{equation}
and so the kernel equation \eqref{eq: kernel_equation} becomes
\begin{equation} \label{eq: Kernel_equation_phi}
    \frac{\phi_n'}{\phi_n} = - \frac{y}{2 P} \left( \frac{2\pi n}{\Delta z} + r A_z + q_{\text{G}} \mathcal{A}_z \right)\,, 
\end{equation}
so near $y=y_1$ we have
\begin{equation} 
    \phi_n (y) \sim (y-y_1)^{\frac{2\pi}{\Delta z} \frac{-y_1}{2(y_1-y_2)(y_1-y_3)}\left(n - r \frac{\eta }{2n_1} + q_{\text{G}} \frac{\mathfrak{m}_1}{n_1} \right)}\,,
\end{equation}
and near $y=y_2$ we have
\begin{equation}
     \phi_n (y) \sim (y_2-y)^{\frac{2\pi}{\Delta z} \frac{-y_2}{2(y_2-y_1)(y_2-y_3)}\left(n -  \frac{r}{2n_{2}} + q_{\text{G}} \frac{\mathfrak{m}_2}{n_2} \right)}\,,
\end{equation}
the analysis at the South Pole is identical for both twist and anti-twist, namely in that one has regularity for 
\begin{equation} \label{eq: inequality_south}
    n \geq \frac{r}{2 n_2} - q_{\text{G}} \frac{\mathfrak{m}_2}{n_2}\,,
\end{equation}
but the North Pole will differ for twist and anti-twist. For the anti-twist case, $y_1 < 0$ and we get 
\begin{equation}
    n \geq \frac{r}{2 n_1} - q_{\text{G}} \frac{\mathfrak{m}_1}{n_1}\,,
\end{equation}
for the twist we have $y_1>0$ and thus we get
\begin{equation} \label{eq: inequality_twist_north}
n \leq - \frac{r}{2n_1} - q_{\text{G}} \frac{\mathfrak{m}_1}{n_1} \,,
\end{equation}
we analysed the anti-twist case rather extensively in \cite{Jeon:2025rfc} and the present analysis shows that the result will not be any different for the anti-twisted spindle when starting from the STU model as opposed to minimal gauged supergravity. We can thus focus on the twist case in computing our one-loop determinants in this section, as this will be the novel calculation. 

\paragraph{Twist case:}

Combining the inequalities \eqref{eq: inequality_south} and \eqref{eq: inequality_twist_north}, we have 
\begin{equation} \label{eq: inequality_combined}
    \frac{r}{2n_2} - q_{\text{G}} \frac{\mathfrak{m}_2}{n_2} \leq n \leq - \frac{r}{2n_1} - q_{\text{G}} \frac{\mathfrak{m}_1}{n_1}\,,
\end{equation}
and if we introduce the notation of \cite{Inglese:2023tyc} 
\begin{equation}
    \mathfrak{p}_{1} = \eta \frac{r}{2} - q_{\text{G}} \mathfrak{m}_{1}\,, \qquad  \mathfrak{p}_{2} =  \frac{r}{2} - q_{\text{G}} \mathfrak{m}_{2}\,,
\end{equation}
then \eqref{eq: inequality_combined} becomes
\begin{equation}
    \frac{\mathfrak{p}_2}{n_2} \leq n \leq \frac{\mathfrak{p}_1}{n_1}\,,
\end{equation}
and since $n \in \mathbb{Z}$ we have
\begin{equation}
    \ceil*{\frac{\mathfrak{p}_2}{n_2}} = - \floor*{-\frac{\mathfrak{p}_2}{n_2}} \leq n \leq \floor*{\frac{\mathfrak{p}_1}{n_1}}.
\end{equation}
In our chosen gauge, the eigenfunctions of $Q^2_{eq}$ take the form
\begin{equation}
    Q^2_{eq} \phi = -\i \frac{L_0}{L} \left( n - q_{\text{G}} \frac{L}{L_0} \Lambda^{\text{G}} \right) \phi\,,
\end{equation}
so this contributes to the one-loop determinant \eqref{eq: 1-loop-det} in a product of the form 
\begin{equation} \label{eq: P_1_one_loop}
   P_1= \sqrt{\frac{1}{\prod\limits_{n = - \floor*{-\frac{\mathfrak{p}_2}{n_2}}}^{\floor*{\frac{\mathfrak{p}_1}{n_1}}} - \i \frac{L_0}{L}  \left( n - q_{\text{G}} \frac{L}{L_0} \Lambda^{\text{G}} \right)  }} 
\end{equation}
which may be part of the one-loop calculation, or not contribute at all, depending on the values of $r, q_{\text{G}}$. We introduce the notation of 
\begin{equation} \label{eq: b_c_defs}
    \mathfrak{b} = 1 -\eta \floor*{- \eta \frac{\mathfrak{p}_1}{n_1}} + \floor*{- \frac{\mathfrak{p}_2}{n_2}}\,, \qquad \mathfrak{c} = \frac{\llbracket - \mathfrak{p}_2 \rrbracket_{n_2}}{n_2} + \eta \frac{\llbracket - \eta \mathfrak{p}_1 \rrbracket_{n_1}}{n_1}\,,
\end{equation}
where we note that $\eta = -1$ is twist and $\eta = 1$ is anti-twist.\footnote{Note this is in slight contrast to \cite{Inglese:2023tyc}, where $\mathfrak{\sigma} = \pm1$ denotes twist and anti-twist respectively.} We also introduce, following \cite{Inglese:2023wky, Inglese:2023tyc, Jeon:2025rfc}
\begin{equation}
\chi_- = \frac{1}{n_1} - \frac{1}{n_2}\,, 
\end{equation}
and thus rewrite the contribution to the one-loop \eqref{eq: P_1_one_loop} as
\begin{equation}
    P_1 = \sqrt{\frac{1}{\prod\limits_{n = 0}^{\mathfrak{b}-1} - \i \frac{L_0}{L}  \left( n + \frac{1}{2} (1-\mathfrak{b} + \mathfrak{c}) - \frac{r}{4} \chi_- - q_{\text{G}} \gamma_{\text{G}} \right)  }}\,,
\end{equation}
which we note will only contribute non-trivially when $\mathfrak{b} \geq 1$. 
We can regularise this product as 
\begin{equation}
     P_1 = \sqrt{\frac{\prod\limits_{\widetilde{n} = 0}^{\infty} - \i \frac{L_0}{L}  \left( \widetilde{n} + \frac{1}{2} (1+\mathfrak{b} + \mathfrak{c}) - \frac{r}{4} \chi_- - q_{\text{G}} \gamma_{\text{G}} \right)  }{\prod\limits_{n = 0}^{\infty} - \i \frac{L_0}{L}  \left(n + \frac{1}{2} (1-\mathfrak{b} + \mathfrak{c}) - \frac{r}{4} \chi_- - q_{\text{G}} \gamma_{\text{G}} \right)  }}\,,
\end{equation}
an expression which we will return to when we finish discussing the one-loop result

 The next kernel equation we look to solve is 
\begin{equation}
    P_+ \widetilde{\phi} = 0\,,
\end{equation}
where 
\begin{equation}
    P_+ = \i \widetilde{\epsilon} \gamma^{\mu} \widetilde{\epsilon} D_{\mu}  \,,
\end{equation}
if we use the ansatz of 
\begin{equation}
    \widetilde{\phi}(y,z) = \exp \left( 2\pi \i m \frac{z}{\Delta z} \right) \widetilde{\phi}_m(y)\,,
\end{equation}
then we find an identical equation to \eqref{eq: Kernel_equation_phi} since $\widetilde{\epsilon} \gamma^{y} \widetilde{\epsilon} = - \epsilon \gamma^y \epsilon$, $\widetilde{\epsilon} \gamma^{z} \widetilde{\epsilon} =  \epsilon \gamma^z \epsilon$ and the R and gauge charges of $\widetilde{\phi}$ are $-r$ and $-q_{\text{G}}$ respectively. The eigenvalues are 
\begin{equation}
    Q^2_{eq} \widetilde{\phi} = \i \frac{L_0}{L} \left( m - q_{\text{G}} \frac{L}{L_0} \Lambda^{\text{G}} \right) \widetilde{\phi}\,,
\end{equation}
and thus we find a potential contribution to the one-loop determinant of the form 
\begin{align}
\begin{split}
\widetilde{P}_1 & = \sqrt{\frac{1}{\prod\limits_{m = - \floor*{-\frac{\mathfrak{p}_2}{n_2}}}^{\floor*{\frac{\mathfrak{p}_1}{n_1}}} \i \frac{L_0}{L}  \left( m - q_{\text{G}} \frac{L}{L_0} \Lambda^{\text{G}} \right)  }} \\ 
& =  \sqrt{\frac{\prod\limits_{\widetilde{m} = 0}^{\infty} \i \frac{L_0}{L}  \left( \widetilde{m} + \frac{1}{2} (1+\mathfrak{b} + \mathfrak{c}) - \frac{r}{4} \chi_- - q_{\text{G}} \gamma_{\text{G}} \right)  }{\prod\limits_{m = 0}^{\infty}  \i \frac{L_0}{L}  \left( m + \frac{1}{2} (1-\mathfrak{b} + \mathfrak{c}) - \frac{r}{4} \chi_- - q_{\text{G}} \gamma_{\text{G}} \right)  }}\,,
\end{split}
\end{align}
which also only contributes to the one-loop determinant for $\mathfrak{b} \geq 1$. We now turn to analyse the contributions from the fermions $\Psi = (\epsilon \psi, \widetilde{\epsilon} \widetilde{\psi})$.

We want to solve
\begin{equation}
    P_- (\widetilde{\epsilon} \widetilde{\psi} ) = 0 \,, \quad P_+(\epsilon \psi) = 0\,,
\end{equation}
with 
\begin{equation}
\widetilde{\epsilon} \widetilde{\psi} = \exp \left( -2 \pi \i l \frac{z}{\Delta z} \right) \widetilde{\epsilon} \widetilde{\psi}_l(y)\, , \quad  \epsilon \psi = \exp \left( -2 \pi \i p \frac{z}{\Delta z} \right) \epsilon \psi_p(y) \,, \quad l, p \in \mathbb{Z}\,,  
\end{equation}
then we find the same equations as \eqref{eq: Kernel_equation_phi} although now $r \rightarrow 2-r$ and $q_{\text{G}} \rightarrow -q_{\text{G}}$. This means that the bounds (e.g. for $l$) become  
\begin{equation}
     1 + \floor*{- \frac{\mathfrak{p}_2}{n_2}} \leq l \leq  -1 - \floor*{ \frac{\mathfrak{p}_1}{n_1}} \,,
\end{equation}
and the eigenvalues are 
\begin{equation}
    Q^2_{eq} ( \widetilde{\epsilon} \widetilde{\psi} ) = - \i \frac{L_0}{L} \left( l + q_{\text{G}} \frac{L}{L_0} \Lambda^{\text{G}} \right) (\widetilde{\epsilon} \widetilde{\psi})\,, \quad Q^2_{eq} ( \epsilon \psi ) = \i \frac{L_0}{L} \left(p + q_{\text{G}} \frac{L}{L_0} \Lambda^{\text{G}} \right) (\widetilde{\epsilon} \widetilde{\psi})\,.
\end{equation}
If we first focus on the contribution from $\widetilde{\epsilon} \widetilde{\psi}$ then we have 
\begin{equation}
    P_2 = \sqrt{\prod\limits_{l = 1+ \floor*{- \frac{\mathfrak{p}_2}{n_2}}}^{-1-\floor*{ \frac{\mathfrak{p}_1}{n_1}}} - \i \frac{L_0}{L}  \left( l + q_{\text{G}} \frac{L}{L_0} \Lambda^{\text{G}} \right) }
\end{equation}
which we can rewrite as
\begin{equation}
    P_2 = \sqrt{\prod\limits_{l = 0}^{-1-\mathfrak{b}} - \i \frac{L_0}{L}  \left( l + \frac{1}{2} (1+ \mathfrak{b} - \mathfrak{c} ) + \frac{r}{4} \chi_- + q_{\text{G}} \gamma_{\text{G}} \right) }\, ,
\end{equation}
which we note only contributes to the one-loop determinant for $\mathfrak{b} \leq -1$. We can then regularise as 
\begin{equation}
    P_2 = \sqrt{\frac{\prod\limits_{l = 0}^{\infty} - \i \frac{L_0}{L}  \left( l + \frac{1}{2} (1+ \mathfrak{b} - \mathfrak{c} ) + \frac{r}{4} \chi_- + q_{\text{G}} \gamma_{\text{G}} \right) }{\prod\limits_{\widetilde{l} = 0}^{\infty} - \i \frac{L_0}{L}  \left( \widetilde{l} + \frac{1}{2} (1- \mathfrak{b} - \mathfrak{c} ) + \frac{r}{4} \chi_- + q_{\text{G}} \gamma_{\text{G}} \right) }}\, .
\end{equation}
Similarly, the contribution from $\epsilon \psi$ can be written as 
\begin{equation}
    \tilde{P}_2 = \sqrt{\frac{\prod\limits_{p = 0}^{\infty}  \i \frac{L_0}{L}  \left( p + \frac{1}{2} (1+ \mathfrak{b} - \mathfrak{c} ) + \frac{r}{4} \chi_- + q_{\text{G}} \gamma_{\text{G}} \right) }{\prod\limits_{\widetilde{p} = 0}^{\infty} \i \frac{L_0}{L}  \left( \widetilde{p} + \frac{1}{2} (1- \mathfrak{b} - \mathfrak{c} ) + \frac{r}{4} \chi_- + q_{\text{G}} \gamma_{\text{G}} \right) }}\, .
\end{equation}

We now want to compute the full one-loop determinant in both the cases $\mathfrak{b} \geq 1$, when only the bosons $\Phi = (\phi, \widetilde{\phi})$ contribute, as well as for $\mathfrak{b} \leq -1$, in which case the fermions $\Psi = (\epsilon \psi, \widetilde{\epsilon} \widetilde{\psi})$ contribute. We start with the bosonic result and write 
\begin{equation} \label{eq: Z_boson}
    Z^{\Phi}_{\text{1-L}} = P_1 \widetilde{P}_1 = \sqrt{\left(\frac{L}{L_0} \right)^{2 \mathfrak{b}} \frac{\Gamma\left(\frac{1}{2} (1-\mathfrak{b} + \mathfrak{c}) - \frac{r}{4} \chi_- - q_{\text{G}} \gamma_{\text{G}}\right)^2}{\Gamma \left( \frac{1}{2} (1+\mathfrak{b} + \mathfrak{c} )- \frac{r}{4} \chi_- - q_{\text{G}} \gamma_{\text{G}} \right)^2} }\,,
\end{equation}
where in doing this we used the zeta function regularisation \cite{e684b263-94c7-31f3-8829-20ce2a1af791} 
\begin{equation}
    \prod\limits_{n = 0}^{\infty} \left( \frac{n+x}{Y} \right) = \frac{Y^{-\frac{1}{2} +x}}{\Gamma(x)}\,.
\end{equation}
We can apply the identical regularisation procedure to the fermionic one-loop determinant and we find
\begin{equation} \label{eq: Z-fermion}
    Z^{\Psi}_{\text{1-L}} = P_2 \widetilde{P}_2 = \sqrt{\left(\frac{L}{L_0} \right)^{2 \mathfrak{b}} \frac{\Gamma\left(\frac{1}{2} (1-\mathfrak{b} - \mathfrak{c}) + \frac{r}{4} \chi_- + q_{\text{G}} \gamma_{\text{G}}\right)^2}{\Gamma \left( \frac{1}{2} (1+\mathfrak{b} - \mathfrak{c} )+ \frac{r}{4} \chi_- + q_{\text{G}} \gamma_{\text{G}} \right)^2} }\,,
\end{equation}
now if one makes use of Euler's reflection formula for gamma functions 
\begin{equation}
    \Gamma(z) \Gamma(1-z) = \frac{\pi}{\sin (\pi z)} \,, \quad z \notin \mathbb{Z} \,, 
\end{equation}
then we find 
\begin{equation} \label{eq: Z_boson=Z_fermion}
    Z^{\Psi}_{\text{1-L}} = Z^{\Phi}_{\text{1-L}} = Z_{\text{1-L}}\,,
\end{equation}
where up to an overall sign we have 
\begin{equation}
    Z_{\text{1-L}} = \left(\frac{L}{L_0} \right)^{\mathfrak{b}} \frac{\Gamma\left(\frac{1}{2} (1-\mathfrak{b} - \mathfrak{c}) + \frac{r}{4} \chi_- + q_{\text{G}} \gamma_{\text{G}}\right)}{\Gamma \left( \frac{1}{2} (1+\mathfrak{b} - \mathfrak{c} )+\frac{r}{4} \chi_- + q_{\text{G}} \gamma_{\text{G}} \right)}\,,
\end{equation}
where we note the ratio of gamma functions matches the result in equation (3.85) of \cite{Inglese:2023tyc} for $\mathfrak{\sigma} =1$, when one uses the dictionary given in equations (5.69), (5.70) and (5.71) of \cite{Jeon:2025rfc}. A novel aspect of our calculation, as in \cite{Jeon:2025rfc}, is that we keep careful track of the $L/L_0$ ratio which enters the one-loop determinant, including the (moduli dependent) exponent $\mathfrak{b}$. We note that this exponent is different to that which appeared in the anti-twist case \cite{Jeon:2025rfc}. This will also have an effect on the final partition function. 

\subsection{Fixed point theorem}

The second way of arriving at the one-loop determinants is to use the fixed point theorem:
\begin{equation} \label{eq: index_orbifold}
    \text{ind}_{\text{orb}}(D_{10}; g) = \sum_{p=1}^2 \frac{1}{n_p} \sum_{w \in \mathbb{Z}_{n_p}} \frac{\text{tr}_{\Gamma_0} (w g) - \text{tr}_{\Gamma_1}( w g)}{\det (1 - W_p J_p)}\,,
\end{equation}
where we use all of the same conventions as \cite{Jeon:2025rfc}.

In order to use the fixed point theorem, we need to work in regular gauges at the poles. This means that the gauge fields take the form
\begin{equation}
    \left. A \right|_{\mathcal{U}_i} = \frac{1}{2} \left[ \sum_{I =1}^3 \left( \frac{y}{h_I(y)}  - \frac{y_i}{h_I(y_i)}\right) \right] \textrm{d}z\,, \qquad \left. \mathcal{A} \right|_{\mathcal{U}_i} =    \frac{L \rho_0}{H^{1/3}} (y_i - y) \textrm{d}z\,,
\end{equation}
and, as we have already computed in \eqref{eq: Lambda_R_regular} and \eqref{eq: Lambda^G_regular},
\begin{equation}
 \left. \Lambda^{\text{R}} \right|_{\mathcal{U}_i} = - \frac{L_0}{L} \frac{\eta_i}{2n_i}\,, \qquad \left. \Lambda^{\text{G}} \right|_{\mathcal{U}_i} = \Lambda^{\text{G}} + \frac{L_0}{L} \frac{\mathfrak{m}_i}{n_i}\,.
\end{equation}
This means that the equivariant supercharge squared is given by 
\begin{equation}
    \left. Q_{eq}^2 \right|_{\mathcal{U}_i} = \frac{L_0}{L} \left( \mathcal{L}_{\partial_\phi} - \i \frac{\widehat{\mathfrak{p}}_i}{n_i} + \i \widehat{q}_G  \frac{L}{L_0} \Lambda^{\text{G}}  \right) \,,
\end{equation}
and hence the map $g$ appearing in the orbifold index \eqref{eq: index_orbifold} is 
\begin{equation}
    \left. g \right|_{\mathcal{U}_i} = \exp \left( t \frac{L_0}{L} \left( \mathcal{L}_{\partial_\phi} - \i \frac{\widehat{\mathfrak{p}}_i}{n_i} + \i \widehat{q}_G  \frac{L}{L_0} \Lambda^{\text{G}} \right) \right) \,,
\end{equation}
which we separate as in \cite{Inglese:2023wky,Inglese:2023tyc, Jeon:2025rfc} into 
\begin{equation}
   \left. g  \right|_{\mathcal{U}_{1,2}} = g_{(0)} \left. g_{\mathbb{\Sigma}} \right|_{\mathcal{U}_{1,2}}  \, ,
\end{equation}
where  
\begin{equation}
    \left. g_{\mathbb{\Sigma}} \right|_{\mathcal{U}_{i}} = e^ { {\color{black}{\frac{L_0}{L}}} t  \left( \partial_{\phi} - \i \frac{\widehat{\mathfrak{p}}_{i}}{n_{i}} \right) }\,, \qquad {g_{(0)} = e^{\i t \widehat{q}_{\text{G}}  \Lambda^{\text{G}}}}\,.
\end{equation}

The calculation now proceeds in identical fashion to that of \cite{Jeon:2025rfc}. In particular, since $\mathfrak{p}_2^{\text{twist}} = \mathfrak{p}_2^{\text{anti-twist}}$ we find that the contribution from the fixed point at the South Pole $y=y_2$ is identical to the result presented in \cite{Jeon:2025rfc}
\begin{equation} \label{eq: south_pole_orb_index}
     I_{\mathcal{U}_2} = \mathfrak{q}^{-q_{\text{G}} \frac{L}{L_0} \Lambda^{\text{G}}} \frac{\mathfrak{q}^{-\floor*{-\mathfrak{p}_2/n_2}}}{1-\mathfrak{q}}-  \mathfrak{q}^{q_{\text{G}} \frac{L}{L_0} \Lambda^{\text{G}}} \frac{\mathfrak{q}^{1+\floor*{-\mathfrak{p}_2/n_2}}}{1-\mathfrak{q}}\,,
\end{equation}
where $\mathfrak{q}= \textrm{e}^{-\i \frac{L_0}{L} t} = q_1^{n_1} = q_2^{n_2}$. The differences in the twist vs anti-twist cases are entirely contained in the contribution from the fixed point at the North Pole of the spindle. We will now derive the twist case since the anti-twist is already given in \cite{Jeon:2025rfc}. 

\paragraph{Twist case:} 
First we give a brief aside on the coordinates used for the spindle near $y=y_1$. From \eqref{eq: metric_near_poles} we have 
\begin{equation}
     \textrm{d}s^2_{\mathbb{\Sigma}} \simeq L^2 \frac{|y_1|^{1/3}}{4|P'(y_1)|} \left( \textrm{d}\varrho_1^2 + \left( \frac{2\pi}{\Delta z } \frac{1}{n_1} \right)^2 \varrho_1^2 \textrm{d}z^2 \right) \,,
\end{equation}
then redefining 
\begin{equation} \label{eq: varrho_rescaling}
    \varrho_1 = \frac{2|P'(y_1)|^{1/2}}{|y_1|^{1/6}} \widetilde{\varrho}_1\,,
\end{equation} and recalling \eqref{eq: killing_phi} for the definition of the $2\pi$-periodic coordinate $\phi$ we have 
\begin{equation}
    \textrm{d}s^2_{\mathbb{\Sigma}} \simeq L^2 \left( \textrm{d}\widetilde{\varrho}_1^2 + \frac{1}{n_1^2} \textrm{d}\phi^2 \right) = L^2 \textrm{d}w_1 \textrm{d}\bar{w}_1\,,  
\end{equation}
where 
\begin{equation} \label{eq: complex_coords_north}
    w_1 = \widetilde{\varrho}_1 \exp(\i \phi/n_1)\, , \quad \bar{w}_1 = \widetilde{\varrho}_1 \exp (-\i \phi/n_1) \,, \qquad w_1 \sim \textrm{e}^{2\pi \i/n_1} w_1 \equiv u_1 w_1\,.
\end{equation}
This form of the coordinates is crucial as it will help us identify how the operator $g_{\mathbb{\Sigma}}$ acts on the coordinates $w_1$. Explicitly, the action on the coordinates is 
\begin{equation}
    \left. g_{\mathbb{\Sigma}} \right|_{\mathcal{U}_{1}} \circ w_1 = q_1 w_1\,, \quad  \left. g_{\mathbb{\Sigma}} \right|_{\mathcal{U}_{1}} \circ \bar{w}_1 = q_1^{-1} \bar{w}_1\,, \qquad q_1 = \textrm{e}^{-\i {\color{black}\frac{L_0}{L}} \frac{t}{n_1}}\,, 
\end{equation} 
and the equivariant action on the bosonic fields $\Phi= \{ \phi, \widetilde{\phi}\}  \in \Gamma_0$ is
\begin{equation}
    \left. g_{\mathbb{\Sigma}} \right|_{\mathcal{U}_{1}} \circ \phi = q_1^{\mathfrak{p}_1} \phi\,, \qquad  \left. g_{\mathbb{\Sigma}} \right|_{\mathcal{U}_{1}} \circ \widetilde{\phi} = q_1^{-\mathfrak{p}_1} \widetilde{\phi}\,,
\end{equation}
where the difference in the signs in the exponents arises from the opposite R and gauge charge signs between $\phi$ and $\widetilde{\phi}$. In order to compute the equivariant action on fields in $\Gamma_1 = D_{10}(\Gamma_0)$, we can use the fact that $D_{10}$ maps from bosonic to fermionic quantities in the cohomological complex. Thus the action on fermions $\Psi = \{\epsilon \psi, \widetilde{\epsilon} \widetilde{\psi} \}$ is
\begin{equation} \label{eq: action_fermions}
     \left. g_{\mathbb{\Sigma}} \right|_{\mathcal{U}_{1}} \circ (\epsilon \psi) = q_1^{\mathfrak{p}_1+1}(\epsilon \psi)\,, \qquad \left. g_{\mathbb{\Sigma}} \right|_{\mathcal{U}_{1}} \circ (\widetilde{\epsilon}\widetilde{\psi}) = q_1^{-(\mathfrak{p}_1+1)}(\widetilde{\epsilon}\widetilde{\psi})\,. \qquad 
\end{equation}
which we see is in contrast to the anti-twist case, since $\mathfrak{p}_1^{\text{twist}} \neq \mathfrak{p}_1^{\text{anti-twist}}$. 

As an aside, we also note that the index of $D_{10}$ is equivalent to that of the operator 
\begin{equation}
    \mathcal{P} = \Biggl( \ba{cc} 0 & P_+  \\ P_- & 0 \ea  \Biggr)\,.
\end{equation}
In order to show this, we consider the form of the operators $P_{\pm}$ near the poles of the spindle
\begin{equation}
   \left. P_- \right|_{\mathcal{U}_1} = \left. \i \epsilon \gamma^{\mu} \epsilon D_{\mu} \right|_{\mathcal{U}_1} = -\frac{2k_0 |y_1|^{1/6}}{L} \partial_{\bar{w}_1}\,, \qquad  \left. P_+ \right|_{\mathcal{U}_1} = \left. \i \widetilde{\epsilon} \gamma^{\mu} \widetilde{\epsilon} D_{\mu} \right|_{\mathcal{U}_1} = \frac{2k_0 |y_1|^{1/6}}{L} \partial_{w_1}\,,
\end{equation}
where for comparison the values in the regular gauges at the South Pole are
\begin{equation}
    \left. P_- \right|_{\mathcal{U}_2} = \left. \i \epsilon \gamma^{\mu} \epsilon D_{\mu} \right|_{\mathcal{U}_2} = \frac{2k_0 |y_2|^{1/6}}{L} \partial_{\bar{w}_2}\,, \qquad  \left. P_+ \right|_{\mathcal{U}_2} = \left. \i \widetilde{\epsilon} \gamma^{\mu} \widetilde{\epsilon} D_{\mu} \right|_{\mathcal{U}_2} = - \frac{2k_0 |y_2|^{1/6}}{L} \partial_{w_2}\,,
\end{equation}
where the complex coordinates in the southern patch are
\begin{equation}
     w_2 = \widetilde{\varrho}_2 \exp(-\i \phi/n_2)\, , \quad \bar{w}_2 = \widetilde{\varrho}_2 \exp (\i \phi/n_2) \,, \qquad w_2 \sim \textrm{e}^{-2\pi \i/n_2} w_2 = u_2^{-1} w_2\,.
\end{equation}
which is the same for both twist and anti-twist. For comparison the values of the operators for the anti-twisted spindle in the northern patch are 
\begin{equation}
    \left. P^{\text{anti-twist}}_- \right|_{\mathcal{U}_1} = - \frac{2 k_0 |y_1|^{1/6}}{L} \partial_{w_1} \,, \qquad  \left. P^{\text{anti-twist}}_+ \right|_{\mathcal{U}_1}  = \frac{2k_0 |y_1|^{1/6}}{L} \partial_{\bar{w}_1} \,,
\end{equation}
all of which can be derived directly using the coordinate maps given in equations \eqref{eq: varrho_rescaling} and \eqref{eq: complex_coords_north}, together with the known facts that $\eta = \mp1$ for twist/anti-twist and $y_1>0$ ($<0$) for twist (anti-twist). Returning to the contribution to the spindle index from the North Pole fixed point for the twisted spindle, we note that the action on the image under $\mathcal{P}$ is
\begin{align}
    \left. g_{\mathbb{\Sigma}} \right|_{\mathcal{U}_{1}} \circ ( P_- \phi) =  -\frac{2k_0 |y_1|^{1/6}}{L} \left. g_{\mathbb{\Sigma}} \right|_{\mathcal{U}_{1}} \circ ( \partial_{\bar{w}_1} \phi) = q_1^{\mathfrak{p}_1+1}( P_- \phi )\,,\\
     \left. g_{\mathbb{\Sigma}} \right|_{\mathcal{U}_{1}} \circ ( P_+ \widetilde{\phi}) =  \frac{2k_0 |y_1|^{1/6}}{L} \left. g_{\mathbb{\Sigma}} \right|_{\mathcal{U}_{1}} \circ ( \partial_{w_1} \widetilde{\phi}) = q_1^{-\mathfrak{p}_1-1}( P_+ \widetilde{\phi} )\,,
\end{align}
which matches the eigenvalues of \eqref{eq: action_fermions} and illustrates that the index computed with respect to $\mathcal{P}$ is equivalent to that of $D_{10}$, see also the similar discussion in \cite{Jeon:2025rfc}. 

Returning to the evaluation of the index of $D_{10}$, we can now proceed with computing the \textit{manifold} contribution to the twisted spindle index at the North Pole fixed point as
\begin{align}
\begin{split}
    I_1 & = \frac{\mathfrak{q}^{q_{\text{G}} \frac{L}{L_0} \Lambda^{\text{G}}} (q_1^{-\mathfrak{p}_1} - q_1^{-1-\mathfrak{p}_1}) + \mathfrak{q}^{-q_{\text{G}} \frac{L}{L_0} \Lambda^{\text{G}}} (q_1^{\mathfrak{p}_1} - q_1^{\mathfrak{p}_1+1})}{(1-q_1)(1-q_1^{-1})} \\
    & = \mathfrak{q}^{q_{\text{G}} \frac{L}{L_0} \Lambda^{\text{G}}}\frac{q_1^{-\mathfrak{p}_1}  }{1-q_1} +  \mathfrak{q}^{-q_{\text{G}} \frac{L}{L_0} \Lambda^{\text{G}}}\frac{q_1^{\mathfrak{p}_1}}{1-q_1^{-1}}\,,
    \end{split}
\end{align}
now upon performing the ``orbifold'' procedure to obtain the contribution to the orbifold index \eqref{eq: index_orbifold} we obtain 
\begin{align}
\begin{split}
I_{\mathcal{U}_1} & = \mathfrak{q}^{q_{\text{G}} \frac{L}{L_0} \Lambda^{\text{G}}} q_1^{-\mathfrak{p}_1} \frac{1}{n_1} \sum_{j=0}^{n_1-1} \frac{u_1^{-j \mathfrak{p}_1}}{1-u_1^{j} q_1} - \mathfrak{q}^{-q_{\text{G}} \frac{L}{L_0} \Lambda^{\text{G}}} q_1^{\mathfrak{p}_1} \frac{1}{n_1} \sum_{j=0}^{n_1-1} \frac{u_1^{j \mathfrak{p}_1}}{1-u_1^{-j} q_1^{-1}} \\
& = \mathfrak{q}^{q_{\text{G}} \frac{L}{L_0} \Lambda^{\text{G}}} \frac{\mathfrak{q}^{-\floor*{\mathfrak{p}_1/n_1}}}{1-\mathfrak{q}} - \mathfrak{q}^{-q_{\text{G}} \frac{L}{L_0} \Lambda^{\text{G}}} \frac{\mathfrak{q}^{1+\floor*{\mathfrak{p}_1/n_1}}}{1-\mathfrak{q}}\,,
\end{split}
\end{align}
where as in \cite{Inglese:2023tyc, Jeon:2025rfc} we made use of the identities
\begin{equation}
    \frac{1}{n_1} \sum_{j=0}^{n_1-1} \frac{u_1^{ja}}{1-u_1^j q} = \frac{1}{n_1} \sum_{j=0}^{n_1-1} \frac{u_1^{-ja}}{1-u_1^{-j} q} = \frac{q^{\llbracket -a \rrbracket_{n_1}}}{1-q^{n_1}}\,, \qquad 
\floor*{\frac{\bullet}{n_{1}}}  = \frac{\bullet}{n_{1}}- \frac{\llbracket \bullet \rrbracket_{n_{1}}}{n_{1}}\,.  
\end{equation}
The contribution from the South Pole is \eqref{eq: south_pole_orb_index} and thus we can combine these two terms in order to write 
\begin{equation}
 \text{ind}_{\text{orb}}(D_{10}; g) = \mathfrak{q}^{q_{\text{G}} \frac{L}{L_0} \Lambda^{\text{G}}} \frac{\mathfrak{q}^{-\floor*{\mathfrak{p}_1/n_1}} - \mathfrak{q}^{1+\floor*{-\mathfrak{p}_2/n_2}}}{1-\mathfrak{q}} - \mathfrak{q}^{-q_{\text{G}} \frac{L}{L_0} \Lambda^{\text{G}}} \frac{\mathfrak{q}^{1+\floor*{\mathfrak{p}_1/n_1}}-\mathfrak{q}^{-\floor*{-\mathfrak{p}_2/n_2}}}{1-\mathfrak{q}}\,.
 \end{equation}
 
We now want to convert the index above into the expression for the one-loop determinants. We start by considering the case of $\mathfrak{b} =0$ i.e. 
\begin{equation}
    0= 1 +\floor*{ \frac{\mathfrak{p}_1}{n_1}} + \floor*{- \frac{\mathfrak{p}_2}{n_2}}\,,
\end{equation}
from which one can immediately see that 
\begin{equation}
    \text{ind}_{\text{orb}}(D_{10}; g) \stackrel{\mathfrak{b}=0}{=} 0\,, 
\end{equation}
which is as expected from the method of unpaired eigenvalues. 

Now we consider the case of $\mathfrak{b} \geq 1$. We can write the index as
\begin{align}
\begin{split} \label{eq: index_b>1}
    \text{ind}_{\text{orb}}(D_{10}; g) & = \left(\mathfrak{q}^{q_{\text{G}} \frac{L}{L_0} \Lambda^{\text{G}}} \mathfrak{q}^{-\floor*{\mathfrak{p}_1/n_1}} + \mathfrak{q}^{-q_{\text{G}} \frac{L}{L_0} \Lambda^{\text{G}}} \mathfrak{q}^{-\floor*{-\mathfrak{p}_2/n_2}}  \right) \frac{1 - \mathfrak{q}^\mathfrak{b}}{1-\mathfrak{q}} \\
    & = \left(\mathfrak{q}^{q_{\text{G}} \frac{L}{L_0} \Lambda^{\text{G}}} \mathfrak{q}^{-\floor*{\mathfrak{p}_1/n_1}} + \mathfrak{q}^{-q_{\text{G}} \frac{L}{L_0} \Lambda^{\text{G}}} \mathfrak{q}^{-\floor*{-\mathfrak{p}_2/n_2}}  \right) (1 + \mathfrak{q} + \ldots + \mathfrak{q}^{\mathfrak{b} -1} )\,,
   \end{split} 
\end{align}
where we note that we expand in powers of $\mathfrak{q}$ at both North and South Poles, rather that $\mathfrak{q}^{-1}$ at the North Pole and $\mathfrak{q}$ at the South Pole as was done in the anti-twist case. Now recalling the original definition of $\mathfrak{q} = \textrm{e}^{-\i \frac{L_0}{L} t}$ we can rewrite the index as 
\begin{align}
\begin{split}
     \text{ind}_{\text{orb}}(D_{10}; g) & = \sum_{n=-\floor*{\mathfrak{p}_1/n_1}}^{\floor*{-\mathfrak{p}_2/n_2}} \textrm{e}^{-\i \frac{L_0}{L} t \left( n+ q_{\text{G}} \frac{L}{L_0} \Lambda^{\text{G}} \right)} + \sum_{m=-\floor*{-\mathfrak{p}_2/n_2}}^{\floor*{\mathfrak{p}_1/n_1}} \textrm{e}^{-\i \frac{L_0}{L} t \left( m- q_{\text{G}} \frac{L}{L_0} \Lambda^{\text{G}} \right)} \\
     & = \sum_{n=-\floor*{-\mathfrak{p}_2/n_2}}^{\floor*{\mathfrak{p}_1/n_1}} \textrm{e}^{\i \frac{L_0}{L} t \left( n- q_{\text{G}} \frac{L}{L_0} \Lambda^{\text{G}} \right)} + \sum_{m=-\floor*{-\mathfrak{p}_2/n_2}}^{\floor*{\mathfrak{p}_1/n_1}} \textrm{e}^{-\i \frac{L_0}{L} t \left( m- q_{\text{G}} \frac{L}{L_0} \Lambda^{\text{G}} \right)}\,,
     \end{split}
\end{align}
and finally using the rule
\begin{equation}
{\rm ind}_{\text{orb}}( D_{10};g) = \sum_n a(n){\rm e}^{ \lambda_n t} \quad \Rightarrow \quad Z^{ \qeq \cV^{\text{c.m.}}}_{1\text{-loop}}   \=\prod_n \lambda_n^{-\frac{1}{2}a(n)}\,,\label{eq:indDeg}
\end{equation}
we are able to write the one-loop determinant as 
\begin{align}
\begin{split}
    \quad Z^{ \qeq \cV^{\text{c.m.}}}_{1\text{-loop}} & \stackrel{\mathfrak{b} \geq 1}{=} \left( \prod\limits_{n=-\floor*{-\mathfrak{p}_2/n_2}}^{\floor*{\mathfrak{p}_1/n_1}} \i \frac{L_0}{L} \left( n - q_{\text{G}} \frac{L}{L_0} \Lambda^{\text{G}} \right) \prod\limits_{m=-\floor*{-\mathfrak{p}_2/n_2}}^{\floor*{\mathfrak{p}_1/n_1}}- \i \frac{L_0}{L} \left( m - q_{\text{G}} \frac{L}{L_0} \Lambda^{\text{G}} \right) \right)^{-1/2} \\
    & = P_1 \widetilde{P}_1\,,
    \end{split}
\end{align}
demonstrating that this method of computing the one-loop determinant precisely matches that of the eigenvalue method for $\mathfrak{b} \geq 1$ \eqref{eq: Z_boson}. We also note that this reproduces the unpaired eigenvalues coming from the bosonic sections $\Phi = (\phi, \widetilde{\phi})$. 

Now we examine the case of $\mathfrak{b} \leq -1$. We can write the index as 
\begin{align}
\begin{split} \label{eq: index_b<-1}
    \text{ind}_{\text{orb}}(D_{10}; g) & = - \left(\mathfrak{q}^{q_{\text{G}} \frac{L}{L_0} \Lambda^{\text{G}}} \mathfrak{q}^{1+\floor*{-\mathfrak{p}_2/n_2}} + \mathfrak{q}^{-q_{\text{G}} \frac{L}{L_0} \Lambda^{\text{G}}} \mathfrak{q}^{1+\floor*{\mathfrak{p}_1/n_1}}  \right) \frac{1 - \mathfrak{q}^{-\mathfrak{b}}}{1-\mathfrak{q}} \\
    & = - \left(\mathfrak{q}^{q_{\text{G}} \frac{L}{L_0} \Lambda^{\text{G}}} \mathfrak{q}^{1+\floor*{-\mathfrak{p}_2/n_2}} + \mathfrak{q}^{-q_{\text{G}} \frac{L}{L_0} \Lambda^{\text{G}}} \mathfrak{q}^{1+\floor*{\mathfrak{p}_1/n_1}}  \right) (1 + \mathfrak{q} + \ldots + \mathfrak{q}^{-\mathfrak{b} -1} )\,,
   \end{split} 
\end{align}
and thus we can convert this into the sum 
\begin{align}
\begin{split}
     \text{ind}_{\text{orb}}(D_{10}; g) & = - \sum_{l=1+\floor*{-\mathfrak{p}_2/n_2}}^{-1-\floor*{\mathfrak{p}_1/n_1}} \textrm{e}^{-\i \frac{L_0}{L} t \left( l+ q_{\text{G}} \frac{L}{L_0} \Lambda^{\text{G}} \right)} - \sum_{p=1+\floor*{-\mathfrak{p}_2/n_2}}^{-1-\floor*{\mathfrak{p}_1/n_1}} \textrm{e}^{\i \frac{L_0}{L} t \left( p+ q_{\text{G}} \frac{L}{L_0} \Lambda^{\text{G}} \right)} \,,
     \end{split}
\end{align}
and finally using the rule \eqref{eq:indDeg} we can convert this into the one-loop determinant 
\begin{align}
\begin{split}
    \quad Z^{ \qeq \cV^{\text{c.m.}}}_{1\text{-loop}} & \stackrel{\mathfrak{b} \leq -1}{=} \left( \prod\limits_{l = 1+\floor*{-\mathfrak{p}_2/n_2}}^{-1-\floor*{\mathfrak{p}_1/n_1}} -\i \frac{L_0}{L} \left( n + q_{\text{G}} \frac{L}{L_0} \Lambda^{\text{G}} \right) \prod\limits_{p=1+\floor*{-\mathfrak{p}_2/n_2}}^{-1-\floor*{\mathfrak{p}_1/n_1}} \i \frac{L_0}{L} \left( p + q_{\text{G}} \frac{L}{L_0} \Lambda^{\text{G}} \right) \right)^{1/2} \\
    & = P_2 \widetilde{P}_2\,,
    \end{split}
\end{align}
which precisely matches the eigenvalue method for $\mathfrak{b} \leq -1$ \eqref{eq: Z-fermion}. We note that in the eigenvalue argument this contribution came entirely from the fermionic sections $\Psi = (\epsilon \psi, \widetilde{\epsilon} \widetilde{\psi})$. As was shown in \eqref{eq: Z_boson=Z_fermion}, we have $P_1 \widetilde{P}_1 = P_2 \widetilde{P}_2$ and thus the chiral one-loop determinant computed via the fixed point theorem agrees precisely with the method of unpaired eigenvalues. 

In summary, the one-loop determinant for the twisted spindle is given by 
    \begin{equation} \label{eq: one_loop_twist}
    Z_{\text{1-L}} = \left(\frac{L}{L_0} \right)^{\mathfrak{b}} \frac{\Gamma\left(\frac{1}{2} (1-\mathfrak{b} - \mathfrak{c}) + \frac{r}{4} \chi_- + q_{\text{G}} \gamma_{\text{G}}\right)}{\Gamma \left( \frac{1}{2} (1+\mathfrak{b} - \mathfrak{c} )+\frac{r}{4} \chi_- + q_{\text{G}} \gamma_{\text{G}} \right)}\,,
\end{equation}
and thus we are almost ready to compute the partition function. Before doing this, we first provide a brief discussion of the analogous vector multiplet one-loop determinant contribution.

\subsection{Comment on the vector multiplet one-loop determinant}

We now turn to computing the vector multiplet one-loop determinant, which will require us taking into account the contribution of zero modes more carefully. First we recall \cite{Inglese:2023tyc} in that the non-equivariant $\mathfrak{q} \rightarrow 1$ limit of the index \eqref{eq: index_b>1} (equivalently \eqref{eq: index_b<-1}) gives 
\begin{equation} \label{eq: RRK_theorem}
    \lim_{\mathfrak{q} \rightarrow 1} \text{ind}_{\text{orb}}(D_{10}; g) = \text{ind}_{\text{orb}}(D_{10}) = 2\mathfrak{b}\,,
\end{equation}
which counts (minus) the number of zero-modes of the operator $D_{10}$ and reproduces the Riemann-Roch-Kawasaki theorem for genus zero orbifolds \cite{zbMATH03629358, Closset:2018ghr, Inglese:2023tyc}. 

When computing the contribution from zero modes in the partition function, we are only concerned about the contribution to the vector multiplet. As explained in \cite{GonzalezLezcano:2023cuh}, the chiral multiplet does not admit any zero modes but the vector multiplet does admit two fermionic zero modes, namely the constant modes of $c, \widetilde{c}$, which are Ghost fields added to the vector multiplet. These zero modes are fermionic and removing these from the cohomological complex we have
\begin{equation}
    (\Phi, Q_{eq} \Phi, \Psi', Q_{eq} \Psi')\,, 
\end{equation}
which has one-to-one correspondence between bosonic and fermionic variables. Following \cite{GonzalezLezcano:2023cuh}, we then compute 
\begin{align}
\begin{split}
    \Tr_{\Psi'} \textrm{e}^{t Q^2_{eq}} - \Tr_{\Phi} \textrm{e}^{t Q^2_{eq}} & = \Tr_{\Psi} \textrm{e}^{t Q^2_{eq}} - \Tr_{\Phi} \textrm{e}^{t Q^2_{eq}} - \Tr_{\Psi_{\text{zm}}} \mathbb{I}  \\
    & = - \text{ind}(D_{10})(t) - n_{\text{zm}}^{\Psi} \,,
    \end{split}
\end{align}
where $\text{ind}(D_{10})(t)$ is what we have just been computing for the chiral multiplet in order to evaluate the chiral multiplet one loop determinants. 

For the vector multiplet we now have 
\begin{equation}
    n_{\text{zm}}^{\Psi} = 2\,,
\end{equation}
which will contribute to the index via the following regularisation
\begin{equation} \label{eq: zero_mode_regularisation}
    - \frac{1}{2} \int_{\bar{\upvarepsilon}}^{\infty} \frac{dt}{t} \left. (-n_{\text{zm}}^{\Psi}) \right|_{\text{reg}} = \left( \frac{L}{L_0} \right)^{\frac{1}{2}n^{\Psi}_{\text{zm}}} \,,
\end{equation}
where the UV cutoff is $\bar{\upvarepsilon} \equiv \upvarepsilon \left(\frac{L_0}{L} \right)$. The full one-loop determinant with the zero modes correctly excluded is thus 
\begin{equation}
Z'^{Q_{eq} \mathcal{V}_{\text{v.m.}}}_{1-\text{loop}} = \left( \frac{L}{L_0} \right)^{\frac{1}{2} n^{\Psi}_{\text{zm}} } \prod_{n} \lambda_n^{-\frac{1}{2} a(n)}\,,
\end{equation}
where we again make use of the expansion of the index as in \eqref{eq:indDeg}. We thus need to evaluate the index for the vector multiplet cohomological complex
\begin{equation}
   \left. \text{ind}_{\text{orb}}(D_{10}; g) \right|_{q_{\text{G}} =0, r =2} = - 2 = a(0)\,,
\end{equation}
which also agrees with \eqref{eq: RRK_theorem} as $\mathfrak{b}|_{r=2, q_{\text{G}} = 0} = -1$. Applying the same regularisation procedure as \eqref{eq: zero_mode_regularisation} for the eigenvalues, we thus find a trivial vector multiplet contribution of the form 
\begin{equation}
    Z'^{Q_{eq} \mathcal{V}_{\text{v.m.}}}_{1-\text{loop}} =  \left( \frac{L}{L_0} \right)^{\frac{1}{2} n^{\Psi}_{\text{zm}} + \frac{1}{2} a(0)} =   \left( \frac{L}{L_0} \right)^{1-1} = 1\,.
\end{equation}
Finally, we note that the same phenomenon occurs for the anti-twisted spindle, and thus we also expect the vector multiplet one-loop determinant to be trivial in that context.  
\section{Partition function} \label{sec: partition_function}
\subsection{Contour integral}

We begin this section with a summary of the remaining integral and sum that one needs to perform after reducing the full partition function to a sum over BPS locus moduli. For both twist and anti-twist cases we have the following general expression for the full partition function
\begin{equation} \label{eq: Z_full}
     Z_{\mathbb{\Sigma}} = \sum_{\mathfrak{m} \in \mathbb{Z}} \int_{-\infty}^{\infty} \frac{\textrm{d}(\sigma_0 {\color{black}L})}{2\pi} \, Z_{\text{class}} (\sigma_0{\color{black}L},\mathfrak{m}) Z^{Q_{eq} \mathcal{V}_{\text{c.m.}}}_{\text{1-loop}}(\sigma_0{\color{black}L},\mathfrak{m}) \,,
\end{equation}
where, as in \cite{Jeon:2025rfc}, we showed that for the anti-twist case
\begin{equation} \label{eq: classical_antitwist}
    Z_{\text{class}} (\sigma_0{\color{black}L},\mathfrak{m}) =\left. Z_{\text{FI} + \text{top}} \right|_{\text{BPS}} = \textrm{e}^{\left. -S_{\text{FI}} - S_{\text{top}} \right|_{\text{BPS
}}} = \textrm{e}^{-4 \pi \xi \gamma_{\text{G}} - \i \theta \mathfrak{f}_{G}}\,,
\end{equation}
and 
\begin{eqnarray} \label{eq: one-loop-eigenmodes}
     Z^{ \qeq \cV}_{1\text{-loop}} = \textcolor{black}{\left( \frac{L}{L_0} \right)^{1-\frac{r}{2} \chi + 2 q_{\text{G}} \gamma_{\text{G}} - \mathfrak{c}}} \frac{\Gamma\left( \frac{\mathfrak{b}+\mathfrak{c}-1}{2} + \frac{r}{4} \chi  - {q_{\text{G}} \gamma_{\text{G}}} \right) }{\Gamma\left( \frac{\mathfrak{b}-\mathfrak{c}+1}{2} - \frac{r}{4} \chi + {q_{\text{G}} \gamma_{\text{G}}} \right) }\,.
\end{eqnarray}
One may be concerned that even though \eqref{eq: classical_antitwist} and \eqref{eq: one-loop-eigenmodes} have an identical structural form to the analogous terms in \cite{Jeon:2025rfc}, they are in fact not the same as those presented there due to the model-dependent term $\gamma_{\text{G}}$ \eqref{eq: gamma_G}, which is different for the STU and minimal gauged solutions. However, upon performing the change of integration variables from $\sigma_0 L$ to $\gamma_{\text{G}}$ using \eqref{eq: S_anti_twist} and recalling from \cite{Jeon:2025rfc} that the contour in the $\sigma_0 L$ plane picks up all poles, one can immediately see that the result matches the anti-twisted spindle partition function when the starting point was $D=5$ minimal gauged supergravity.\footnote{Equivalently, one can also show via explicit computation that the partition function is independent of $\beta$ in \eqref{eq: Lambda^G}, demonstrating that the STU and minimal gauged results will be equal.} 
 
The computation for the anti-twist case was performed in detail in \cite{Jeon:2025rfc} and thus we will not repeat it here. We will instead focus on the twist case for the remainder of this section.

\paragraph{Twist case:}

In this article we have computed the twist values to be 
\begin{equation} \label{eq: Z_class_twist}
    Z_{\text{class}} (\sigma_0{\color{black}L},\mathfrak{m}) =\left. Z_{\text{FI} + \text{top}} \right|_{\text{BPS}} = \textrm{e}^{-\frac{\mathfrak{m}}{n_1n_2}(2\pi \xi + \i \theta)} = \textrm{e}^{-\frac{\mathfrak{m}}{n_1n_2}\widetilde{\xi}} \,,
\end{equation}
and 
\begin{equation} \label{eq: Z_1-loop_twist}
 Z^{Q_{eq} \mathcal{V}}_{\text{1-loop}} = \left(\frac{L}{L_0} \right)^{\mathfrak{b}} \frac{\Gamma\left(\frac{1-\mathfrak{b} - \mathfrak{c}}{2}  + \frac{r}{4} \chi_- + q_{\text{G}} \gamma_{\text{G}}\right)}{\Gamma \left( \frac{1+\mathfrak{b} - \mathfrak{c}  }{2} +\frac{r}{4} \chi_- + q_{\text{G}} \gamma_{\text{G}} \right)}\,.
\end{equation}

We can thus compute the twist partition function by putting the results \eqref{eq: Z_class_twist} and \eqref{eq: Z_1-loop_twist} into \eqref{eq: Z_full} and computing the remaining integral and sum over moduli. We start as in \cite{Jeon:2025rfc} by considering the integrand in the partition function, namely 
\begin{equation}
    I_0 = \frac{\Gamma\left(\frac{1-\mathfrak{b} - \mathfrak{c}}{2}  + \frac{r}{4} \chi_- + q_{\text{G}} \gamma_{\text{G}}\right)}{\Gamma \left( \frac{1+\mathfrak{b} - \mathfrak{c}  }{2} +\frac{r}{4} \chi_- + q_{\text{G}} \gamma_{\text{G}} \right)}\,,
\end{equation}
and we want to consider the behaviour of this quantity for large values of $\sigma_0$. Recalling the relation between $\gamma_{\text{G}}$ and $\sigma_0$ via \eqref{eq: gamma_G}, we find for $\mathfrak{b} \sim \mathcal{O}(1)$ the asymptotic behaviour is 
\begin{equation}
\lim_{|\sigma_0| \rightarrow \infty} I_0 = 1\,,
\end{equation}
so we cannot use the same type of semi-circular contour that we chose for the anti-twist case since the integrand does not vanish on the circular portion of the contour. Instead one should choose the contour of integration according to the Jeffrey-Kirwan prescription \cite{JEFFREY1995291, 1999math......3178B, Szenes:2004cpv}, which needs to be identified for the case of the twisted spindle. In order to choose the contour, we use the reference \cite{Closset:2015rna}, which considers localisation computations on the $\Omega$-deformed sphere $S^2_{\Omega}$. In particular, the supergravity background for $S^2_{\Omega}$ is characterised by one unit of flux for the $U(1)_{\text{R}}$ gauge field
\begin{equation}
    \frac{1}{2\pi} \int_{S^2} \textrm{d} A^{(R)} = -1\,,
\end{equation}
which we see matches the limit of \eqref{eq: Q^R} as we take $n_1 = n_2 =1$. We also note that setting $n_1=n_2=1$ in the one-loop determinant \eqref{eq: Z_1-loop_twist} recovers the chiral multiplet one-loop determinant of \cite{Closset:2015rna}, see appendix \ref{app: one_loop_comparisons} for more discussion on this point.

In choosing our contour, we follow the arguments of \cite{Closset:2015rna}, in particular the example of the abelian Higgs model on $S^2_{\Omega}$. As in \cite{Witten:1993yc}, see also \cite{Closset:2015rna}, we define the effective FI parameter on the Coulomb branch \textit{at infinity}\footnote{Note that while \eqref{eq: xi_UV} looks structurally similar to the renormalisation of $\xi$ found in the anti-twist case ((6.2) in \cite{Jeon:2025rfc}), we would like to point out that these two phenomena are different kinds of renormalisation. This one is due to the running of the FI parameter at ininifty on the Coulomb branch whereas the other is a renormalisation with respect to the scale $L/L_0$, used to partially absorb the dependence of $\sigma_0 L$ in the integral.}  via 
\begin{equation} \label{eq: xi_UV}
    \xi^{\text{UV}}_{\text{eff}} = \xi + \frac{q_{\text{G}}}{2\pi} \lim_{R \rightarrow \infty} \log R\,,
\end{equation}
and note that for $q_{\text{G}} >0$ which we can take w.l.o.g. then we have $\xi^\text{UV}_{\text{eff}} = + \infty$ and this forces a non-vanishing VEV for the chiral multiplet $\Phi$. Following the JK prescription of \cite{Closset:2014pda}, we thus conclude that the contour takes into account the residues at \textit{all} poles of the integrand in \eqref{eq: Z_full} and we should thus compute all of the residues in order to arrive at the twisted spindle partition function.

In order to identify all poles we first rewrite the ratio of gamma functions as
\begin{equation} \label{eq: gamma_ratio_twist}
  I_0 = \frac{\Gamma\left(\frac{1-\mathfrak{b} - \mathfrak{c}}{2}  + \frac{r}{4} \chi_- + q_{\text{G}} \gamma_{\text{G}}\right)}{\Gamma \left( \frac{\mathfrak{b} - \mathfrak{c} +1 }{2} +\frac{r}{4} \chi_- + q_{\text{G}} \gamma_{\text{G}} \right)} = \frac{\Gamma(\widetilde{\sigma}_0)}{\Gamma(\widetilde{\sigma}_0+\mathfrak{b})}\,,
\end{equation}
where we used the change of variable
\begin{equation}
    \widetilde{\sigma}_0 = \frac{1-\mathfrak{b} - \mathfrak{c}}{2}  + \frac{r}{4} \chi_- + q_{\text{G}} \gamma_{\text{G}}\,,
\end{equation} 
which corresponds to shifting the contour of integration in $\mathbb{C}$ and rescaling the integration variable via
\begin{equation}
    \textrm{d}(\sigma_0 L) = - \frac{\i}{q_{\text{G}}} \textrm{d} \widetilde{\sigma}_0 \,,
\end{equation}
Using the elementary properties of the gamma functions, we see that there are poles in the numerator of \eqref{eq: gamma_ratio_twist} at
\begin{equation}
    \widetilde{\sigma}_0 = -n\,, \qquad n = 0, 1,2,\ldots\,.
\end{equation}
However, the argument of the gamma function in the denominator evaluated at these potential poles is $\mathfrak{b}-n$ which would correspond to poles of the denominator (and cancel those of the numerator) for 
\begin{equation}
      \mathfrak{b}-n  = -h\,, \qquad h=0,1,2, \ldots\,,
\end{equation} i.e., for
$\mathfrak{b} \leq n  \,.$ We can thus conclude that the one-loop integrand has uncancelled  poles for $\mathfrak{b}>n_{\textrm{min}}=0$ and no poles otherwise. Recalling the definition of $\mathfrak{b}$ in \eqref{eq: b_c_defs} we have
\begin{equation} \label{eq: b>0}
    \mathfrak{b} > 0 \quad \Leftrightarrow \quad  \floor*{ \frac{\mathfrak{p}_1}{n_1}} + \floor*{- \frac{\mathfrak{p}_2}{n_2}} \geq 0\,,
\end{equation}
so the final sum over $\mathfrak{m}$ in \eqref{eq: Z_full} will be subject to the constraint above since the values of $\mathfrak{m}$ which satisfy this constraint will contribute poles. This constraint is the spindle version of the phenomenon observed in \cite{Closset:2015rna} where the summation over $\mathfrak{m}$ reduces to the closure of the cone dual to the cone in FI parameter space that defines the UV phase of the gauged linear sigma model under consideration.\footnote{Note that for $n_1=n_2=1$, $r =0$ \eqref{eq: b>0} reduces to $q_{\text{G}} \mathfrak{m} > 0$, as in \cite{Closset:2015rna}.}  The (JK)-residues of the integrand at these poles are 
\begin{equation}
  \underset{\widetilde{\sigma} = -n}{\text{Res}} I_0 = \frac{(-1)^{n}}{n!}\frac{1}{(\mathfrak{b}-n-1)!}\,, \qquad n = 0, 1, 2,\ldots, \mathfrak{b}-1\,,
\end{equation}
and with this in hand we can compute the partition function \eqref{eq: Z_full} to be 
\begin{align}
\begin{split} \label{eq: twist_partition_intermed}
    Z_{\mathbb{\Sigma}} & = \frac{1}{q_{\text{G}}} \sum_{\mathfrak{m} \in \mathbb{Z}} \left(\frac{L}{L_0} \right)^\mathfrak{b} \textrm{e}^{-\frac{\mathfrak{m}}{n_1n_2}\widetilde{\xi}}  \int \frac{-\i \, \textrm{d}\widetilde{\sigma}_0}{2\pi} \frac{\Gamma(\widetilde{\sigma}_0)}{\Gamma(\widetilde{\sigma}_0+ \mathfrak{b})} \\
    &= \frac{1}{q_{\text{G}}} \sum_{\mathfrak{m} \in \mathbb{Z}} \left(\frac{L}{L_0} \right)^\mathfrak{b} \textrm{e}^{-\frac{\mathfrak{m}}{n_1n_2}\widetilde{\xi}} \sum_{n=0}^{\mathfrak{b}-1} \frac{(-1)^{n}}{n!}\frac{1}{(\mathfrak{b}-n-1)!} \\
    & = \frac{1}{q_{\text{G}}} \left(\frac{L}{L_0} \right)  \sum_{\mathfrak{m} \in \mathbb{Z}: \mathfrak{b}=1}  \textrm{e}^{-\frac{\mathfrak{m}}{n_1n_2}\widetilde{\xi}} \,, 
    \end{split}
\end{align}
where in moving from the second to the third line we used
\begin{eqnarray}
\sum_{n=0}^{\mathfrak{b}-1} \frac{(-1)^{n}}{n!}\frac{1}{(\mathfrak{b}-n-1)!} =\delta_{\mathfrak{b},1}\,\,,
\end{eqnarray}
which arises as a simple consequence of the binomial theorem. 

We now need to find a way to perform the remaining summation in \eqref{eq: twist_partition_intermed}, in particular we need to take careful care of the constraint $\mathfrak{b} =1$. In order to do this, it is useful to write 
\begin{equation}
    \mathfrak{m} = n_1 n_2 \mathfrak{m}' + \mathfrak{l}\,, \qquad \mathfrak{m}' \in \mathbb{Z} \,, \quad \mathfrak{l} = 0,1, \ldots, n_1n_2-1 \,,
\end{equation}
and then the equation $\mathfrak{b} =1$ may be written as
\begin{equation} \label{eq: b=1_constraint}
    1= \mathfrak{b}(\mathfrak{m}) = q_{\text{G}} \mathfrak{m}' + \mathfrak{b}(\mathfrak{l})\,, 
\end{equation}
and thus the partition function \eqref{eq: twist_partition_intermed} becomes 
\begin{equation}
     Z_{\mathbb{\Sigma}} =  \frac{1}{q_{\text{G}}} \left(\frac{L}{L_0} \right)  \sum_{\mathfrak{m}' \in \mathbb{Z}: \mathfrak{b} =1} \sum_{\mathfrak{l} = 0}^{n_1n_2-1}   \textrm{e}^{-\left(\mathfrak{m}' + \frac{\mathfrak{l}}{n_1n_2}\right)\widetilde{\xi}} = \frac{1}{q_{\text{G}}} \left(\frac{L}{L_0} \right)  \sum_{\mathfrak{l} = 0}^{n_1n_2-1}   \textrm{e}^{-\left(\frac{1-\mathfrak{b}(\mathfrak{l})}{q_{\text{G}}} + \frac{\mathfrak{l}}{n_1n_2}\right)\widetilde{\xi}}\,,
\end{equation}
where we dealt with the summation over $\mathfrak{m}'$ using \eqref{eq: b=1_constraint}. We can now manipulate the right hand side of the equation above into the form 
\begin{equation}
    Z_{\mathbb{\Sigma}} =  \frac{1}{q_{\text{G}}} \left(\frac{L}{L_0} \right) \textrm{e}^{-\frac{r}{2} \chi \widetilde{\xi} \frac{1}{q_{\text{G}}}} \sum_{\mathfrak{l} = 0}^{n_1n_2-1}   \textrm{e}^{\left( - \frac{\llbracket -\frac{r}{2} - q_{\text{G}} a_1 \mathfrak{l} \rrbracket_{n_1}}{q_{\text{G}} n_1} - \frac{\llbracket - \frac{r}{2} + q_{\text{G}} a_2 \mathfrak{l} \rrbracket_{n_2}}{q_{\text{G}} n_2} \right)\widetilde{\xi}}\,,
\end{equation}
and now for computational viability (in a similar vein to \cite{Jeon:2025rfc}) we set $q_{\text{G}}=1$: 
\begin{equation}
    \left. Z_{\mathbb{\Sigma}} \right|_{q_{\text{G}} =1} =  \left(\frac{L}{L_0} \right) \textrm{e}^{-\frac{r}{2} \chi \widetilde{\xi}} \sum_{\mathfrak{l} = 0}^{n_1n_2-1}   \textrm{e}^{\left( - \frac{\llbracket -\frac{r}{2} -  a_1 \mathfrak{l} \rrbracket_{n_1}}{ n_1} - \frac{\llbracket - \frac{r}{2} + a_2 \mathfrak{l} \rrbracket_{n_2}}{ n_2} \right)\widetilde{\xi}}\,,
\end{equation}
then writing 
\begin{equation}
    \mathfrak{l} = n_1 \mathfrak{l}' + \mathfrak{o}\,, \qquad \mathfrak{l}' = 0,1,\ldots, n_2-1\,, \quad \mathfrak{o} = 0,1,\ldots n_1-1\,,
\end{equation}
we can perform the remaining summation explicitly as 
\begin{align}
\begin{split} \label{eq: partition_function_final}
     \left. Z_{\mathbb{\Sigma}} \right|_{q_{\text{G}} =1} &  =  \left(\frac{L}{L_0} \right) \textrm{e}^{-\frac{r}{2} \chi \widetilde{\xi}} \sum_{\mathfrak{l} = 0}^{n_1n_2-1}   \textrm{e}^{\left( - \frac{\llbracket -\frac{r}{2} -  a_1 \mathfrak{o} \rrbracket_{n_1}}{ n_1} - \frac{\llbracket - \frac{r}{2} + a_2 \mathfrak{o} + \mathfrak{l}' \rrbracket_{n_2}}{ n_2} \right)\widetilde{\xi}} \\
     & =  \left(\frac{L}{L_0} \right) \textrm{e}^{-\frac{r}{2} \chi \widetilde{\xi}} \sum_{\mathfrak{l}' = 0}^{n_2-1} \textrm{e}^{ - \frac{\mathfrak{l}'}{ n_2} \widetilde{\xi}} \sum_{\mathfrak{o}=0}^{n_1-1}  \textrm{e}^{ - \frac{\mathfrak{o}}{n_1} \widetilde{\xi}} \\
     & =  \left(\frac{L}{L_0} \right) \textrm{e}^{-\frac{r}{2} \chi \widetilde{\xi}} \left( \frac{1-\textrm{e}^{-\widetilde{\xi}}}{1-\textrm{e}^{-\widetilde{\xi}/n_1}} \right)\left( \frac{1-\textrm{e}^{-\widetilde{\xi}}}{1-\textrm{e}^{-\widetilde{\xi}/n_2}} \right)\,,
     \end{split}
\end{align}
completing the derivation of the partition function for an abelian theory defined on the spindle which preserves supersymmetry via the twist mechanism.

\subsection{Comments on the spindle partition function: twist and anti-twist cases}
The result \eqref{eq: partition_function_final} warrants several comments which we make below: 

\paragraph{Factorised form:} We observe \eqref{eq: partition_function_final} to have a factorised form between contributions arising from the North ($n_1$) and South Poles ($n_2$) of the spindle, much in the same way the anti-twisted partition function does as was shown in \cite{Jeon:2025rfc}. Such a structure is reminiscent of various related phenomena in both two \cite{Benini:2012ui, Doroud:2012xw} and three dimensions \cite{Pasquetti:2011fj} where the partition function factorises into contributions arising from the fixed points. This structure would be expected from the alternative scheme of Higgs branch localisation \cite{Benini:2012ui, Closset:2015rna}, where the localisation locus consists of vortex solutions on the fixed points of the background. Verifying our result either via direct Higgs branch localisation on the spindle, or via gluing the partition functions of singular discs \cite{Bah:2021hei, Kim:2025ziz}, would be interesting directions to pursue in the future.  

\paragraph{$S^2_{\Omega}$ limit:}We note that 
\begin{equation}
    \lim_{n_{1,2} \rightarrow 1} \left. Z_{\mathbb{\Sigma}} \right|_{q_{\text{G}} =1} = \left(\frac{L}{L_0} \right) \textrm{e}^{-r \widetilde{\xi}}\,, 
\end{equation}
and upon setting $r=0$, we recover the result for the partition function of the abelian Higgs model on $S^2_{\Omega}$ 
\begin{equation}
\left. Z_{S^2_{\Omega}} \right|_{q_{\text{G}} =1} = \frac{L}{L_0}\,, 
\end{equation}
which matches \cite{Closset:2015rna} up to the unimportant numerical factor of the ratio $L/L_0$, which is merely due to a different scaling of the integration in that work.

\paragraph{Vacuum degeneracy:} An alternative limit that one can take is the vacuum limit, i.e. the vanishing of both the FI and topological parameters
\begin{equation}
    \widetilde{\xi} \rightarrow 0\,.  
\end{equation}
and taking this limit directly upon \eqref{eq: partition_function_final} we find 
\begin{equation}
     \lim_{\widetilde{\xi} \rightarrow 0} \left. Z_{\mathbb{\Sigma}} \right|_{q_{\text{G}} =1} = \left(\frac{L}{L_0} \right) n_1n_2\,,
\end{equation}
which counts a degeneracy a $n_1n_2$ vacua of the theory. In a similar vein to \cite{Inglese:2023tyc, Jeon:2025rfc}, the reason for this degeneracy is the fact that the constraint \eqref{eq: b=1_constraint} of $\mathfrak{b}(\mathfrak{m}) =1$ has $n_1 n_2$ solutions for $\mathfrak{m} \in \mathbb{Z}$.

\paragraph{Contrast with the anti-twist result:} By making use of the parameter $\eta$, we can combine the main result of this work with our main result of \cite{Jeon:2025rfc} and write the partition function for both twist \& anti-twist in a unified way as 
\begin{align}
\begin{split} \label{eq: twists_comparison}
    \left. Z^{\eta}_{\mathbb{\Sigma}} \right|_{q_{\text{G}} =1} & = \frac{L}{L_0} \textrm{e}^{-\frac{r}{2} \left( \frac{2\pi \xi - \i \eta \theta}{n_1} +  \frac{2\pi \xi + \i  \theta}{n_2} \right)} \textrm{e}^{-\i (1+\eta) \frac{L}{L_0} \textrm{e}^{-2\pi \xi} \sin \theta} \\
  &  \qquad \times  \left( \frac{1-\textrm{e}^{-(2\pi \xi - \i \eta \theta)}}{1-\textrm{e}^{-(2\pi \xi - \i \eta \theta)/n_1}} \right)\left( \frac{1-\textrm{e}^{-(2\pi \xi + \i \theta)}}{1-\textrm{e}^{-(2\pi \xi + \i \theta)/n_2}} \right)\,.
  \end{split}
\end{align}
We note that while the twist case depends entirely on the ``holomorphic'' complexified parameter $\widetilde{\xi}$, the anti-twist case depends on $\widetilde{\xi}$ where the South Pole integer $n_2$ appears, but the ``anti-holomorphic'' parameter $\widetilde{\xi}^*$ in the presence of the North Pole integer $n_1$. This observation is somewhat consistent with the fact that the contribution to the one-loop determinants arising from the South Pole are the same for both twists, but the North pole contributions differ. 

Globally, we note that the twist case depends on the parameter $\widetilde{\xi}$, 
\begin{equation}
      \left. Z^{\eta=-1}_{\mathbb{\Sigma}} \right|_{q_{\text{G}} =1}  = Z(\widetilde{\xi})\,,
\end{equation} 
whereas the anti-twist case factorises into a product of functions of $\widetilde{\xi}, \widetilde{\xi}^*$
\begin{equation}
      \left. Z^{\eta=1}_{\mathbb{\Sigma}} \right|_{q_{\text{G}} =1} = Z_{n_2}(\widetilde{\xi} )Z_{n_1}(\widetilde{\xi}^*)\,,
\end{equation}
which one can see by writing the second exponential factor in \eqref{eq: twists_comparison} as
\begin{equation}
    \textrm{e}^{- \i (1+\eta)  \frac{L}{L_0} \textrm{e}^{-2\pi \xi} \sin \theta} = \textrm{e}^{\frac{L}{L_0} \frac{1+\eta}{2} \left( \textrm{e}^{-\widetilde{\xi}} - \textrm{e}^{-\widetilde{\xi}^*}\right) }\,.
\end{equation} 
The meaning of this, and whether there may be a connection with dualities such as mirror symmetry \cite{Hori:2000kt, Benini:2012ui} is another topic ripe for further exploration. 

\paragraph{Model independence:} In performing the calculation of localisation on spindles of both twists, we started with the spindle solutions of $D=5$ STU gauged supergravity \eqref{eq: STU_Lagrangian}. As discussed below \eqref{eq: Delta_z_spindle_data}, these solutions are parameterised by the flux and spindle data $(p_I, n_i, \eta)$ together with the constraint \eqref{eq: Q^R}. When constructing the two-dimensional $\mathcal{N} = (2,2)$ theory on $\mathbb{\Sigma}$, our intermediate expressions such as the one-loop determinants \eqref{eq: one_loop_twist} thus have dependence on all of  parameters of the data, however the final result of the partition function \eqref{eq: twists_comparison} only depends on the parameters $(n_i, \eta)$. This must be the case as the two-dimensional theory only comes equipped with the $U(1)_{\text{R}}$ gauge field, and is insensitive to the $U(1)^3$ structure of the STU model. Another way to see this model independence is that the expression for the partition function of the theory on the anti-twisted spindle is the same whether one starts with STU gauged supergravity, as in this work, or minimal gauged supergravity, as in \cite{Jeon:2025rfc}. 

We emphasise again that the starting point of the spindle solutions to $D=5$ STU gauged supergravity is a useful computational tool in order to solve the Killing spinor equations \eqref{kse} and not an essential step in performing localisation calculations on the spindle. In future, it would be interesting to revisit these calculations without choosing an explicit metric on $\mathbb{\Sigma}$, an approach which has been followed in the related works \cite{Inglese:2023wky, Inglese:2023tyc, Pittelli:2024ugf}.

\section{Conclusions and outlook} \label{sec: Conclusions}

In this work we have applied the technique of supersymmetric localisation in order to compute the partition functions of $\mathcal{N} = (2,2)$ theories defined on the spindle $\mathbb{\Sigma}$. The main feature of this work is the application to theories which admit $U(1)_{\text{R}}$ flux of both twist and anti-twist type, extending the earlier work \cite{Jeon:2025rfc} which only encompassed anti-twist. 

In order to achieve this new result, we began by considering the spindle solutions \cite{Ferrero:2021etw} of $D=5$ STU gauged supergravity, a theory which is suitably general as to allow solutions of both twists (which can be differentiated between via use of a single sign parameter $\eta$). We used these supergravity solutions to construct a 2d $\mathcal{N} = (2,2)$ theory on $\mathbb{\Sigma}$, in particular by providing the choice of metric and R-symmetry gauge field, as well as the functions $\mathcal{G},\mathcal{H}$ which appear in the 2d Killing spinor equations \eqref{kse}. We then considered a supersymmetric theory \eqref{eq: classical_action} whose field content consisted of an abelian vector multiplet and a charged chiral multiplet, together with the inclusion of FI and topological terms. We utilised the technique of supersymmetric localisation in order to compute the partition function of such a theory, with the final result being a unified presentation of the partition function for both twists, given in \eqref{eq: twists_comparison}.  

Many of the intermediate results reported in this work, such as the one-loop determinants, can be verified via comparison with \cite{Inglese:2023wky, Inglese:2023tyc}. In that work, the expressions for the $\mathcal{N}=(2,2)$ one-loop determinants on $\mathbb{\Sigma}$ were obtained via reduction of their 3d results on $\mathbb{\Sigma} \times S^1$ and agree with our results derived directly from the 2d perspective. As in \cite{Jeon:2025rfc}, a difference in our approaches is that we consider theories with entirely real fields and do not find the need to resort to the complex configurations considered in \cite{Inglese:2023tyc}. It is not clear whether this is because we fix a particular (real) choice of metric and R-symmetry gauge field on $\mathbb{\Sigma}$ and perhaps working with generic, complex fields would allow for more general application of our 2d results.

Further extensions include applications to abelian theory consisting of $N_f > 1$ chiral multiplets as has been performed in the $S^2$ cases \cite{Benini:2012ui, Doroud:2012xw, Gomis:2012wy, Closset:2015rna}, including application to cases where the GLSM defined on the spindle flows to a non-trivial SCFT in the IR, for which 
\begin{equation}
    \xi^{\text{UV}} = \xi \,, \qquad \sum_{I} q_I = 0\,,
\end{equation}
a condition which would also allow for application to non-linear sigma models defined on Calabi-Yau manifolds, such as the quintic threefold as studied in various localisation contexts \cite{Benini:2013nda, Closset:2015rna}.

Even more pressing would be the extension to study non-abelian gauge theories defined on the spindle. For the abelian theory in this work, we utilised the same residue prescription as that for the abelian Higgs model on $S^2_{\Omega}$ as considered in \cite{Closset:2015rna}. Despite this, we currently lack a deeper understanding of the integration contour and how the JK residue prescription applies more generally for the twisted spindle. The prescription has already been derived in various 2d contexts: the A-twisted  theories as studied in \cite{Closset:2015rna}, and in calculating elliptic genera \cite{Benini:2013nda, Benini:2013xpa}. Similar considerations also arise in 1d index computations \cite{Cordova:2014oxa, Hori:2014tda} and in topologically twisted theories in 3d \cite{Benini:2015noa}. The derivation of the residue prescription appears to rely on singularities arising from the zero modes of either the underlying theory or the BPS equations \cite{Closset:2019hyt}, modes whose existence may be highly sensitive to the structure of the spindle orbifold, see e.g. \cite{Bawane:2014uka} for a discussion of the role of the manifold (specifically, its Betti number) and the number of fermionic zero modes of the Laplacian operator. This issue requires further examination on the spindle and we hope to return to it in future work. It would also be of interest to investigate whether non-abelian localisation techniques \cite{Witten:1992xu} could be applied to theories on the spindle, potentially giving a localisation formula which circumvents the need for the JK contour. For some related recent work considering such a localisation scheme for A-twisted theories on $S^2$, see \cite{Leeb-Lundberg:2023jsj, Leeb-Lundberg:2025baz}.

These directions will allow for a more complete presentation of the spindle partition function and will allow for more general investigations into the nature of such an object. It would be interesting to examine whether this object would provide a tool to compute orbifold Gromov-Witten invariants \cite{Witten:1988xj, Gromov, chen2001orbifoldgromovwittentheory, Coates2013AMT, Cheong2014OrbifoldQT}, whether it satisfies supersymmetric dualities \cite{Hori:2000kt, Hori:2002fa, Hori:2006dk, Benini:2016qnm}, and whether the ``large-$N$'' limit of the index has an interesting holographic interpretations in terms of three dimensional gravity as the $\mathbb{\Sigma} \times S^1$ index was shown to in \cite{Colombo:2024mts}. Application to SCFT may allow one to construct ``2+2 = 4'' dualities \cite{Rastelli:2025nyn} involving the spindle, potentially via utilisation of our results for the twist case together with those of \cite{Pittelli:2024ugf} for the localisation calculation on $\mathbb{\Sigma} \times T^2$.  

Finally, we note that all of the discussions above are for $\mathcal{N} = (2,2)$ theories, and it would also be interesting to apply localisation techniques to theories defined on the spindle with different degrees of supersymmetry e.g. $\mathcal{N} = (0,2)$ theories.

\acknowledgments

We would like to thank Edoardo Colombo, Hee-Cheol Kim, Heeyeon Kim, Seok Kim and Enrico Turetta for discussions. This work is supported by the National Research Foundation of Korea under the grants: RS-2023-00243491 (H.K.), RS-2025-00518906 (N.K. and A.P.), RS-2025-25457100 (A.P.), 2021R1A2C2012350 (A.R.), and  partially  by the Asia Pacific Center for Theoretical Physics (APCTP). A.P. and A.R. respectively acknowledge the conferences ``Recent Progress in Field Theory and Gravity'' at Kyung Hee University and ``Aspects of Supersymmetric Quantum Field Theory 2025'' at Seoul National University, where preliminary results from this work were presented.  I.J. acknowledges the
warm hospitality of Heribertus Bayu Hartanto, Hyun-Sik Jeong and Patricio Salgado-Rebolledo during his visit to the APCTP.

\appendix

\section{Definitions and conventions} \label{app: Defs}
\subsection{Indices}
We set five-dimensional curved spacetime indices as $M,N,\ldots$ and local indices as $A,B,\ldots$. We denote the indices corresponding to $AdS_3$ direction and $\mathbb{\Sigma}$ direction as  
\begin{equation}
    M = (\alpha\,, \mu)\,, \qquad A = (a\,,m )\,.
\end{equation}

\subsection{Gamma matrix conventions}
In five dimensions with Lorentzian signature (mostly plus signature), a consistent choice of gamma matrix satisfies the following relations:
\begin{equation}\label{5dgammaLor}
\begin{array}{llll}
\Gamma_{A}^{\dagger}\=-A\Gamma_{A}A^{-1}\,,~&~A\=\Gamma_{0}\,,~&~A^{\dagger}\=A^{-1}\=-\Gamma_{0}\,,~&\\
\Gamma_{A}^{T}\={\cal{C}}\Gamma_{A}{\cal{C}}^{-1}\,,~&~{\cal{C}}^{T}\=-{\cal{C}}\,,~&~{\cal{C}}^{\dagger}={\cal{C}}^{-1}\,,~&\\
\Gamma_{A}^{*}\=-B\Gamma_{A}B^{-1}\,,~&~B^{T}\={\cal{C}}A^{-1}\,,~&~B^{\dagger}\=B^{-1}\,,~~~~B^{*}B\=-1\,.
\end{array}\end{equation}
This is followed by the property, regarding the charge conjugation matrix $\cC$, 
\begin{equation}
({\cal{C}}\Gamma_{A_1 A_2\cdots A_p})^{T}= - (-)^{p(p-1)/2}{\cal{C}}\Gamma_{A_1 A_2\cdots A_p}\,.
\end{equation} 
Due to the property of the charge conjugation matrix, we can use the spinor representation satisfying the symplectic-Majorana condition
\be\label{SMLor}
(\psi^i)^\dagger \gamma_0 \=  \varepsilon_{ij}(\psi^j)^T \mathcal{C}\,,\quad \Leftrightarrow \quad (\psi^i)^\ast = \varepsilon_{ij} B \psi^j\,, 
\ee
where~$i$ is an SU(2)$_\text{R}$ index
with $\varepsilon_{ij}$ being the SU(2) symplectic metric $\varepsilon_{12} = -\varepsilon_{21} = 1$.

\subsection{Comparison of gamma matrix conventions with other literature} \label{App: Gamma-matrix_comparison}

Our presentation of the Killing spinor equations \eqref{eq: Gravitino_KSE} and \eqref{eq: Gaugino_KSE} is consistent with our previous work \cite{Jeon:2025rfc}, which concerned the localisation of anti-twisted spindles arising from $D=5$ minimal gauged supergravity. In order to recover $D=5$ minimal gauged supergravity from $D=5$
STU supergravity, one sets all of the constants $q_I$ to be equal, i.e. \begin{equation} \label{eq: minimal_limit}
    q_I = q\,, \qquad I = 1,2,3\,.
\end{equation}

It is important to understand how our conventions relate to those of the works \cite{Ferrero:2021etw} ($D=5$ STU supergravity) and \cite{Ferrero:2020laf} ($D=5$ minimal gauged supergravity). In this section we will give the explicit relations between our conventions and those of \cite{Ferrero:2020laf, Ferrero:2021etw}, including a demonstration of how the Killing spinors of each work are related. 

\paragraph{Comparison with ``Supersymmetric spindles'' \cite{Ferrero:2021etw}}  Firstly, we note a difference in normalisation of the gauge fields $A_{M}^{(I)}$ between this work and \cite{Ferrero:2021etw}: as discussed below \eqref{eq: STU_R_symmetry_gf}, we want to fix our convention on the flux to match the convention in \cite{Jeon:2025rfc}. This means that relative to \cite{Ferrero:2021etw}, we have $2 A_{M, \text{here}}^{(I)} = A_{M, \text{there}}^{(I)}$ and thus the terms containing gauge fields in \eqref{eq: Gravitino_KSE}, \eqref{eq: Gaugino_KSE} are scaled by a factor of 2. We also perform constant gauge transformations in \eqref{eq: 5d_gauge_field} relative to \cite{Ferrero:2021etw}, for reasons relating to the $z$-dependence present in the Killing spinors. 

The second difference is in the signs of the terms involving gamma matrices. Terms with odd powers of gamma matrices have the opposite sign relative to \cite{Ferrero:2021etw}. This difference can be explained via our differences in conventions regarding the choice of gamma matrices. Starting from \eqref{eq: 5d_Gamma}, in order to match with \cite{Ferrero:2021etw}, we should instead make the choice 
\begin{equation} \label{eq: STU_gamma_choice}
        \rho_0 = -\i \sigma_2\,,\qquad \rho_1 = - \sigma_1 \,, \qquad \rho_2 = - \sigma_3 \,,\qquad  \gamma_1 = - \sigma_2\,, \qquad \gamma_2 = \sigma_1\,,
\end{equation}
    which brings the gamma matrix convention into the form
\begin{equation}
         \Gamma^a = - \beta^a \otimes \sigma_3\,, \quad a = \{ 0,1,2\} \,, \qquad \Gamma^3 = - 1 \otimes \sigma_1\,, \qquad \Gamma^4 = - 1 \otimes \sigma_2\,,
\end{equation}
where 
\begin{equation}
    \beta_0 = \i \sigma_2, \qquad \beta_1 = \sigma_1,  \qquad \beta_3 = \sigma_3\,,
\end{equation} 
as in \cite{Ferrero:2021etw}. We see that this convention matches \cite{Ferrero:2021etw} up to an overall minus sign on all of the gamma matrices.\footnote{Note that this also explains why we find the opposite sign for the AdS$_3$ Killing spinor relative to \cite{Ferrero:2021etw}, i.e. $\nabla_a \vartheta = -\frac{1}{2L} \rho_a \vartheta$.} This means that the choice \eqref{eq: STU_gamma_choice} brings us precisely to the form of the Killing spinor equations as presented in \cite{Ferrero:2021etw}, since odd products of gamma matrices carry opposite sign to ours in \eqref{eq: Gravitino_KSE}, \eqref{eq: Gaugino_KSE}. It will be important to note that to move between these choices of gamma matrices along the spindle directions we can utilise the transformations 
    \begin{equation}
        S \sigma_1 S^{-1} = - \sigma_2\,, \quad S \sigma_2 S^{-1} = \sigma_1\,, \qquad S = \begin{pmatrix}
1 & 0 \\
0 & -\i 
\end{pmatrix}\,,
    \end{equation}
and thus the spindle spinor solution to \eqref{eq: Gravitino_KSE}, \eqref{eq: Gaugino_KSE} is related to the one presented in \cite{Ferrero:2021etw} via 
\begin{equation}
\chi_{\text{here}} = S^{-1} \chi_{\text{there}} = H^{1/12} \begin{pmatrix}
1 & 0 \\
0 & \i 
\end{pmatrix} \begin{pmatrix}
\sin \frac{\alpha}{2} \\
- \cos \frac{\alpha}{2}
\end{pmatrix} = H^{1/12} \begin{pmatrix}
\sin \frac{\alpha}{2} \\
-\i \cos \frac{\alpha}{2}
\end{pmatrix}\,,
\end{equation}
which explains equation \eqref{eq: chi_kse} in the main text. Note that our spinor is missing the factor of $\textrm{e}^{\i z}$ which appears in \cite{Ferrero:2021etw}. This is due to the constant gauge transformation that we performed in \eqref{eq: 5d_R_gauge_field} relative to that work.

\paragraph{Comparison with ``D3-branes wrapped on a spindle'' \cite{Ferrero:2020laf}}

Upon taking the minimal limit \eqref{eq: minimal_limit} of \eqref{eq: Gravitino_KSE} we recover the Killing spinor equation as written in equation (2.9) of \cite{Jeon:2025rfc}. We note that this equation is the structurally the same as that of \cite{Ferrero:2020laf}, but different choices of gamma matrices between the two works make the relation between their respective Killing spinors somewhat subtle. We will now explicitly derive this relationship, and in doing so, we correct a typo in footnote 4 of \cite{Jeon:2025rfc}, where an incorrect mapping was given between the conventions of \cite{Jeon:2025rfc} and \cite{Ferrero:2020laf}.

In order to match with \cite{Ferrero:2020laf}, we should make the choice of 
\begin{equation} \label{eq: gamma_minimal}
    \rho_0 = -\i \sigma_2\,,\qquad \rho_1 = - \sigma_1 \,, \qquad \rho_2 = - \sigma_3 \,,\qquad \gamma_1 = \sigma_1 \,, \quad \gamma_2 = - \sigma_2\,,
\end{equation}
bringing the gamma matrix convention  into the form 
\begin{equation}
    \Gamma^a = \beta^a \otimes \sigma_3\,, \quad a = \{ 0,1,2\} \,, \qquad \Gamma^3 = 1 \otimes \sigma_2\,, \qquad \Gamma^4 =  1 \otimes \sigma_1\,,
\end{equation}
which matches the gamma matrix conventions of \cite{Ferrero:2020laf}. Along the spindle directions, the relation between our choice in \cite{Jeon:2025rfc} and \eqref{eq: gamma_minimal} is 
\begin{equation}
    T \sigma_1 T^{-1} = \sigma_1\,, \quad T \sigma_2 T^{-1} = - \sigma_2 \,, \qquad T = \begin{pmatrix}
0 & \i \\
\i & 0 
\end{pmatrix}\,,
\end{equation}
where one can check that $T$ maps the Killing spinors in \cite{Jeon:2025rfc} to those of \cite{Ferrero:2020laf} (up to overall normalisation), explicitly 
\begin{equation}
    T \chi = \begin{pmatrix}
0 & \i \\
\i & 0 
\end{pmatrix} \begin{pmatrix}
\sqrt{\frac{q_2(y)}{y}} \\
-\i \sqrt{\frac{q_1(y)}{y}}
\end{pmatrix}  =  \begin{pmatrix}
\sqrt{\frac{q_1(y)}{y}} \\
\i \sqrt{\frac{q_2(y)}{y}}
\end{pmatrix}\,,
\end{equation}
where $\chi$ is the Killing spinor given in Equation (2.13) of \cite{Jeon:2025rfc} and the r.h.s. is the Killing spinor in Equation (9) of \cite{Ferrero:2020laf}. 

\section{Minimal gauged supergravity as a limit of the STU model} \label{app: STU_to_minimal}

Here we demonstrate that the spindle solution of minimal gauged supergravity can be recovered via a suitable limit of the spindle solutions of STU gauged supergravity.
\paragraph{Metric:} We begin with the metric \eqref{eq: 5d_spindle_metric} and apply the limit \eqref{eq: stu_min_par_limt}, preserving the form of the metric
\begin{equation} \label{eq: min_stu_coords}
    \textrm{d}s^2 = L^2 H^{1/3} \bigg[\textrm{d}s^2_{\textrm{AdS}} + \frac{1}{4 P} \textrm{d}y^2 + \frac{P}{H} \textrm{d}z^2 \bigg]\,,
\end{equation}
except now instead of \eqref{eq: P-H_defs} one has
\begin{equation}
    h = y + q\,, \qquad H = h^3\,, \qquad P = H - y^2\,.
\end{equation}
The form of the line element \eqref{eq: min_stu_coords} is not immediately equivalent to that of the minimal gauged supergravity solution of \cite{Ferrero:2020laf}, where the metric was written as\footnote{We have included the additional conformal factor of $L^2$, as in \cite{Jeon:2025rfc}.}     
\begin{equation} \label{eq: minimal_metric}
  \textrm{d}s^2 = L^2 \left( \frac{4 \tilde y}{9} \textrm{d}s^2_{\textrm{AdS}} + \frac{\tilde y}{ \tilde{q}(\tilde{y})} \textrm{d} \tilde y^2 + \frac{\tilde{q}(\tilde{y})}{36 \tilde y^2} \textrm{d} \tilde z^2 \right) \,,
\end{equation}
where 
\begin{equation}
    \tilde{q}(\tilde{y}) = 4 \tilde{y}^3 - 9 \tilde{y}^2 + 6 a \tilde{y} - a^2\,,
\end{equation}
for more details concerning the constant parameter $a$, see e.g. \cite{Ferrero:2020laf, Jeon:2025rfc}. We observe that the line elements \eqref{eq: min_stu_coords} and \eqref{eq: minimal_metric} are equivalent via the following coordinate transformations
\begin{equation}
    \tilde{y} = \frac{9}{4} \left(y+q\right)\,, \quad \tilde{z} = 2 z\,,
\end{equation}
and the identification of parameters 
\begin{equation}
    a = \frac{27}{4} q\,.
\end{equation}

\paragraph{R-symmetry gauge field:}
The R-symmetry gauge fields in the STU model are given by \eqref{eq: 5d_gauge_field}, which in the minimal limit of \eqref{eq: stu_min_par_limt} become
\begin{eqnarray}
    A^{(I)} = \frac{1}{3} A^{\text{R}} = \frac{L}{2} \frac{y}{h} dz\,,\qquad \forall I=1,2,3\,.
\end{eqnarray} With the above data, we can read off the R-symmetry gauge field in minimal gauged SUGRA via the relation
\begin{eqnarray} \label{eq: STU_R_symmetry_gf}
     A^{\text{R}} = \frac{L}{2} \left( \frac{3y}{y+q} -2 \right) dz = \frac{L}{4} \left(1-\frac{a}{\tilde y}\right) d\tilde{z} \,.
\end{eqnarray} 
which is equivalent to the R-symmetry gauge field given in \cite{Ferrero:2020laf, Jeon:2025rfc}. Finally, we note that in the minimal limit of the STU model, one immediately recovers the trivial scalar profile of \cite{Ferrero:2020laf} through \eqref{eq: stu_to_minimal}.

\section{Graviphoton fluxes}

We have taken the metric \eqref{eq: metric_spindle} and the $U(1)_{\text{R}}$ gauge field \eqref{eq: A_2d} directly from the solution \eqref{eq: 5d_spindle_metric}, \eqref{eq: 5d_gauge_field} as discussed in \cite{Hosseini:2021fge, Boido:2021szx, Ferrero:2021etw}. However, in general one does not need the supergravity background to be on-shell in order to examine the supersymmetric field theory placed on a curved space \cite{Festuccia:2011ws, Closset:2014pda}. One may consider the off-shell supergravity background 
\begin{equation}
    (g_{\mu \nu}\,, A_{\mu} \,, C_{\mu}\,, \widetilde{C}_{\mu} )\,,
\end{equation}
where $C_{\mu}$, $\widetilde{C}_{\mu}$ form a complex graviphoton which couples to the conserved current for the central charge $Z, \widetilde{Z}$. One can use the functions $\mathcal{H}, \mathcal{G}$ which appear in the Killing spinor equations \eqref{kse} to determine the graviphoton field strengths via 
\begin{equation} \label{eq: graviphoton_field_strengths}
    - \i \epsilon^{\mu \nu} \partial_{\mu} C_{\nu} = \mathcal{H} + \i \mathcal{G}\,, \qquad - \i \epsilon^{\mu \nu} \partial_{\mu} \widetilde{C}_{\nu} = \mathcal{H} - \i \mathcal{G}\,,
\end{equation}
and one may also do this for the spindle. We first assume 
\begin{equation}
    C_{\mu} \textrm{d}x^\mu = C_{\mu}(y) \textrm{d}x^\mu =  C_z(y) \textrm{d} z\,, \qquad   \widetilde{C}_{\mu} \textrm{d}x^\mu = \widetilde{C}_{\mu}(y) \textrm{d}x^\mu =  \widetilde{C}_z(y) \textrm{d} z\,, 
\end{equation} 
which is natural as $\partial_z$ is a Killing direction on the spindle and we can remove the components along the $\textrm{d}y$ direction via a gauge transformation. We can then consider the graviphoton fluxes through the spindle 
\begin{equation} \label{eq: graviphoton_fluxes}
    \mathfrak{f}_C = \frac{1}{2\pi} \int_{\mathbb{\Sigma}} \textrm{d} C = \frac{\Delta z}{2\pi} \int_{y_1}^{y_2} \partial_y C_z \, \textrm{d}y\,, \qquad \mathfrak{f}_{\widetilde{C}} = \frac{1}{2\pi} \int_{\mathbb{\Sigma}} \textrm{d} \widetilde{C} = \frac{\Delta z}{2\pi} \int_{y_1}^{y_2} \partial_y \widetilde{C}_z \, \textrm{d}y\,,
\end{equation}
we can now proceed with computing the explicit expressions for the fluxes using \eqref{eq: graviphoton_field_strengths} 
\begin{equation}
    \partial_{y} C_{z} = \sqrt{g}_{\mathbb{\Sigma}} ( \i \mathcal{H} - \mathcal{G} )\,, \qquad \partial_{y} \widetilde{C}_{z} = \sqrt{g}_{\mathbb{\Sigma}} ( \i \mathcal{H} + \mathcal{G} )\,,
\end{equation}
we find 
\begin{equation}
    C_z = - 4 L ( y H^{-1/3} - H^{1/6}  )\,, \qquad \widetilde{C}_z = - 4 L (y H^{-1/3}+H^{1/6} )\,,
\end{equation}
and thus the graviphoton flux $\mathfrak{f}_C$ \eqref{eq: graviphoton_fluxes} becomes 
\begin{equation}
    \mathfrak{f}_{C} = \left. - 4L \frac{\Delta z}{2\pi} ( y H^{-1/3} - H^{1/6}  ) \right|^{y=y_2}_{y=y_1} = \left. 4L \frac{\Delta z}{2\pi} ( y H^{-1/3} - H^{1/6}  ) \right|_{y=y_1}\,,
\end{equation}
where we used the fact that $H(y_i)=y_i^2$ and $y_2>0$ in order to remove the contribution from the South Pole. The North Pole contribution now splits into two different values depending on whether the spindle is of the twist or anti-twist class. For twist we have 
\begin{equation}
    \mathfrak{f}^{\text{twist}}_C = 0\,,
\end{equation}
since $y_1>0$. For anti-twist we have 
\begin{equation}
    \mathfrak{f}^{\text{anti-twist}}_{C} = 8 L \frac{\Delta z}{2\pi} y_1^{1/3}\,,
\end{equation}
which can be written explicitly in terms of the flux and spindle data using \eqref{eq: y_spindle_data} and \eqref{eq: Delta_z_spindle_data}. 

Similarly, we can evaluate the graviphoton flux $\mathfrak{f}_{\widetilde{C}}$ 
\begin{equation}
    \mathfrak{f}_{\widetilde{C}} = \left.  - 4 L \frac{\Delta z}{2\pi} (y H^{-1/3}+H^{1/6} ) \right|^{y=y_2}_{y=y_1} \,,
\end{equation}
so again the expression for this flux can be separated into twist and anti-twist cases. 
\begin{equation}
     \mathfrak{f}_{\widetilde{C}}^{\text{twist}} = 8 L \frac{\Delta z}{2\pi}( y_1^{1/3} - y_2^{1/3} )\, ,
\end{equation}
and 
\begin{equation}
     \mathfrak{f}_{\widetilde{C}}^{\text{anti-twist}} =- 8 L \frac{\Delta z}{2\pi}y_2^{1/3}\, ,
\end{equation}
which can again be written in terms of the flux and spindle data using \eqref{eq: y_spindle_data} and \eqref{eq: Delta_z_spindle_data}. The fact that the fluxes $\mathfrak{f}_{C}, \mathfrak{f}_{\widetilde{C}}$ are non-vanishing tells us that the central charges $Z, \tilde{Z}$ will generically undergo a particular quantisation condition on the spindle.

\section{Multiplets and Supersymmetry transformations} \label{app: multiplets}
In this subsection we give details of the supersymmetry multiplets for our theory living on the spindle. We will consider a theory consisting of one vector and one chiral multiplet, giving the basic details of the various fields in these multiplets, as well as their supersymmetry transformations. 

\subsection{Vector multiplet} The vector multiplet consists of a vector, two real scalars, two Dirac spinors and an auxiliary real scalar 
\begin{equation} \label{eq: vector_multiplet}
    \text{Vector} : \{ \mathcal{A}_{\mu}, \sigma, \rho, \lambda, \widetilde{\lambda}, \widehat{D}\}\,,
\end{equation}
where the R-charge assignment is $(0,0,0,-1,1,0)$.

Let us denote the  equivariant supercharge $\qeq$ by combining the supercharge with the BRST charge for the $G=U(1)$ gauge symmetry as
\be\label{qeqDef}
\qeq \;\equiv \;Q + Q_{\text{brst}}\,.
\ee
where we denote the supercharge as  $Q \equiv Q_\epsilon + Q_{\widetilde{\epsilon}}$.  Under this equivariant supercharge, the transformations of  the vector multiplet fields are given by
\be\ba{lll}
\label{eq:deltaA}
\qeq \mathcal{A}_\mu &=& -\i \half (\widetilde\epsilon \gamma_\mu \lambda + \epsilon \gamma_\mu \widetilde\lambda) + \partial_\mu c\,, \\
\qeq \S  &=& -\half  (\widetilde\epsilon \lambda - \epsilon  \widetilde\lambda)\,,\\
\qeq \P  &=& -\i \half  (\widetilde\epsilon \gamma_3 \lambda + \epsilon \gamma_3 \widetilde\lambda)\,,\\
\qeq \lambda &=& \i \gamma_3 \epsilon \mathcal{F} -\widehat{D} \epsilon - \i \gamma^\mu \epsilon \,\partial_\mu \S  - \gamma_3 \gamma^\mu \epsilon \,\partial_\mu \P  + (\i \mathcal{H} - \mathcal{G}\gamma_3)\epsilon \,\S  +(\mathcal{H} \gamma_3 +\i \mathcal{G}) \epsilon\,\P 
\\
&=& \i \gamma_3 \epsilon \mathcal{F} -\widehat{D} \epsilon - \i \gamma^\mu D_\mu (\epsilon \S ) - \gamma_3 \gamma^\mu D_\mu (\epsilon \P  )\,,
\\
\qeq \widetilde\lambda &=& \i \gamma_3 \widetilde\epsilon \mathcal{F} +\widehat{D} \widetilde\epsilon +
 \i \gamma^\mu \widetilde\epsilon  \,\partial_\mu  \S  - \gamma_3 \gamma^\mu \widetilde\epsilon \,\partial_\mu   \P  -(\i \mathcal{H} +\mathcal{G}\gamma_3)\widetilde\epsilon  \,\S  + (\mathcal{H} \gamma_3 -\i \mathcal{G}) \widetilde\epsilon  \,\P  
 \\
 &=& \i \gamma_3 \widetilde\epsilon \mathcal{F} +\widehat{D} \widetilde\epsilon +
 \i \gamma^\mu D_\mu( \widetilde\epsilon  \S ) - \gamma_3 \gamma^\mu D_\mu ( \widetilde\epsilon  \P  )\,,
 \\
\qeq \widehat{D}&=& - \i\half  \widetilde\epsilon \gamma^\mu  D_\mu  \lambda + \i \half \epsilon \gamma^\mu  D_\mu  \widetilde\lambda   - \i\half  \bar\epsilon( \mathcal{H}+\i \mathcal{G} \gamma_3) \lambda + \i \half  \epsilon( \mathcal{H} -\i \mathcal{G}\gamma_3) \widetilde\lambda  

\\
&=& - \i\half  D_\mu (\widetilde\epsilon \gamma^\mu \lambda) + \i \half D_\mu (\epsilon \gamma^\mu  \widetilde\lambda  )
\,,
\ea\ee
where $\mathcal{F}= \half \epsilon^{\mu\nu}\mathcal{F}_{\mu\nu} = \epsilon^{\mu \nu} \partial_{\mu} \mathcal{A}_{\nu}$ and $c$ is an additional ghost field required for systematic treatment of gauge fixing using BRST quantisation \cite{GonzalezLezcano:2023cuh}.  

The transformation of the ghost $c$ is 
\be \label{eq: Ghost_transformation}
\qeq c \= -\Lambda^{\text{G}} +\Lambda^{\text{G}}_0\,,  \qquad \Lambda^{\text{G}} \equiv - \i \widetilde{\epsilon}\gamma^\mu \epsilon \mathcal{A}_\mu - \widetilde\epsilon \epsilon \sigma - \i \widetilde{\epsilon}\gamma_3 \epsilon \rho\,,
\ee
where  $\Lambda^{\text{G}}_0$ is the constant part of the field dependent parameter $\Lambda^{\text{G}}$.  
Here, the covariant derivative on each field is summarised as
\begin{equation}
    D_\mu \= \nabla_\mu -\i \widehat{q}_{\text{R}} A_{\mu}  - \i \widehat{q}_{\text{G}} \mathcal{A}_{\mu}\,,
\end{equation} 
where  $\widehat{q}_{\text{R}}$ is the $R$-charge and $\widehat{q}_{\text{G}}$ is the gauge charge of the field upon which $D_{\mu}$ acts.

As in \cite{GonzalezLezcano:2023cuh, Jeon:2025rfc}, we take the reality conditions upon the vector multiplet fields to be 
\begin{equation} \label{eq: reality_conditions_vector}
    \mathcal{A}^* = \mathcal{A}\,, \quad \sigma^* = \sigma\,, \quad \rho^* = \rho\,, \quad D^* = D\,,
\end{equation}
which allows for the functional integration of the Euclidean theory to be well-defined, at the price of giving up commutativity of $Q_{eq}$ and complex conjugation. 

\subsection{Chiral multiplet} The chiral multiplet consists of two complex scalars, two Dirac fermions and two auxiliary bosonic fields
\be \label{eq: chiral_multiplet}
\text{Chiral} : \{\phi \,, \wt{\phi}\,,\psi \,,\wt{\psi}\,,\mathfrak{F}\,,\wt{\mathfrak{F}}\}\,,
\ee
where the R-charge assignment is $(r, -r, r-1, -r+1,r-2, -r+2 )$. We also consider abelian gauge coupling via the $U(1)_{\text{G}}$ gauge field $\mathcal{A}$ in \eqref{eq: vector_multiplet}, with gauge charge assignment $ (1\,,-1\,,1\,,-1\,,1\,,-1)$.\footnote{In the main text we will take the charge of the chiral multiplet to be the more generic $q_{\text{G}}$. Such a choice modifies the supersymmetry transformations \eqref{deltachiral} by rescaling the coupling between vector multiplet and chiral multiplet fields by $q_{\text{G}}$.} The supersymmetry transformations of the chiral multiplet fields are given by 
\be\ba{l} \label{deltachiral}
 \qeq \phi = \widetilde\epsilon \psi+ \i c \phi \,,\\
\qeq \wt\phi = \epsilon \wt\psi - \i c\wt\phi \,,\\
\qeq \psi = \i \gamma^\mu \epsilon D_\mu \phi  
-\i \epsilon ( \S +\frac{\r}{2} \mathcal{H})\phi   +\gamma_3 \epsilon (\P +\frac{\r}{2}\mathcal{G})\phi +\widetilde\epsilon \mathfrak{F}
+ \i c \psi \,,\\
\qeq \wt\psi = \i \gamma^\mu \widetilde\epsilon D_\mu \wt\phi  
-\i \widetilde\epsilon ( \S +\frac{\r}{2} \mathcal{H})\wt\phi   -\gamma_3 \widetilde\epsilon ( \P +\frac{\r}{2}\mathcal{G})\wt\phi +\epsilon \wt{\mathfrak{F}} - \i c \wt{\psi}\,,
\\
\qeq \mathfrak{F} =    \i \epsilon\gamma^\mu D_\mu \psi  +\i  \epsilon\psi ( \S +\frac{\r}{2}\mathcal{H})+\epsilon\gamma_3 \psi ( \P +\frac{\r}{2}\mathcal{G})   -\i  \epsilon\lambda \phi + \i  c \mathfrak{F}\,,
\\
\qeq \wt{\mathfrak{F}} =    \i \widetilde\epsilon\gamma^\mu D_\mu \wt\psi  +\i  \widetilde\epsilon\wt\psi (  \S +\frac{\r}{2}\mathcal{H} )-\wt\epsilon\gamma_3 \wt\psi ( \P +\frac{\r}{2}\mathcal{G})     +\i  \widetilde\epsilon \wt\lambda \,\wt\phi- \i   c \wt{\mathfrak{F}} \,,
\ea\ee 
and we choose reality conditions as in \cite{GonzalezLezcano:2023cuh, Jeon:2025rfc}  of 
\begin{equation} \label{eq: chiral_bosonic_reality}
    \phi^\ast = \widetilde{\phi}\,,\qquad \mathfrak{F}^\ast = \widetilde{\mathfrak{F}}\,,
\end{equation}
which together with \eqref{eq: reality_conditions_vector} allows us to compute 
\begin{align} \label{RealityofQPsi}
\begin{split}
    ( \qeq (\epsilon \psi ))^\dagger &=  \widetilde{\epsilon}\epsilon \widetilde{\mathfrak{F}} - \i \widetilde\epsilon \gamma^\mu \widetilde\epsilon D_\mu \widetilde\phi  +\widetilde\epsilon \gamma_3 \widetilde\epsilon \left( \rho + \frac{r}{2}\mathcal{G}\right)\widetilde\phi - \i c^\ast (\epsilon \psi)^\ast 
    \\
        &=\qeq (\widetilde\epsilon \widetilde\psi ) -2 \left[ \i \widetilde\epsilon \gamma^\mu \widetilde\epsilon D_\mu   - \widetilde{\epsilon}\gamma_3\widetilde\epsilon \left( \rho +\frac{r}{2}\mathcal{G} \right)\right]\widetilde\phi + \i c (\widetilde\epsilon \widetilde\psi) -\i c^\ast (\epsilon \psi)^\ast\,,
        \\
 ( \qeq (\widetilde\epsilon \widetilde\psi ))^\dagger &=    
  \epsilon \widetilde{\epsilon} \mathfrak{F} - \i \epsilon \gamma^\mu \epsilon D_\mu \phi - \epsilon \gamma_3 \epsilon \left( \rho + \frac{r}{2}\mathcal{G} \right)\phi + \i c^\ast (\widetilde\epsilon \widetilde\psi )^\ast\,
 \\
  &=\qeq (\epsilon \psi)- 2 \left[ \i \epsilon \gamma^\mu \epsilon D_\mu   + {\epsilon}\gamma_3\epsilon \left( \rho +\frac{r}{2}\mathcal{G} \right)\right]\phi    - \i c (\epsilon \psi) + \i c^\ast (\widetilde\epsilon \widetilde\psi)^\ast\,.
\end{split}
\end{align} 
where we also used $\mathcal{G}=\mathcal{G}^*$, the reality properties of the bilinears in \eqref{eq: spinor_bilinears}, and the reality of the components of the Killing vector \eqref{eq: Killing_z}.  
We do not specify the reality condition for $c$, $\epsilon \psi$ and $\widetilde{\epsilon}\wt{\psi}$ as these will not be relevant for the computation.

\section{Comparison of \texorpdfstring{$S^2_{\Omega}$}{} limits of one-loop determinants} \label{app: one_loop_comparisons}

A natural limit of the chiral multiplet one-loop determinant on the twisted spindle is to take $n_{1,2} \rightarrow 1$, in which case one would expect to recover the chiral multiplet one-loop determinant on the $\Omega$-deformed sphere $S^2_{\Omega}$ as studied in \cite{Closset:2015rna}. Here we take this limit, demonstrate that our one-loop determinant does indeed recover the $S^2_{\Omega}$ result, and provide a comparison of notation with the related work \cite{Inglese:2023tyc}.

First we start with the expression for the chiral multiplet one-loop determinant given in equation (4.6) of \cite{Closset:2015rna} (which has $q_{\text{G}} =1$). We consider the specific case of $\mathbf{G} = U(1)$, and also set $m^F = 0$. The one-loop determinant is 
\begin{equation}
    Z^{\Phi}_k(\hat{\sigma}, \bm{\epsilon}_{\Omega}) = \bm{\epsilon}_{\Omega}^{r-k-1} \frac{\Gamma(\bm{\epsilon}_{\Omega}^{-1} \hat{\sigma} + \frac{r-k}{2})}{\Gamma(\bm{\epsilon}_{\Omega}^{-1} \hat{\sigma} - \frac{r-k}{2} +1)}\,,
\end{equation}
then writing $k=\mathfrak{m}$ and $\mathfrak{r} = r - \mathfrak{m}$ we obtain 
\begin{equation} \label{eq: Z_one_loop_chiral_CCP}
    Z^{\Phi}_k(\hat{\sigma}, \bm{\epsilon}_{\Omega}) = \bm{\epsilon}_{\Omega}^{\mathfrak{r}-1} \frac{\Gamma(\bm{\epsilon}_{\Omega}^{-1} \hat{\sigma} + \frac{\mathfrak{r}}{2})}{\Gamma(\bm{\epsilon}_{\Omega}^{-1} \hat{\sigma} + 1 - \frac{\mathfrak{r}}{2})} = \bm{\epsilon}_{\Omega}^{\mathfrak{r}-1} \left(1 - \frac{\mathfrak{r}}{2} + \frac{\hat{\sigma}}{\bm{\epsilon}_{\Omega}}\right)_{\mathfrak{r}-1}\,,
\end{equation}
where we used the definition of the Pochhammer symbol 
\begin{equation}
    (z)_m = \prod_{\ell =0}^{m-1} (\ell + z) = \Gamma(z+m)/\Gamma(z)\,,
\end{equation}
in writing the final equality of \eqref{eq: Z_one_loop_chiral_CCP}. 

We can compare this result with the limit of our expression for the chiral one-loop determinant on the twisted spindle \eqref{eq: Z_1-loop_twist}. We can take the limit of this quantity (again with $q_{\text{G}} =1$ for the sake of comparison) as
\begin{equation}
    \lim_{n_{1,2}\rightarrow 1} \left(\frac{L}{L_0} \right)^{\mathfrak{b}} \frac{\Gamma\left(\frac{1-\mathfrak{b} - \mathfrak{c}}{2}  + \frac{r}{4} \chi_- + q_{\text{G}} \gamma_{\text{G}}\right)}{\Gamma \left( \frac{\mathfrak{b} - \mathfrak{c} +1 }{2} +\frac{r}{4} \chi_- + q_{\text{G}} \gamma_{\text{G}} \right)} = \left(\frac{L}{L_0} \right)^{1-r+\mathfrak{m}} \frac{\Gamma\left(\frac{r-\mathfrak{m}}{2}   + \gamma_{\text{G}}\right)}{\Gamma \left( 1 - \frac{r - \mathfrak{m}}{2} + \gamma_{\text{G}} \right)}\,,
\end{equation}
which allows us to make the natural identifications of 
\begin{equation} \label{eq: identifications_with_CCP}
    \bm{\epsilon}_{\Omega} = \frac{L_0}{L}\,, \qquad \hat{\sigma} = \bm{\epsilon}_{\Omega} \gamma_{\text{G}} =  \Lambda^{\text{G}} + \frac{1}{2} \frac{L_0}{L}  \left(\mathfrak{m}_1 + \mathfrak{m}_2 \right) \,,
\end{equation}
and agrees with the definition of $\hat{\sigma}$ as given in \cite{Closset:2015rna} of 
\begin{equation}
    \hat{\sigma} = \frac{1}{2}(\sigma_N + \sigma_S)\,,
\end{equation}
since the square of the supercharge in that work is given by the algebra 
\begin{equation} \label{eq: CCP_supercharge_algebra}
    \{ \delta, \widetilde{\delta}\} = 2 \bm{\epsilon}_{\Omega} \mathcal{L}^{(a)}_{V} + 2\i \sigma\,, 
\end{equation}
which when compared with \eqref{eq: Q^2} provides further justification for the identifications \eqref{eq: identifications_with_CCP}.\footnote{Note that in \cite{Closset:2015rna} $V = \partial_\phi = (L/L_0)v$ and that schematically we also observe that $\frac{1}{2}(\left. \Lambda^{\text{G}} \right|_{\mathcal{U}_1} + \left. \Lambda^{\text{G}} \right|_{\mathcal{U}_2} + \left. \Lambda^{\text{R}} \right|_{\mathcal{U}_1}  + \left. \Lambda^{\text{R}} \right|_{\mathcal{U}_2}) =  \Lambda^{\text{G}} + \frac{1}{2} \frac{L_0}{L}  \left(\mathfrak{m}_1 + \mathfrak{m}_2 \right)$ for $n_{1,2}=1$ in the twist case. One has to be careful with the sphere limit here since $\Lambda^{\text{G}}$ is singular in such a limit. The factor of 2 present in \eqref{eq: CCP_supercharge_algebra} comes from the differences in definitions of the SUSY transformations (\cite{Closset:2015rna} has a extra factor of $\sqrt{2}$ for both $\delta, \widetilde{\delta}$ which accounts for the overall factor).}  

Finally, we note that \cite{Inglese:2023tyc} also recovers the chiral multiplet one-loop determinant of \cite{Closset:2015rna}, assuming a specific choice of Killing vector normalisation. The limit given in \cite{Inglese:2023tyc} is 
\begin{equation}
    Z^{\text{CM},(+1)}_{S^2} = \omega^{\mathfrak{r}-1} \left( 1 - \frac{\mathfrak{r}}{2} - \omega^{-1} \gamma_{\text{G}}^{\text{2d}} \right)_{\mathfrak{r}-1}\,,
\end{equation}
which identifies $\omega = \bm{\epsilon}_{\Omega}$ instead. Using the dictionary given in equation (5.71) of \cite{Jeon:2025rfc}, we see that this is only equivalent to our \eqref{eq: identifications_with_CCP} when $k_{0}^{\text{(IMP)}}=1$. This was argued to be a fine choice for the overall Killing vector normalisation in \cite{Inglese:2023tyc}. The advantage of our generic $L_0$ is that it allows us to work with arbitrary normalisation instead.
 
\bibliographystyle{JHEP}
\bibliography{spindle.bib}
\end{document}